\documentclass[runningheads]{llncs}
\usepackage{graphicx}

\usepackage{tikz}
\usepackage{comment}
\usepackage{amsmath,amssymb} 
\usepackage{color}
\usepackage{orcidlink}

\usepackage[accsupp]{axessibility}  

\usepackage{graphicx}
\usepackage{amsmath}
\usepackage{amssymb}
\usepackage{booktabs}
\usepackage{wrapfig}

\usepackage{caption}
\usepackage{subcaption}

\input{header.tex}

\usepackage[capitalize]{cleveref}
\crefname{section}{Sec.}{Secs.}
\Crefname{section}{Section}{Sections}
\Crefname{table}{Table}{Tables}
\crefname{table}{Tab.}{Tabs.}

\begin{document}

\pagestyle{headings}
\mainmatter

\title{A Repulsive Force Unit for Garment Collision Handling in Neural Networks} 

\titlerunning{A Repulsive Force Unit for Garment Collision Handling in Neural Networks}
%
\author{Qingyang Tan\inst{1}\orcidlink{0000-0002-9269-5289}
\and
Yi Zhou\inst{2}
\and
Tuanfeng Wang\inst{2}
\and
Duygu Ceylan\inst{2}
\and
Xin Sun\inst{2}
\and
Dinesh Manocha\inst{1}
}
\authorrunning{Tan et al.}
%
\institute{Department of Computer Science, University of Maryland at College Park 
\email{\{qytan,dmanocha\}@umd.edu}\\
\and
Adobe Research
\email{\{yizho,yangtwan,ceylan,xinsun\}@adobe.com}
\url{https://gamma.umd.edu/researchdirections/mlphysics/refu/}}

\maketitle

\begin{abstract}
    Despite recent success, deep learning-based methods for predicting 3D garment deformation under body motion  suffer from interpenetration problems between the garment and the body. To address this problem, we propose a novel collision handling neural network layer called Repulsive Force Unit (ReFU). Based on the signed distance function (SDF) of the underlying body and the current garment vertex positions, ReFU predicts the per-vertex offsets that push any interpenetrating vertex to a collision-free configuration while preserving the fine geometric details. We show that ReFU is differentiable with trainable parameters and can be integrated into different network backbones that predict 3D garment deformations. Our experiments show that ReFU significantly reduces the number of collisions between the body and the garment and better preserves geometric details compared to  prior methods based on collision loss or post-processing optimization.
\end{abstract}

\begin{figure}

		\centering
		\includegraphics[width=0.98\linewidth]{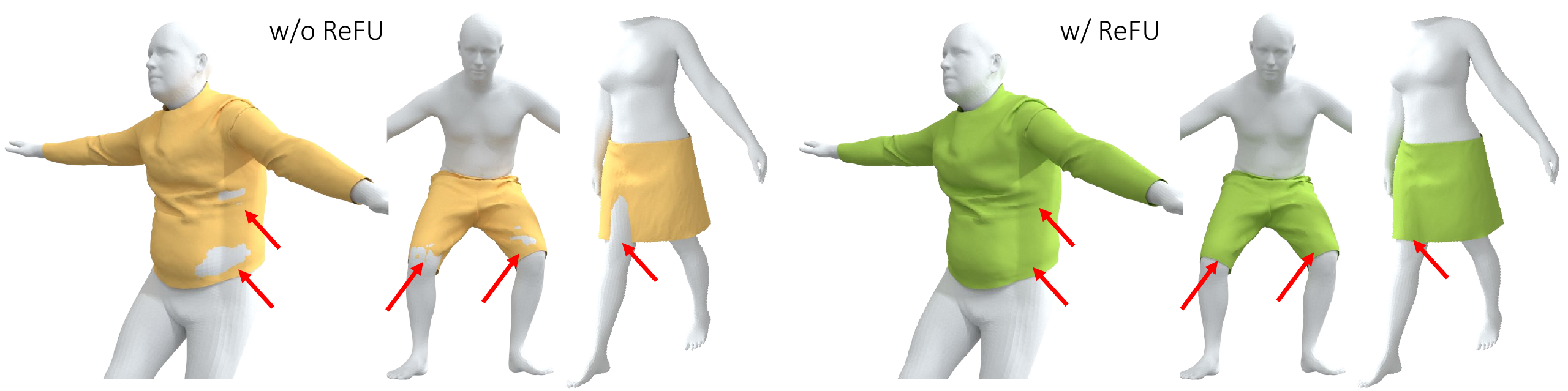}

		\caption{
		\label{fig:teaser}Collisions solved by applying ReFU in garment prediction neural networks. Our approach reduces the number of artifacts. }
		
\end{figure}

\section{Introduction}

Predicting how a 3D garment deforms in response to the underlying 3D body motion is essential for many applications, including realistically dressed human body reconstruction~\cite{bhatnagar2019multi}, interactive garment design \cite{garmentdesign_Wang_SA18}, virtual try-on 
\cite{santesteban2019learning}, and robotics control \cite{tan2020realtime}. 
To generate accurate cloth deformations, most techniques are based on physically-based simulation (PBS). 
Common physically-based models include the mass-spring system~\cite{baraff1998large,choi2005stable}, the finite element approach~\cite{Narain:2012:AAR,larson2013finite}, the thin-shell model \cite{grinspun2003discrete}, etc.
However, these methods tend to be computationally intensive since they typically involve solving large linear systems and handling collisions.   In particular, robust collision handling based on collision detection and response computation is a critical component of cloth or garment simulation. Even a single missed collision can considerably affect the accuracy of the overall simulator~\cite{Bridson02,GovindarajuVR05}. The most accurate physically-based simulators run at $0.5$ seconds per frame on commodity GPUs~\cite{tang-siga18}, where collision handling can take 
 50-80$\%$ of total simulation time. As a result, these simulators are unable to provide real-time performance for interactive applications such as gaming and virtual try-on.

Machine learning methods provide a promising direction to dramatically reduce the computational cost of cloth simulators. Hence, in recent years, various neural network methods have been proposed to predict 3D cloth deformations. However, a common setback of such methods is the lack of efficient handling of collisions between the garments and the body surface as shown in the yellow garments in~\prettyref{fig:teaser}. In our experiments (\prettyref{sec:results}), we observe that only $49\%$ of garments predicted from TailorNet~\cite{patel2020tailornet}, a state-of-the-art neural network based 3D garment prediction method, are collision-free. For some tight clothes like shirts, only $12\%$ of models are collision-free. Thus, the resulting state of the cloth mesh can collide with the body mesh, which affects the reliability and usefulness of these methods for many applications related to rendering, simulation, and animation~\cite{Brochu12,Provot97,Wang14}. As a result, it is important to design learning methods that can significantly reduce or eliminate such collisions.

One option to address the body-cloth collision problem is to perform post processing optimizations~\cite{guan2012drape}. However, these optimization approaches can take considerable CPU time (around 0.6-0.8s per frame), which can be too expensive for interactive applications. A more common practice is to apply specialized collision loss functions during training~\cite{bertiche2020pbns,bertiche2021deepsd,bertiche2020cloth3d,gundogdu2019garnet,santesteban2019learning}. However, this only provides a soft constraint to avoid collisions for network training, and the network still cannot handle the penetrated vertices when collisions happen during inference.

\noindent{\bf Main Results:}
To let the network learn to solve the collisions through inference, we propose a novel neural network layer called Repulsive Force Unit (ReFU). ReFU is fully differentiable and can be plugged into different garment prediction backbone networks, trained either through fine-tuning or from scratch.

Our design of ReFU is inspired by physically-based simulators~\cite{macklin2020local,teng2014simulating,fisher2001fast}, which use
 a scheme that collects repulsive forces, friction forces, and adhesion forces as part of time integration. Our goal is to design a learning scheme that can model the effects of repulsive forces and can easily cope with existing 3D garment prediction networks. We compute the force based on the implicit field of the body geometry to quickly detect the set of penetrated garment vertices and the repulsive direction. The repulsive strength is predicted by the neural network inside the ReFU layer. Instead of simply pushing the problematic garment vertices to the body surface, ReFU applies a flexible offset to move them. This  improves the overall collision handling performance, avoids additional Edge-Edge (EE) collisions that normally cannot be detected by the signed distance of the vertices, and overcomes the artifacts in the estimated implicit functions of the human body. To achieve real-time performance for the whole garment prediction system, we leverage the power of neural implicit surface representation and use a neural network to quickly estimate the approximate Signed Distance Function (SDF) of the human body under different poses~\cite{gropp2020implicit}.

To evaluate ReFU with different backbones, we train it with TailorNet \cite{patel2020tailornet}, the state-of-the-art 3D garment prediction network, and a 3D mesh convolutional neural network~\cite{zhou2020fully}. Our experiments show that backbone networks trained with ReFU can significantly reduce the number of body-cloth collisions while achieving real-time performance for the whole garment prediction system during testing. The ReFU layer and SDF network will only add 2 milliseconds of inference time to the overall system. Overall, our method achieves much better results in terms of the number of interpenetrating vertices, the number of collision-free 3D garment models, and the reconstruction error of the generated garments. Compared to prior learning-based methods, the use of ReFU results in garment meshes with fewer artifacts and higher visual quality.

\section{Related Work}

\subsection{Physically-Based Collision Handling}
 Many accurate techniques have been proposed for collision detection between discrete time intervals using continuous methods (CCD), which reduce to solving cubic polynomials for linear interpolating motion~\cite{Brochu12,Provot97,Tang14,Wang14}. There is extensive literature on collision response computation based on constraint solvers~\cite{Otaduy09}, impulse responses~\cite{Bridson02},  and impact zone methods~\cite{Provot97,Harmon08}. These methods have been used to develop robust physics-based simulators that are widely used in animation and VR applications. Current learning-based methods are significantly faster than these physics-based simulators but cannot offer the same level of accuracy or robustness. Other techniques for collision handling are based on optimization-based refinement~\cite{guan2012drape} but have computational overhead.
 
\subsection{Cloth Prediction using Machine Learning}
Many fast techniques have been used to predict cloth deformation in 3D.
These include  simple linear models such as \cite{guan2012drape,de2010stable} and motion graph methods~\cite{Kim:2013:NEP}. More recent works use neural networks~\cite{bertiche2021deepsd,bertiche2020cloth3d,gundogdu2019garnet,santesteban2019learning,santesteban2021self,patel2020tailornet,bertiche2020pbns}. Many of these learning methods have been designed for SMPL-based~\cite{loper2015smpl} parametric obstacle models, including human shapes.  Other methods are designed for general triangle mesh-based obstacles~\cite{holden2019subspace}. The GarNet network architecture~\cite{gundogdu2019garnet} can  predict the cloth deformation from the target posture with DQS pre-processing. However, these learning methods do not explicitly account for cloth-obstacle collisions. 

\subsection{Learning-Based Collision Handling}
Several techniques have been proposed to handle collisions in machine learning methods. Recently Tan et. al.~\cite{tan2020lcollision} estimated a collision-free subspace for 3D human models, but it is not practical to compute a similar subspace for deforming garments as they have a lot of degrees of freedom. Its performance can be improved using active learning~\cite{tan2021active}, but this approach is limited by the use of numerical optimization algorithm. These methods can't be combined with general garment prediction backbone networks. 
Many recent works~\cite{bertiche2021deepsd,gundogdu2019garnet,bertiche2020pbns,bertiche2020cloth3d} use collision loss to penalize penetrated garment-body pairs during training. However, these methods have no component or feature in the network that can resolve these penetrations during inference.

\begin{wrapfigure}{l}{0.18\textwidth}
\centering
\includegraphics[width=0.17\textwidth]{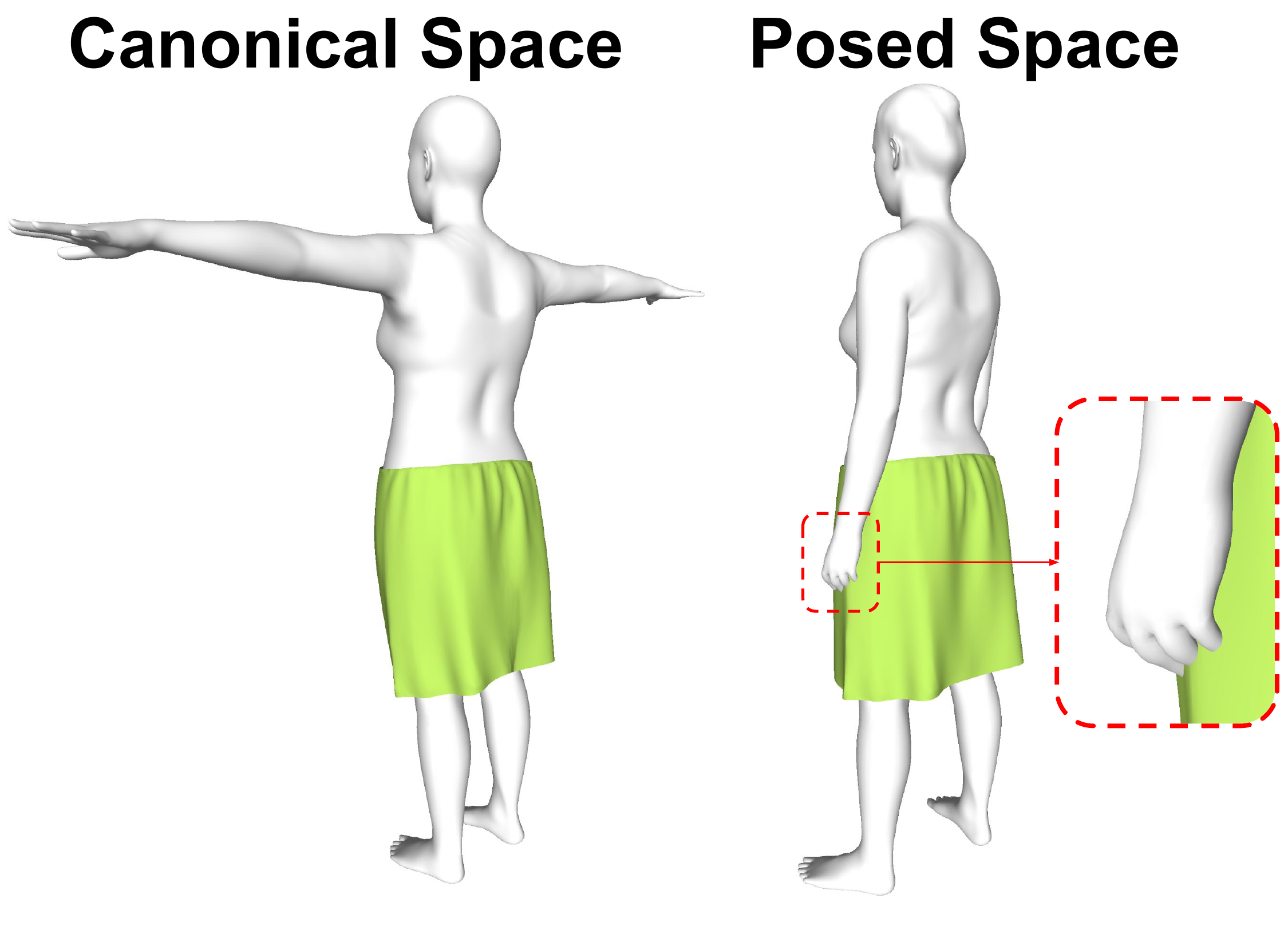}
\end{wrapfigure}
Recently, Santesteban et al.~\cite{santesteban2021self} proposed a self-supervised collision handling method, but it requires strict additional restrictions on the training data, i.e., that both garments in the global space and the canonical space must be collision-free. However, most public garment datasets cannot satisfy this restriction. We show an example on the left from TailorNet dataset which is collision-free in canonical space, but not in the posed space.

To solve the collision problems for the testing set, a simple approach is to perform post-processing on the predicted cloth by detecting the vertices inside the human body and moving them directly  to the nearest point on the body surface~\cite{santesteban2019learning,holden2019subspace}. However, there are two issues with these approaches: first, computing nearest points on the body surface is time-consuming; second, simply moving the garment vertices may generate abrupt cloth movements and lose some properties of the original garment. 

\begin{figure*}[h]
    \centering
    \includegraphics[width=0.98\textwidth]{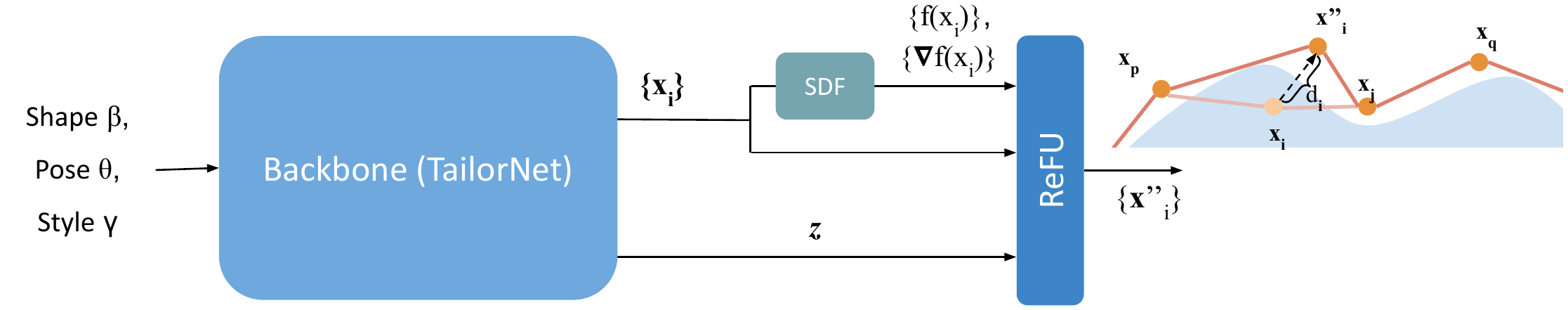}

    \caption{\label{fig:pipeline}A garment inference model with ReFU: Given body shape $\vbeta$, pose $\vtheta$, and garment style $\vgamma$ parameters, the backbone network is used generate a deformed garment with potential body-cloth collisions. Our ReFU layer is attached to the backbone and processes every point $\{\x_i\}_{i=1}^N$ in the garment. This layer first checks whether the point is inside the human body or not based on the SDF value $f(\x_i)$. If it is outside, ReFU directly outputs the point. Otherwise, ReFU will apply a repulsive force along the direction of the gradient of the SDF $\nabla f(\x_i)$ with a predicted amount of movement $d_i$. Finally, we collect all the vertices passed through the ReFU layer and obtain a 3D garment mesh with fewer collisions.}

\end{figure*}

\section{Collision Handling using ReFU}\label{sec:method}
In this section, we will first explain the formulation of the Repulsive Force Unit (ReFU) (\prettyref{sec:refu}) then describe how to apply and train it in the garment prediction backbone network as an additional network layer using the example of TailorNet~\cite{patel2020tailornet} (\prettyref{sec:finetune}). Finally, we will present how we train a neural network to quickly estimate the SDF for the human body (\prettyref{sec:SDF}). Our overall learning-based pipeline is shown in \prettyref{fig:pipeline}. With a neural network-based SDF approximation, our system can achieve real-time performance for garment prediction given the body pose and shape parameters.

\subsection{ReFU: Repulsive Force Unit}\label{sec:refu}
The goal of body-cloth collision handling is to find the penetrating garment vertices and move them to the proper positions to resolve the collision while preserving original wrinkles and other details on the garments. Let  $N_c$ be the number of vertices that need to be moved. The degrees of freedom of this set of vertices result in a large solution space of dimension $3N_c$. Our goal is to reduce the dimension of the solution space. Inspired by prior work in physically-based simulation~\cite{fisher2001fast,teng2014simulating}, we design the Repulsive Force Unit (ReFU) to move the vertices only along a \emph{repulsion} direction, which is toward the closest point on the body surface. Our formulation of ReFU can be seen as applying a virtual repulsive force to move the penetrated vertex outside the human body.

To find the repulsion direction, ReFU uses the implicit representation of the body, i.e., the signed distance function (SDF).
Given a query point $\x$, the SDF function $f$ returns its distance to the  closest point on the corresponding surface, and its sign is associated with whether the point is inside (negative) or outside (positive) the surface:

\begin{align}
    f(\x) = s,\quad \x\in \mathbb{R}^3, s\in \mathbb{R}.
\end{align}

The zero-level set of $f(\x)$ indicates the surface.

Given the nature of SDF, we can quickly determine whether a vertex $\x_i$ on the garment mesh is inside or outside the body. For $\x_i$ with negative SDF value, the gradient of the SDF at $\x_i$ is pointing towards the nearest point on the surface along the normal direction. 
Thus, we formulate ReFU as

\begin{align}\label{eq:refu}
    \ReFU(\x_i) &= \left\{
        \begin{array}{ll}
            \x_i - d_i \widehat{\nabla_{\x_i} f(\x_i)}, &f(\x_i)<0;\\
            \x_i, & otherwise,
        \end{array}
    \right.
\end{align}

where $d_i$ is a predicted offset scalar indicating the amount of movement, and $\widehat{\nabla_{\x} f(\cdot)}$ is the normalized gradient of $f$ at $\x$, indicating the direction of movement, calculated as below:

\begin{align}\label{eq:gradient}
    \widehat{\nabla_{\x} f(\x)} &= \frac{\nabla_{\x} f(\x)}{\|\nabla_{\x} f(\x)\|_2}
\end{align}

Although the gradient of accurate SDF should be a unit vector \cite{crandall1983viscosity,gropp2020implicit}, the approximated SDF by neural networks~\cite{ma2020neural,sitzmann2020metasdf} might not strictly satisfy this property. Thus, we need to normalize the gradient in practice.

\subsubsection{A Learned Moving Offset}\label{sec:scale}
\begin{figure}[h]
    
    \centering
    \includegraphics[width=0.65\textwidth]{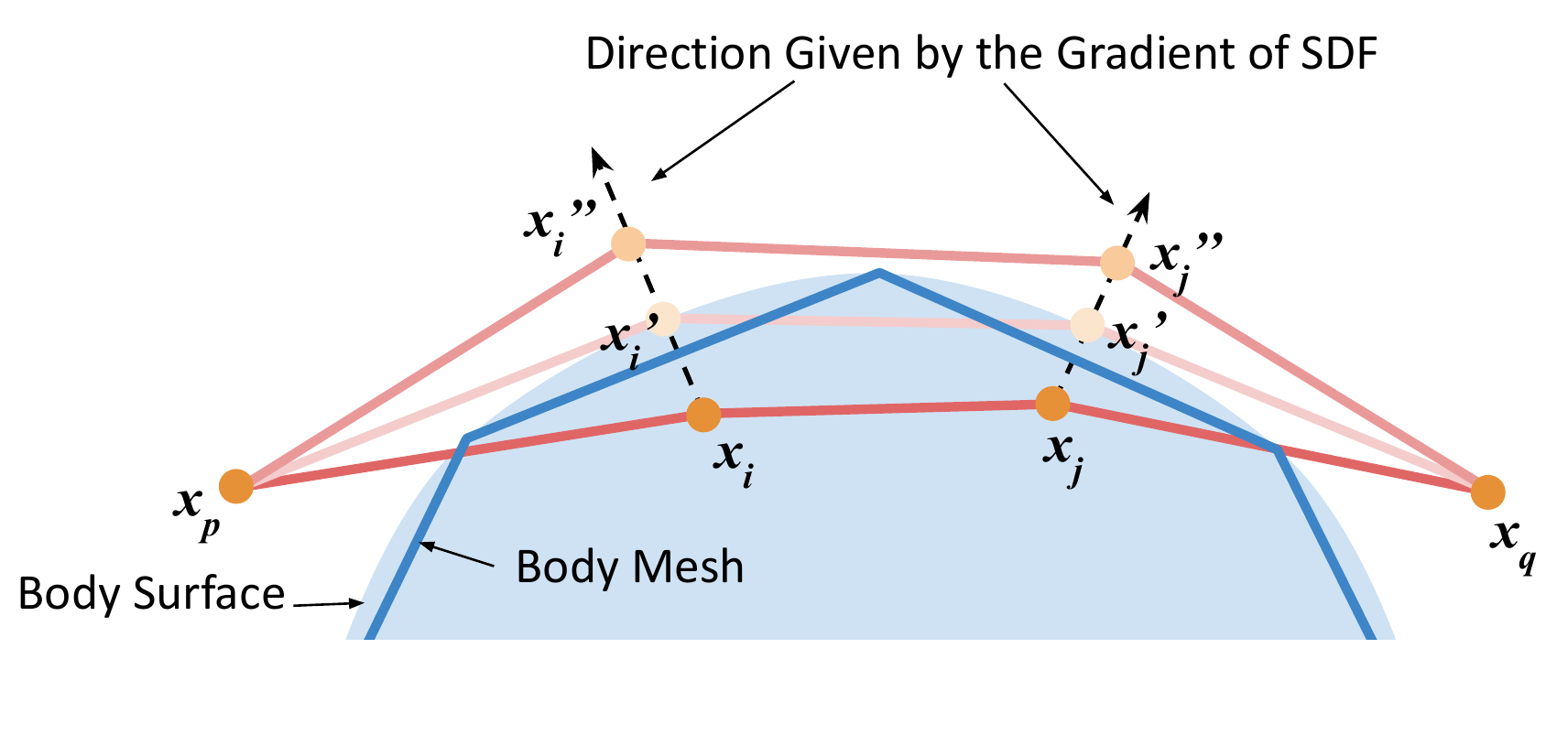}
    \caption{\label{fig:ee}We show an example of resolving vertex-face (VF) and edge-edge (EE) collisions. Assume the blue polyline is a part of the human body mesh, and the light blue region is the body represented by the SDF estimator $f$. We show four garment vertices $\x_{i,j,p,q}$, where $\x_{i,j}$ are inside the human body and $\x_{p,q}$ are outside of it. The dotted arrows are the repulsive force directions given by the gradient of $f$. If we set $\alpha_{i,j}=1$ and the moving offset to be $|f(\x_{i,j})|$, the vertices will moved to $\x_{i,j}'$. While the collision along the edges $\overline{\x_p\x_i'}$ and $\overline{\x_q\x_j'}$ are resolved, the edge $\overline{\x_i'\x_j'}$ will still induce a collision. By allowing $\alpha_{i,j}>=1$, we can move the vertices to $\x_{i,j}''$, resolving all the VF and EE collisions.}

\end{figure}

A straightforward way to decide the moving offset is to use the SDF value directly. However, as pointed out in the context of physically-based simulators~\cite{tang-siga18}, this is only guaranteed to solve the Vertex-Face (VF) collisions, but not the Edge-Edge (EE) collisions. As shown in Figure~\ref{fig:ee}, we push the two neighboring garment vertices further outside to resolve the EE collisions. To compute this extra offset based on the mesh representation, we need to use a global optimizer in an iterative manner. However, this computation can be time consuming and the resulting configuration may not match the groundtruth. Instead, we use neural networks to predict $\alpha_i$, the scale of movement, and multiply it with the SDF value to compute the final offset as: 

\begin{align}\label{eq:offset}
    d_i = \alpha_i f(\x_i), \alpha \in \mathbb{R}.
\end{align}

$\alpha_i$ is predicted based on the global latent feature $\z$ of the whole garment and the SDF value of vertex $\x_i$.

\begin{align}
    \alpha_i = g(k(\z)_i, f(\x_i)), \z\in \mathbb{R}^M,
\end{align}

 where $k: \mathbb{R}^M \rightarrow \mathbb{R}^{N\times D}$ is a topology-dependent Multilayer Perception (MLP) network that infers the latent vector for every vertex from the global feature $\z$, and $k(\z)_i\in\mathbb{R}^D$ is for $i$-th vertex $\x_i$. $g$ is another MLP that outputs the movement scale for $\x_i$. Both $g(\cdot,\cdot)$ and $k(\cdot)$ are  jointly trained with the backbone network in an end-to-end manner. We choose $M=1024$ and $D=10$ for our experiments.

The global garment latent vector $\z$ can be obtained from the backbone network. In terms of using TailorNet as the backbone network, $\z$ is computed from the predefined body pose and shape parameters $\vbeta, \vtheta$ and the garment style parameter $\vgamma$ (as formulated in Tailornet~\cite{patel2020tailornet} for the definition of $\vbeta$, $\vtheta$, and $\vgamma$) with an MLP function $h$.

\begin{align}
    \z=h(\vbeta, \vtheta, \vgamma).
\end{align}

When the accurate groundtruth SDF is given, the output range of $g$ can be set as $[1, +\infty)$ so it ensures all penetrated vertices will be pushed outside. When using a neural network based approximate SDF, we relax the range to be $[0, +\infty)$ to account for inaccuracies in the SDF prediction. Implementation details of the MLPs are given in \prettyref{sec:implementation}.

Using the learned offset has several benefits. First, a flexible extent can handle the collision more naturally and let the moved vertices blend more smoothly with neighboring vertices. Second, it can help resolve additional Edge-Edge (EE) collisions. Finally, the flexible offset can cope with the inaccuracies  of the neural network approximated SDF. For example, when the absolute SDF value predicted by $f(x)$ is smaller than the ground truth, directly using the $f(x)$ as the offset could leave the penetrated vertex inside the human body. On the other hand, if $f(x)$ predicts larger distance than the ground truth, the resulting vertex may be pushed too far away from the human body, resulting in a ``pump out" artifact, as shown in \prettyref{fig:pump}. However, when using ReFU during training, the backbone network and the MLPs in the ReFU layer will foresee the quality of the SDF function. As a result, it will learn to adjust the prediction of both $x_i$ and $\alpha_i$ and thereby result in  a more accurate final output garment with fewer collisions.

\subsection{Train Backbone with ReFU}\label{sec:finetune}
ReFU can be easily plugged into current  neural network frameworks for garment prediction. Here we show how to train the backbone network with ReFU with the example of TailorNet~\cite{patel2020tailornet}. 
When using ReFU, we assume that the colliding vertices are not far from the body surface so that the repulsive force can be estimated through the SDF. Thus, we train the ReFU layer with the backbone network in the fine-tuning stage. One could also train the ReFU with the garment network from scratch. We will explain the use of Graph Convolutional Neural Network (GCNN)~\cite{zhou2020fully} in the supplementary material.

The original TailorNet trains different frequency focused components separately to allow the different components realize the difference between low-frequency and high-frequency deformations. However, the groundtruth representation of the garment, obtained after the frequency division, may not satisfy the collision-free condition, i.e., the high-frequency pieces may have deeper penetrations inside the human body. If we plug in ReFU after computing the high-frequency output, the new vertices will all be moved outside the body, and their coordinates will be different from their high-frequency groundtruth. Thus, we propose training ReFU as a finetuning process together with all the components of TailorNet, which have the summation of all the frequencies with no collisions in the groundtruth data.

We attached a ReFU layer to the end of the pre-trained TailorNet so that it receives the raw output of $\{\x_i\}_{i=1}^N$. Assuming the predicted garment vertex positions after the ReFU layer is $\{\x'_i\}_{i=1}^N$ and the corresponding groundtruth is $\{\tilde{\x}_i\}_{i=1}^N$, we use the following loss terms to train the backbone network and the ReFU layer:

\begin{align}
    \mathcal{L} &= \lambda_1 \mathcal{L}_r + \lambda_2\mathcal{L}_c,\\
    \mathcal{L}_{r} &= \sum_{i=1}^N\|\x_i' - \tilde{\x}_i\|_2^2,\\
    \mathcal{L}_{c} &=\sum_{i=1}^N|\max(-f(\x'_i), 0)|,\label{eq:collisionloss}
\end{align}

where $\mathcal{L}_{r}$ is the reconstruction loss and $\mathcal{L}_{c}$ is the collision loss to cover missed penetrated vertices. $\lambda_{1,2}$ are weights to balance the loss terms.

We train the network with groundtruth collision-free garment data so the reconstruction loss will guide the prediction of the $\x_i$ and $\alpha_i$ to move $\x'_i$ to the position with no EE collisions. In this manner, it better preserves the local smoothness and details. 

It turns out that the post-processing methods in previous works~\cite{santesteban2019learning,holden2019subspace} can only move the vertex along the gradient direction of the SDF of $\x_i$. With our method, however, although the adjustment space of each collided vertex is also one degree-of-freedom (DOF) during inference, through the training, both the backbone and the ReFU layer are fine-tuned so the adjustment space extends to 4 DOFs, i.e., the movement scale $\alpha_i$ and the original network output $\x_i$. With higher adjustment capability, our method achieves performance on par with complex optimization-based post-processing such as \cite{guan2012drape}. Detailed comparisons with previous methods are described in \prettyref{sec:comparison}.

\subsection{Neural Network for Body SDF} \label{sec:SDF}
In physically-based simulation methods, the SDF values are computed using analytical methods and accelerated using spatial data structures like KD-trees. Even with these acceleration data structures, the SDF computation is far from real-time as shown in the running time comparison in the supplementary material. Recent works \cite{park2019deepsdf,gropp2020implicit} show that the implicit function of a 3D geometry can be approximated by a neural network. As a result, one can use the trained network to quickly estimate the SDF values of a 3D point set. We design the network to predict SDF conditioned on the SMPL \cite{loper2015smpl} parameters, please see more details in the supplementary material.



\begin{figure}[ht]
    
    \centering
    \includegraphics[width=.98\textwidth]{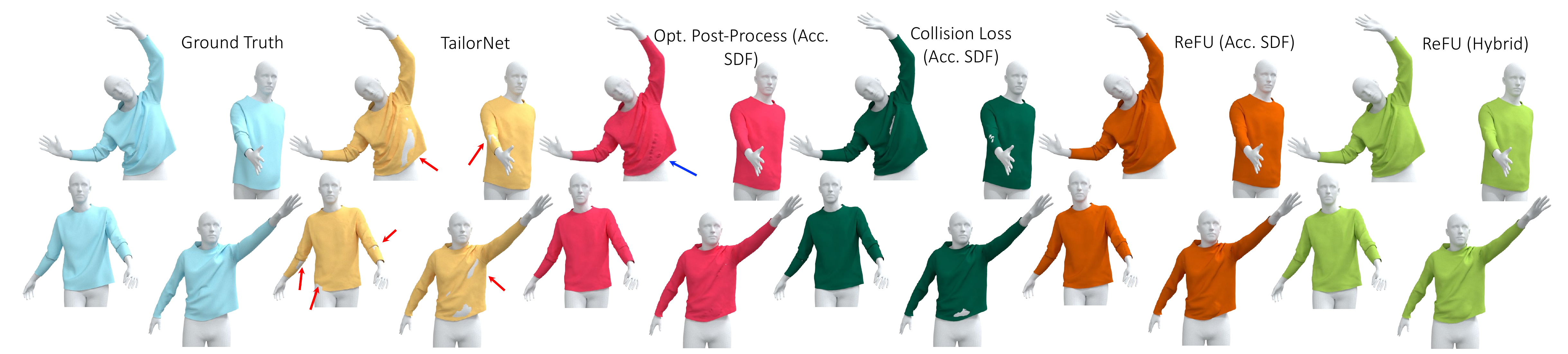}

    \caption{\label{fig:visual} Benefits of ReFU: The baseline models from TailorNet have collisions (red arrow), optimization-based post-processing results in non-smooth regions (blue arrow); collision loss cannot resolve the collisions well; our method based on ReFU resolves the collisions and preserves the original shape with plausible wrinkles in both offline (using accurate SDF) and real-time (``Hybrid'') modes. The results obtained using approximated SDF (``Hybrid'') are similar to those obtained using accurate SDF.}

\end{figure}

\begin{figure}[h]
    
    \centering
    \includegraphics[width=.75\textwidth]{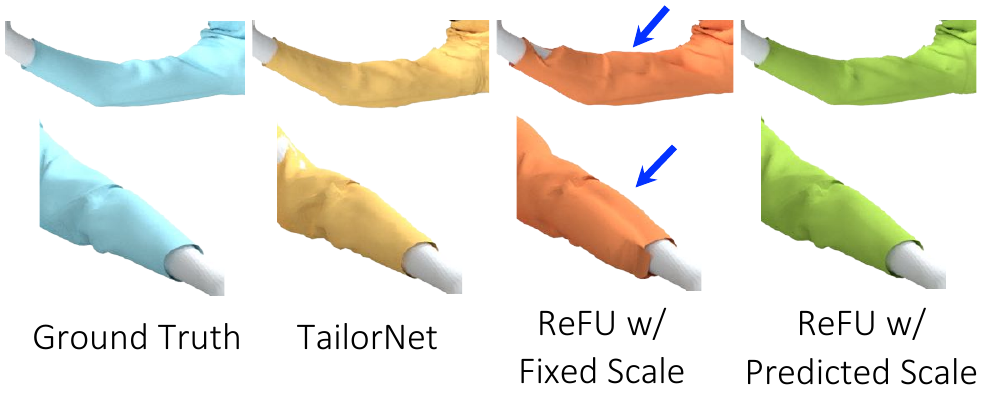}

    \caption{\label{fig:pump} We zoom in to highlight the ``pump out'' artifacts with fixed moving scale and approximated SDF. When $\alpha_i=1$, the ReFU layer may wrongly push out some regions  (blue arrow) with higher approximated SDF values than the ground truth. However, with predicted scale, the network learns how to cope with the inaccuracies of the approximated SDF and generates smooth results. This also reduces the number of VF and EE collisions (Table 3).}

\end{figure}

\section{Experimental Results}\label{sec:results}

In this section, we will quantitatively and qualitatively evaluate our method. Given our goal of achieving real-time performance for the whole garment prediction system, we use an approximated neural network predicted SDF (abbreviated as Approx. SDF) and combine it with ReFU. We also want to see the upper bound of our method, so we combine with an approach that computes accurate SDF (abbreviated as Acc. SDF).

\begin{table}[h]
\centering
\caption{\label{table:dataset} We highlight the details of the datasets used to evaluate our approach. We have selected collision-free subsets from the datasets in \cite{patel2020tailornet}. The resulting datasets include different genders and garment types.}

\setlength{\tabcolsep}{4pt}
\begin{tabular}{lccc}
\toprule
Dataset & \# of vertices & \# of Training Models & \# of Testing Models\\
\midrule
Shirt Male & 9723 & 28360 & 3197\\
T-shirt Male & 7702 & 21097 & 2688\\
Short-pant Male & 2710 & 14555 & 2691\\
Skirt Female & 7130 & 15664 & 3525\\
\bottomrule

\end{tabular}

\end{table}

\subsection{Datasets, Metrics, and Settings}

We utilize the datasets from TailorNet \cite{patel2020tailornet} to evaluate our method. We use  an exact collision detection algorithm available as part of FCL~\cite{pan2012fcl} to select a subset of the garments from different genders and garment types that have no  garment-body collisions or any self-penetrations in the  3D garment mesh. Except for this modification, we use the same train-test split as in \cite{patel2020tailornet} for both garment networks and SDF networks on different garment types to perform  a fair comparison.
The resulting dataset is summarized in \prettyref{table:dataset}. 
``All garments'' in~\prettyref{table:comparison} correspond to the weighted-average performance for four types of datasets.
We include more results on the dataset from Santesteban et al.'s \cite{santesteban2021self} in the supplementary material and highlight the benefits of our approach.

We use the following metrics in our comparisons with prior methods:

\textbf{MPVE}~\cite{pavlakos2018learning} (Mean per-vertex error): Euclidean distance between the ground-truth and predicted garment vertices. It indicates the reconstruction error of the predicted garments. We use millimeters as the underlying unit.

\textbf{VFCP}~\cite{bertiche2021deepsd} (Vertex-face collision percentage): The percentage of vertices on the garment that are inside the body surface.

\textbf{CFMP}~\cite{tan2020lcollision} (Collision-free models percentage): The percentage of garment models that are body-cloth collision-free in both types of VF and EE collisions.

For training and testing, we have three settings:

\textbf{Approx. SDF}: Always use the neural network approximated SDF for training, testing and post-processing.

\textbf{Acc. SDF}: Always use the accurate SDF for training, testing and post-processing. For discretized 3D models such as the human body represented by the SMPL parameters, we compute the SDF value for one query point. This accurate formulation can compute closest point/face and returns the distance on the normal direction, thought it is time consuming (shown in the supplementary material.)

\textbf{Hybrid}: Use the accurate SDF for computing the collision loss during training but the approximate SDF for ReFU during both training and testing.

\subsection{Implementation}\label{sec:implementation}
We implement our method using PyTorch \cite{paszke2019pytorch}. All the training and testing are performed on a server machine with a 96-core CPU, 740GB memory, and 4 NVIDIA V100 GPU with 32GB memory. To estimate the body surface SDF, we design the neural network $f$ with nine hidden layers each with $1024$ neurons. Between each layer, we use a Softplus  activation layer \cite{7280459} with $\beta=100$. We include a skip connection in the fourth layer, i.e., concatenating the query point coordinate with the hidden vector. To train $f$, we feed a batch with 32 human body models each with  SDF value of $4000$ random sampled points. For the network predicting the scale $\alpha_i$, we use three $1024$-dimension layer for $h$, one $N\times10$-dimension layers for $k$, and two $10$-dimension layers for $g$. They all use ReLU as the activation layer in between. We set the weights of the loss terms as $\lambda_1=1.5$ and $\lambda_2=0.5$. For all the training, we use an Adam optimizer with a learning rate of $1\times 10^{-5}$.

\subsection{Performance}
We first evaluate how ReFU performs when using the accurate body SDF during both training and testing. As listed in Table~\ref{table:comparison}, only $49.09\%$ of the garments generated by the original TailorNet are collision-free. After fine-tuning TailorNet with ReFU, the collision-free models increase significantly to $76.77\%$. For difficult cases like men's shirts, which have dense meshes and are tight fitting on the body, collision free models increase significantly by $5.3\times$, from $11.92\%$ to $63.06\%$. Since we set $\alpha>=1$, all the interpenetrating vertices are pushed outside, and the VFCP drops from $0.6\%$ to $0\%$. In the meantime, for MPVE, the reconstruction error also decreases from $8.89$ to $8.64$. However, querying the SDF values is computationally expensive, adding around 0.106 seconds of time cost to each garment inference (see in the supplementary material.).

Then we train and test how ReFU performs when using the neural network approximated body SDF. In other words, we train and test ReFU with imperfect but much faster SDF. The result shows that although this strategy cannot achieve the same level of collision handling as using accurate SDF, it still improves the collision and reconstruction accuracy of the original TailorNet a considerable level. The collision-free models increase to $58.52\%$, MPVE drops to $8.66$, and VFCP decreases by half as shown in the third to last column in Table~\ref{table:comparison}. The inference time for the ReFU layer with approximated SDF is 2.00 milliseconds, and the inference time for the whole garment prediction system is 22.57 milliseconds.

To improve the collision handling capability while maintaining the real-time performance during testing, we experiment with using accurate SDF values when computing the collision loss during training while feeding the approximate SDF into the ReFU layer during both training and testing. The network is guided with accurate collision loss during training while learning to accommodate the errors in the approximated SDF, which will be provided at test time. Now the result is much closer to the case of using accurate SDF. We call this a ``hybrid" mode for simplicity. As shown in the last column in Table~\ref{table:comparison}, the collision-free models achieve $68.71\%$, VFCP improves to $0.24\%$, and the reconstruction error is nearly the same.

From the last two columns in \prettyref{fig:visual}, we can visually observe how using ReFU improves the body-cloth collision problems in the original TailorNet. When trained and tested with Accurate SDF, the body penetrations are almost solved and the predicted garments look much more like the groundtruth. In the hybrid mode, although the network cannot achieve the same quantitative results as in the accurate SDF mode, the visual quality is quite similar. This is impressive for a real-time garment prediction system.

\subsection{Ablation Study}

\begin{table}
\centering 
\caption{\label{table:ablation} Ablation study on the predicted scale in ReFU. As we mentioned in \prettyref{sec:scale}, we use networks to predict the scale $\alpha_i$ to determine the moving offset of collided vertices. Here, we show that if we use only fixed scale ($\alpha_i$ always equals to 1), the performances are worse than the predicted ones with either approximated or accurate SDF. These experimental results show that our approach can also reduce the number of   EE collisions.}

\setlength{\tabcolsep}{3pt}
\begin{tabular}{lcccccc}
\toprule[1pt]
\multirow{3}{*}{Metric}&\multicolumn{6}{c}{Method} \\
\cmidrule(r){2-7}
  & \multirow{2}{*}{TailorNet}&\multicolumn{2}{c}{w/ Fixed Scale} & \multicolumn{3}{c}{w/ Predicted Scale} \\
& & Approx. SDF & Acc. SDF&  Approx. SDF & Acc. SDF & Hybrid\\
\midrule
 MPVE & $11.27$& $10.84$ & $10.60$ &  $10.59$ & $10.56$ & $10.57$\\
\cmidrule(r){1-7}
\metrica &$1.18\%$& $0.58\%$ & $0.00\%$ &  $0.62\%$ & $0.00\%$ & $0.51\%$\\
\cmidrule(r){1-7}
\metricb& $11.92\%$&$4.88\%$ & $37.22\%$&  $26.9\%$ & $63.06\%$ & $49.32\%$\\
\cmidrule(r){1-7}
Avg. VF & 180.36 & 287.85 & 1.94 & 110.12 & 1.45 & 78.21\\
\cmidrule(r){1-7}
Avg. EE & 11.56 & 19.60 & 8.89 &   8.6      &         6.98       &           7.39\\
\bottomrule[1pt]
\end{tabular}

\end{table}
In ReFU, we design a network-predicted scale $\alpha_i$ to determine the offset length $d_i$ for each vertex. Theoretically it can help solve EE collisions and improve the reconstruction quality. Here we perform the ablation study to evaluate the effect of using predicted scale. We report the experimental results for male shirt in Table~\ref{table:ablation}. As listed in the third and fourth columns, when only using $f(\x_i)$ without scaling, the quantitative results for all metrics are worse than with the predicted scaling. The percentage of collision-free models drops by two times in the case of using accurate SDF and over 5 times in the case of using the approximate SDF. Figure~\ref{fig:pump} further shows that when using the approximate SDF resulted in some pump out artifacts. With the fixed scale, the predicted cloth covers the pump out region of the approximate body surface (where approximated SDF values are larger than the accurate ones) in an awkward way. However, when trained with the predicted offset scale, the network learns to handle the imperfectness in the SDF and generates a smoother and more plausible result. 

We also include the average number of colliding triangles on the garment, based on VF and EE contacts, denoted as ``Avg. VF" and ``Avg. EE". The results in~\prettyref{table:ablation} shows that our predicted $\alpha_i$ can further reduce the number of EE collisions, as compared with a fixed value.
We also include another ablation study related to the design choices and other components of our networks, highlighted in~\prettyref{eq:offset_network}, in the supplementary material.

\subsection{Comparisons}\label{sec:comparison}

\begin{table*}

\label{moreevaluation}
\caption{\label{table:comparison} Comparison with baseline and different collision handling methods, including naïve post-processing~\cite{santesteban2019learning,holden2019subspace}, optimization-based post-processing~\cite{guan2012drape}, and soft collision loss~\cite{bertiche2021deepsd,gundogdu2019garnet,bertiche2020pbns,bertiche2020cloth3d}. We include an additional row called ``Trend'' to show the rate of change for each method compared to TailorNet, the baseline model. The report shows all methods perform better with accurate SDF than with approximate SDF. In either situation, ReFU has the best results among all the methods. Overall, ReFU with ``hybrid'' SDF works better than approximate SDF, and ReFU with accurate SDF achieves the best results.
}

\centering 

\resizebox{0.98\textwidth}{!}{
\setlength{\tabcolsep}{3pt}
\begin{tabular}{llcccccccccc}
\toprule[1pt]
\multirow{3}{*}{Dataset}&\multirow{3}{*}{Metric}&\multicolumn{9}{c}{Method} \\
\cmidrule(r){3-12}
& & \multirow{2}{*}{TailorNet}  &\multicolumn{2}{c}{w/ Naïve Post-Process} &\multicolumn{2}{c}{w/ Opt. Post-Process}& \multicolumn{2}{c}{w/Collision Loss} & \multicolumn{3}{c}{w/ ReFU}\\
& & & Approx. SDF & Acc. SDF& Approx. SDF & Acc. SDF& Approx. SDF & Acc. SDF& Approx. SDF & Acc. SDF & Hybrid\\
\midrule
\multirow{3}{*}{Shirt Male}& MPVE & $11.27$ &$11.45$ &$11.26$& $11.30$ & $11.26$ &$10.85$ & $10.61$ & $10.59$ & $\mathbf{10.56}$& $10.57$ \\
\cmidrule(r){2-12}
&\metrica& $1.18\%$ & $0.69\%$ & $0.00\%$ & $0.41\%$ & $0.00\%$ &$1.28\%$ & $0.68\%$ & $0.62\%$ & $\mathbf{0.00\%}$& $0.51\%$  \\
\cmidrule(r){2-12}
&\metricb& $11.92\%$ & $3.1\%$ & $27.5\%$ & $25.21\%$ & $50.36\%$&$20.21\%$ & $39.35\%$ & $26.9\%$ & $\mathbf{63.06\%}$ & $49.32\%$ \\
\midrule
\multirow{3}{*}{T-Shirt Male}& MPVE & $10.77$ & $10.81$& $10.75$& $10.76$ &$10.75$ &$10.60$ & $10.59$ & $10.58$ & $\mathbf{10.57}$& $10.56$ \\
\cmidrule(r){2-12}
&\metrica & $0.95\%$&$0.56\%$ & $0.00\%$ & $0.39\%$ & $0.00\%$ &$0.85\%$ & $0.86\%$ & $0.53\%$ & $\mathbf{0.00\%}$& $0.34\%$  \\
\cmidrule(r){2-12}
&\metricb& $23.96\%$ & $12.01\%$ &$38.02\%$ & $36.57\%$ &$51.60\%$ &$28.98\%$ & $28.76\%$ & $32.18\%$ & $\mathbf{58.77\%}$& $47.25\%$ \\
\midrule
\multirow{3}{*}{Short-pant Male}& MPVE & $6.81$ & $6.87$& $6.83$& $6.82$ &$6.81$ &$6.79$ & $6.80$ & $6.79$ & $\mathbf{6.76}$& $6.77$ \\
\cmidrule(r){2-12}
&\metrica & $0.27\%$& $0.83\%$ & $0.07\%$& $0.11\%$ & $0.00\%$ &$0.19\%$ & $0.21\%$ & $0.13\%$ & $\mathbf{0.00\%}$& $0.13\%$  \\
\cmidrule(r){2-12}
&\metricb& $60.35\%$ & $20.07\%$ &$73.47\%$ & $68.60\%$ &$81.42\%$ &$68.38\%$ & $66.78\%$ & $73.28\%$ & $\mathbf{83.05\%}$& $76.89\%$ \\
\midrule
\multirow{3}{*}{Skirt Female}& MPVE & $6.90$ & $9.35$ &$6.90$& $6.98$ &$6.90$ &$6.92$ & $6.90$ & $6.90$ & $\mathbf{6.87}$& $6.89$ \\
\cmidrule(r){2-12}
&\metrica & $0.067\%$&$0.04\%$ & $0.00\%$ & $0.015\%$ & $0.000\%$ &$0.059\%$ & $0.047\%$ & $0.028\%$ & $\mathbf{0.00\%}$& $0.027\%$  \\
\cmidrule(r){2-12}
&\metricb& $93.36\%$ &$78.78\%$ & $93.87\%$ & $94.01\%$ &$96.45\%$ &$93.99\%$ & $94.44\%$ & $96.03\%$ & $\mathbf{98.16\%}$& $96.43\%$ \\
\midrule
\multirow{3}{*}{All Garments}& MPVE & $8.89$& $9.67$ &	$8.89$ &	$8.92$&	$8.88$&	$8.74$&	$8.67$&	$8.66$&	$\mathbf{8.64}$&	$8.65$\\
\cmidrule(r){2-12}
&\metrica & $0.60\%$& $0.50\%$	& $0.01\%$ &$0.22\%$& $0.00\%$& $0.58\%$& $0.43\%$& $0.31\%$& $\mathbf{0.00}\%$& $0.24\%$  \\
\cmidrule(r){2-12}
&\metricb& $49.09\%$ & $30.89\%$& $59.39\%$ & $57.42\%$ & $70.97\%$ & $54.36\%$ & $59.15\%$ & $58.52\%$ & $\mathbf{76.77\%}$ & $68.71\%$\\

\midrule
\multirow{3}{*}{Trend}& MPVE & - & $+8.77\%$ & $0.00\%$ & $+0.34\%$ & $-0.11\%$ & $-1.69\%$ & $-2.47\%$ & $-2.59\%$ & $\mathbf{-2.81\%}$ & $-2.70\%$ \\
\cmidrule(r){2-12}
&\metrica & -& $-16.67\%$ & $-99.90\%$ & $-97.72\%$ & $-100.00\%$ & $-94.00\%$ & $-95.55\%$ & $-96.79\%$ & $\mathbf{-100.00\%}$ & $-97.52\%$  \\
\cmidrule(r){2-12}
&\metricb& - & $-37.07\%$ & $+20.98\%$ & $+16.97\%$ & $+44.57\%$ & $+10.74\%$ & $+20.49\%$ & $+19.21\%$ & $\mathbf{+56.39\%}$ & $+39.97\%$\\
\bottomrule[1pt]
\end{tabular}
}

\end{table*}
As reported in Table~\ref{table:comparison}, we compared our method with other collision handling approaches. A typical approach is to employ a post-processing step. A naïve method is to move the penetrating vertices directly to the body surface along the SDF gradient direction using $f(\x_i)\widehat{\nabla_{\x_i} f(\x_i)}$~\cite{santesteban2019learning,holden2019subspace} to eliminate the VF-collision. As shown in Figure~\ref{fig:ee}, this method does not handle  EE-collisions and might introduce new artifacts. Our experiments verify that this naïve approach increases the collision-free models only marginally from $49.0\%$ to $59.39\%$ . When tested with approximate SDF, the collision-free models drop to $30.89\%$ and MPVE increases to $9.67$, compared with 8.89 in TailorNet.

Guan et. al.~\cite{guan2012drape} propose a more advanced technique to post-process the collisions using optimization. While this technique is more effective than the naïve approach, it cannot beat the performance of our ReFU-based method in terms of both quantitative results and computational speed.

\begin{wrapfigure}{l}{0.15\textwidth} 

    \centering
    \includegraphics[width=0.15\textwidth]{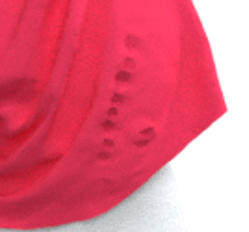}

\end{wrapfigure}
Given the accurate SDF value, it eliminates the VF collisions and increases the ratio of collision-free models to $70.97\%$. Examples are shown in the third column in \prettyref{fig:visual}. 
Although this approach attempts to preserve the original details of the garments, it still results in higher reconstruction error and the resulting garments are not smooth, as shown by the blue arrow in \prettyref{fig:visual} (a close-up view of the corresponding area is shown in the inset figure).
When tested with  approximate SDF, the accuracy with respect to the reconstruction error and collision handling drops significantly. Furthermore, the optimization requires $0.6-0.8$s per frame even with approximated SDF.

Instead of performing post-processing optimization, some techniques~\cite{bertiche2021deepsd,gundogdu2019garnet,bertiche2020pbns,bertiche2020cloth3d} apply the collision loss in \prettyref{eq:collisionloss} in the garment prediction models. The collision loss provides a soft constraint during network training. According to our benchmarks in~\prettyref{table:comparison}, adding collision loss to TailorNet can reduce the overall collision artifacts and improve the reconstruction accuracy for garment samples close to the training set, however, it introduces even more collisions for testing samples that are farther away from the training set (see detail analysis in the supplementary material).  After fine-tuning the TailorNet network with the collision loss using accurate SDF values, the collision-free garment models are around $56\%$ for both approximate SDF and accurate SDF. Moreover, the reconstruction errors are around $8.7$. The fourth column in \prettyref{fig:visual} shows visible collisions in the predicted garments. 
Overall, our approach based on ReFU offers improved accuracy (i.e., fewer collisions), real-time performance, and higher visual quality.

\section{Conclusion, Limitations, and Future Work}
We propose ReFU for handling  body-cloth collisions in neural networks for 3D garment prediction. Our method is inspired by the use of repulsive forces in physics-based simulators. Specifically, ReFU predicts the repulsion direction and the magnitude based on the SDF of the collided vertices and the global latent feature of the garment. ReFU can be combined with different backbone networks, and we highlight the benefits with state-of-the-art learning methods.

While our experiments show that using ReFU for training can significantly reduce the body-cloth collisions  and improve the reconstruction accuracy, our approach has some limitations. To achieve live performance for the whole system, we must use a neural network approximated human body SDF for real-time computation. However, the accuracies in the SDF network prediction affects ReFU's capability of collision handling. But using more advanced neural SDF methods such as the articulated occupancy networks~\cite{deng2020neural,LEAP:CVPR:21,Alldieck_2021_ICCV} can improve ReFU's overall performance. Furthermore, our computation of the moving offset may not fully resolve all collisions, especially all EE collisions.

{
\bibliographystyle{splncs04}
\bibliography{references,collision}
}

\newpage

\appendix

\chapter*{Supplementary Material for A Repulsive Force Unit for Garment Collision Handling in Neural Networks}

\section{SDF Network}

In this section, we will give details about the math formulation of the SDF network, the sampling scheme, and the loss functions we use for training the human body SDF network.  

To train a generalized SDF network that can predict the implicit function of human bodies with different shapes and poses in real-time,  we design the network to predict SDF conditioned on the SMPL \cite{loper2015smpl} parameters. 
\begin{align}
    f(\x,\vbeta,\vtheta) \approx \SDF^{M(\vbeta, \vtheta)}(x).
\end{align}
SMPL is a PCA model computed from a large human shape data. $\vbeta$ and $\vtheta$ are its shape and pose parameters. $M(\vbeta, \vtheta)$ is the human shape reconstructed from  $\vbeta$ and $\vtheta$. 

To train the SDF network, we combine both the regression loss on sampled points in the space and the geometric regularization loss on the gradient as proposed by Park et al. and Gropp et al.~\cite{park2019deepsdf,gropp2020implicit}. For each garment-body pair in the TailorNet dataset, we collect three categories of SDF value samples:
\begin{enumerate}
    \item Randomly sampled points from the body surface, with or without Gaussian disturbance. For samples right on the body surface, we also collect their normals. Note that, we can only get correct SDF gradients for the surface points which are their normals. For other points, we can estimate their gradients through analytic methods.
    \item Randomly sampled points from the garment surface, with or without Gaussian disturbance.
    \item Randomly sampled points inside the bounding box of the body. We use a general bounding box for all the samples with size $4m\times 4m\times 4m$, centering at $[0,0,0]$.
\end{enumerate}

For points from the body surface without disturbance, we denote them as $\{\x_i\}_{i\in I_S}$, their normals as $\{\n_i\}_{i\in I_S}$. For other points, we denote them as $\{\x_j\}_{j\in I_E}$. The ground truth SDF values for all the points are $\{s_i\}_{i\in I_S\cup I_E}$. We compute the loss for training SDF as:
\begin{align}
    \mathcal{L}_{SDF} &= \lambda_a \mathcal{L}_v + \lambda_b \mathcal{L}_{sg} + \lambda_c  \mathcal{L}_{se}\\
    \mathcal{L}_v &= \mathbb{E}_{i\in I_S\cup I_E}(|f(\x_i) - s_i|)\label{eq:abs_error}\\
    \mathcal{L}_{sg} &= \mathbb{E}_{i\in I_S}(\|\nabla_{\x}  f(\x_i) - \n_i\|)\\
    \mathcal{L}_{se} &= \mathbb{E}_{i\in I_E}(\|\nabla_{\x} f(\x_i)\|-1)^2,
\end{align}
where $\mathcal{L}_v$ is a regression loss for the values \cite{park2019deepsdf}, $\mathcal{L}_{sg}$ and $\mathcal{L}_{se}$ are losses for the gradients \cite{gropp2020implicit}. Specifically, $\mathcal{L}_{se}$ is based on the Eikonal equation\cite{crandall1983viscosity}. We set the weights to balance each term as $\lambda_a = 2, \lambda_b = 1, \lambda_c = 0.1$.

We include the performance for the approximated SDF on the datasets we used in \prettyref{table:sdf}. We use two metrics: 

\textbf{Mean Absolute Error} defined in \prettyref{eq:abs_error};

\textbf{Mean Relative Error} defined as
\begin{align}
    \mathbb{E}_{i\in I_S\cup I_E}\left(\left|\frac{f(\x_i) - s_i|}{s_i}\right|\cdot 100\%\right).
\end{align}

Using those loss functions, we can have supervision on the absolute values for the SDF samples, but no supervision on the norm of the gradient for vertices that are not on the body surfaces. Consequently, the mean relative error is much worse than the mean absolute error. Thus, in the main paper, we use the predicted offset scale to help ReFU improve its collision handling ability using the approximated SDF.

\begin{table}[h]
\caption{\label{table:sdf} Mean absolute error and mean relative error of the SDF network.} 
\centering
\setlength{\tabcolsep}{4pt}
\begin{tabular}{lcc}
\toprule
Dataset & Mean Absolute Error & Mean Relative Error\\
\midrule
Shirt Male & 2.38mm & $28.22\%$\\
T-shirt Male & 2.37mm & $25.85\%$\\
Short-pant Male & 2.46mm & $31.66\%$\\
Skirt Female & 3.10mm & $32.64\%$\\
\bottomrule
\end{tabular}
\end{table}

\section{Penetration Energy Histogram}
\begin{figure*}[ht]
    
    \centering
    \includegraphics[width=.98\textwidth]{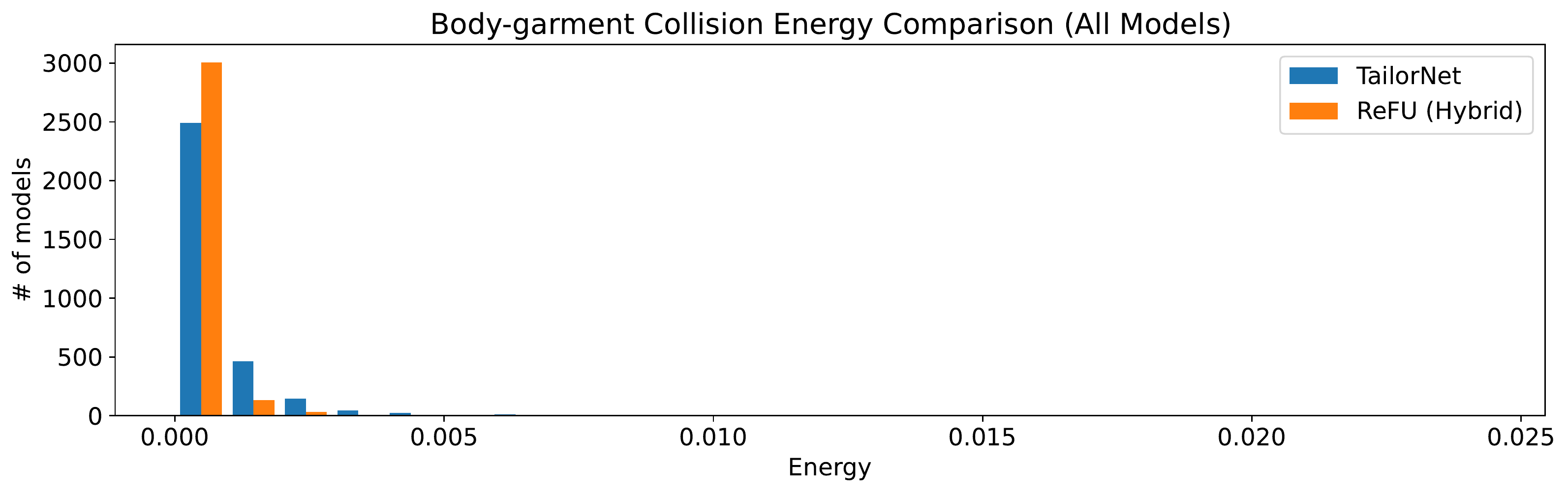}
    \caption{\label{fig:hist}We show the histogram of the collision energy for TailorNet without ReFU (blue) and TailorNet with ReFU trained in the ``Hybrid'' SDF mode (yellow). It shows with ReFU, the network can produce  much more garments with low collision energy.}
\end{figure*}

\begin{figure*}[ht]
    \centering
    \includegraphics[width=.98\textwidth]{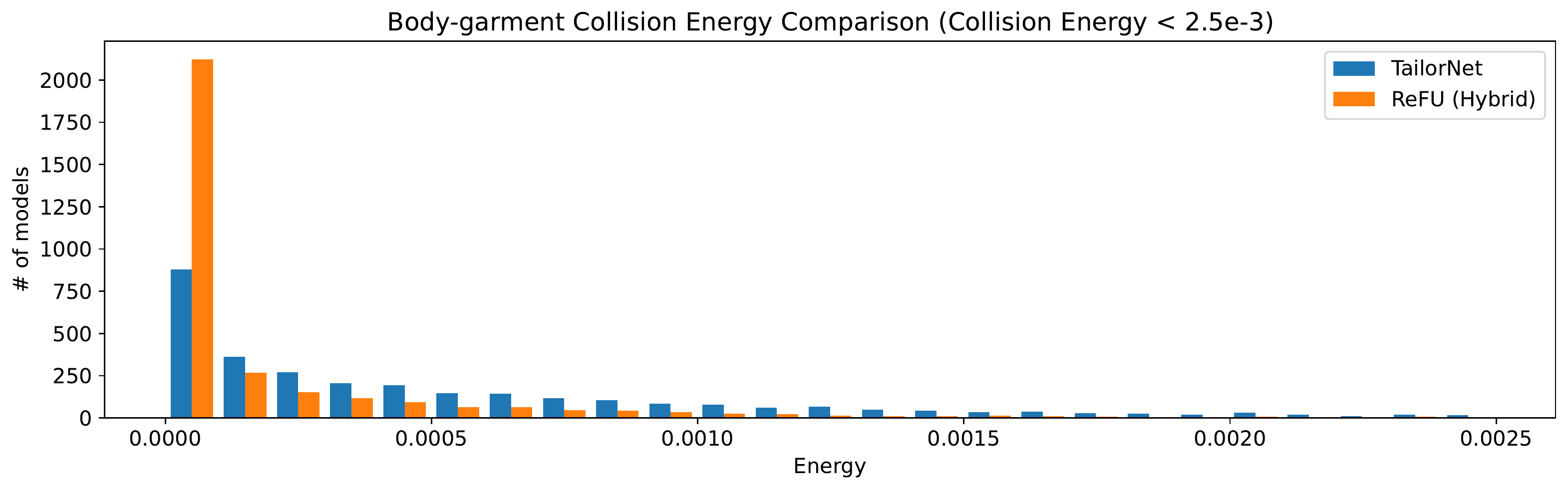}
    \caption{\label{fig:zoom_in_hist}Here is a zoomed-version of \prettyref{fig:hist} for output garments with collision energy less than $2.5\times 10^{-3}$.}
\end{figure*}

Although our method cannot eliminate all the collisions when using the neural network approximated SDF due to the inaccuracies of SDF, it brings a significant decrease in the overall penetration energy as shown in the distribution histogram in \prettyref{fig:hist} and \prettyref{fig:zoom_in_hist}. \prettyref{fig:hist} shows all the results and \prettyref{fig:zoom_in_hist} shows the zoomed-in results with collision energy less than $2.5\times 10^{-3}$. We compute the penetration energy as the way described in \cite{tan2020lcollision}.

\section{Results with GCNN}
In this section, we show how ReFU works when applied in a backbone network based on a graph convolutional neural network (GCNN) \cite{zhou2020fully} for general 3D models. This GCNN does not have a frequency division process as in TailorNet, thus we can directly plug in the ReFU and train from scratch. We mentioned in Sec. 3.2. of the main paper, that SDF for collision handling is only useful when the points are near the surface, so we may better train with ReFU in the refining stage when the predicted cloth already satisfies this condition. If we train from scratch, the initial network prediction may not satisfy the condition. Nevertheless, we still find the network can learn to cope with these, and have good results when the network becomes steady as shown in \prettyref{table:gcnn}. The results demonstrate significant improvements in reducing collisions by training ReFU with GCNN from scratch. This experiment illustrates that ReFU can work with different kinds of backbone networks.

\begin{table}
\caption{\label{table:gcnn} Garment prediction results of the GCNN network trained with or without the ReFU layer.}
\centering 
\setlength{\tabcolsep}{4pt}
\begin{tabular}{lcccc}
\toprule[1pt]
\multirow{3}{*}{Metric}&\multicolumn{4}{c}{Method} \\
\cmidrule(r){2-5}
& \multirow{2}{*}{GCNN\cite{zhou2020fully}}   & \multicolumn{3}{c}{w/ ReFU}\\
& & Approx. SDF & Acc. SDF & Hybrid\\
\midrule
MPVE & $14.05$ & $13.05$ & $12.82$ & $12.93$\\
\cmidrule(r){1-5}
\metrica & $2.26\%$ & $0.81\%$ & $0.00\%$ & $0.72\%$\\
\cmidrule(r){1-5}
\metricb& $8.70\%$ & $19.92\%$ & $40.98\%$ & $36.78\%$\\
\bottomrule[1pt]
\end{tabular}

\end{table}

\section{Results on VTO Dataset \cite{santesteban2021self}}

VTO dataset from~\cite{santesteban2021self} is a new public garment-body dataset with more complex human poses. The dataset contains 17 different shapes and several different motion sequences, including walking, running, jumping, dancing, etc. We evaluate our results on test sets that include four unseen sequences similar to the original paper~\cite{santesteban2021self}. We choose two options for the number of shapes: one with five shapes resulting in a similar amount of samples to the TairlorNet dataset and another with all 17 shapes. We use a 5-layer Multilayer Perception (MLP) as our baseline. We show the results in~\prettyref{table:vto}. Our method works well in this new dataset.

\begin{table}
\caption{\label{table:vto} Results on the new VTO dataset \cite{santesteban2021self}}
\centering 

\resizebox{0.98\textwidth}{!}{
\setlength{\tabcolsep}{4pt}
\begin{tabular}{cccccc}
\toprule[1pt]
\multirow{2}{*}{Shape Num.}  & \multirow{2}{*}{Train Num.}  & \multirow{2}{*}{Test Num.} & \multirow{2}{*}{Metric} & \multicolumn{2}{c}{Method}\\
&&&&Baseline & ReFU (Approx. SDF)\\
\midrule
\multirow{3}{*}{5} & \multirow{3}{*}{33525} & \multirow{3}{*}{2060}& MPVE & 15.36  & 13.15\\
\cmidrule(r){4-6}
&&&VFCP & $1.78\%$ & $0.56\%$\\
\cmidrule(r){4-6}
&&&CFMP & $18.69$& $52.53\%$\\
\midrule
\multirow{3}{*}{17} & \multirow{3}{*}{113985} & \multirow{3}{*}{7004}& MPVE & 18.23 & 16.12\\
\cmidrule(r){4-6}
&&&VFCP & $3.18\%$ & $0.86\%$\\
\cmidrule(r){4-6}
&&&CFMP & $13.30\%$ & $38.39\%$\\
\bottomrule[1pt]
\end{tabular}
}
\end{table}

\section{Ablation Study for Networks Computing $\alpha_i$}

We compare alternative options for computing the scale $\alpha_i$. In the final ReFU structure, we use the following networks to compute:
\begin{align}\label{eq:offset_network}
    \alpha_i = g(k(\z)_i, f(\x_i)), \z\in \mathbb{R}^M,
\end{align}
with $k: \mathbb{R}^M \rightarrow \mathbb{R}^{N\times D}$ as a topology-dependent MLP network that infers the latent vector for every vertex from the global feature $\z$. 

There are two additional possible choices. The first one (``Alt. 1''):
\begin{align}
    \alpha_i = g(k'(\z), f(\x_i)), \z\in \mathbb{R}^M,
\end{align}
with $k: \mathbb{R}^M \rightarrow \mathbb{R}^{D'}$ as another MLP inferring one shared latent vector from $z$. Here we let $D'\propto N\times D$ to maintain the parameter size of the whole network and ensure a fair comparison.

The second one (``Alt. 2''):
\begin{align}
    \alpha_i = g'( f(\x_i)), \z\in \mathbb{R}^M,
\end{align}
where $g'$ is a network directly predicting the scale from each vertex's SDF value.

We show the comparison results on ``Shirt Male'' dataset in \prettyref{table:alpha_ablation}. For all the experiments, we use approximate SDF. The results show that our final choice in \prettyref{eq:offset_network} achieves better results since it considers each vertex's information to compute the final scale.

\begin{table}
\caption{\label{table:alpha_ablation} Results trained with different $\alpha_i$ computing networks.}
\centering 
\setlength{\tabcolsep}{4pt}
\begin{tabular}{lcccc}
\toprule[1pt]
\multirow{2}{*}{Metric}&\multicolumn{4}{c}{Method} \\
\cmidrule(r){2-5}
& Baseline  & Alt. 1 & Alt. 2 & ReFU\\
\midrule
MPVE & $11.27$ & $10.65$ & $10.66$ & $10.59$\\
\cmidrule(r){1-5}
\metrica & $1.18\%$ & $0.76\%$ & $0.79\%$ & $0.62\%$\\
\cmidrule(r){1-5}
\metricb& $11.92\%$ & $21.36\%$ & $20.81\%$ & $26.9\%$\\
\bottomrule[1pt]
\end{tabular}

\end{table}

\begin{table}[h]
\centering

\caption{\label{table:time} Per-frame running time, including approximated SDF query, accurate SDF query computed using spatial data structures,  ReFU layer inference, and the backbone network based on TailorNet inference~\cite{patel2020tailornet}.}

\setlength{\tabcolsep}{4pt}
\begin{tabular}{lcccc}
\toprule
\multirow{2}{*}{Dataset}&\multicolumn{4}{c}{Component}\\
\cmidrule(r){2-5}
 & Approx. SDF & Acc. SDF & ReFU & Backbone \\
\midrule
Shirt Male & 1.97ms  &121.96ms & 0.29ms & 22.14ms\\
T-Shirt Male & 1.77ms  &99.27ms & 0.28ms & 21.51ms\\
Short-pant Male & 1.58ms  & 89.69ms & 0.21ms & 18.82ms\\
Skirt Female & 1.67ms & 107.38ms & 0.23ms & 19.76ms\\
\cmidrule(r){1-5}
All Garments & 1.75ms & 105.50ms & 0.25ms & 20.57ms\\
\bottomrule
\end{tabular}

\end{table}

\section{Running Time}
We include the running time for SDF, ReFU layer, and the backbone network TailorNet in \prettyref{table:time}.

\section{Moving Offset Analysis}

\begin{figure}
    \centering
    \includegraphics[width=0.65\textwidth]{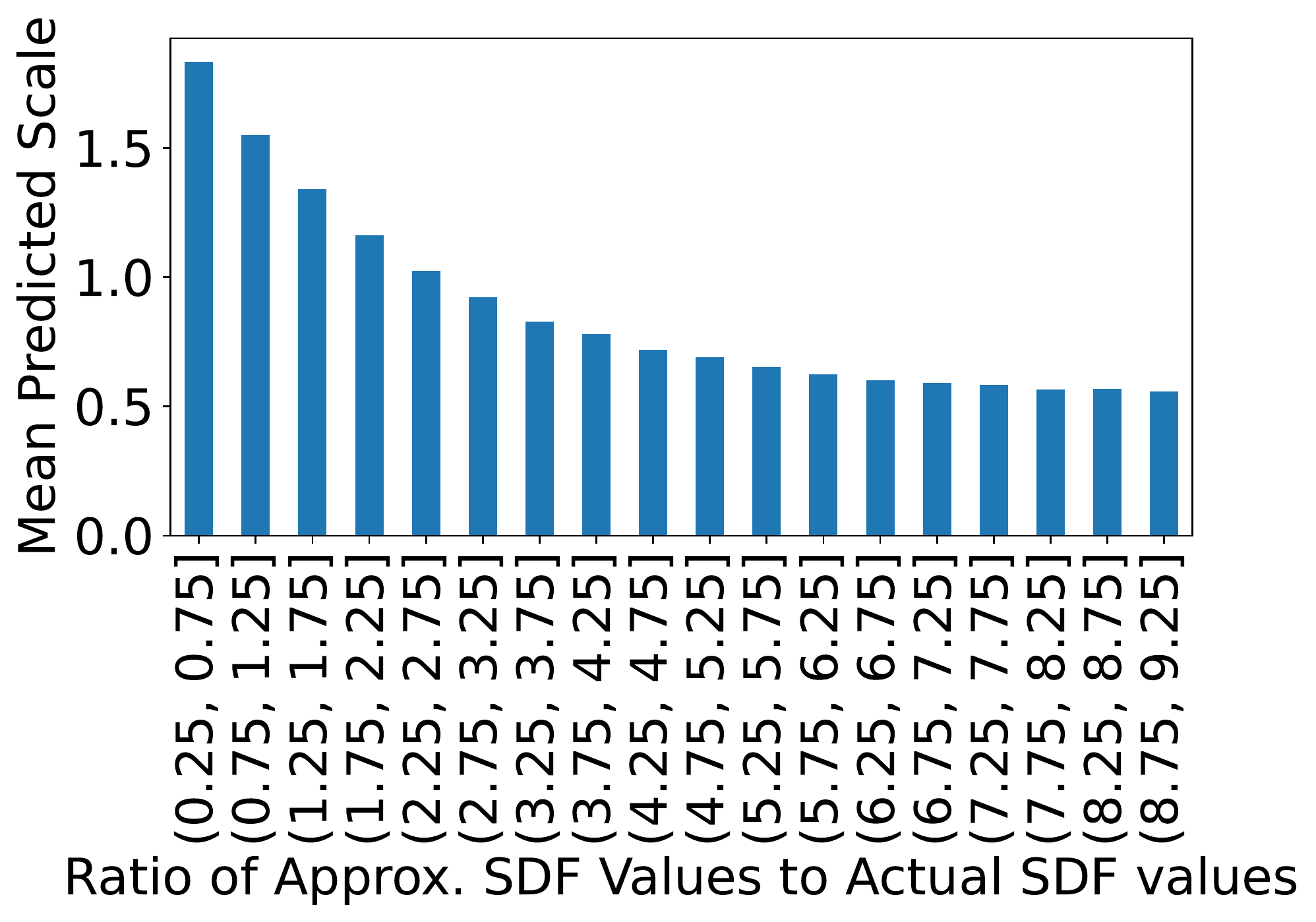}

    \caption{We show the bar plot of the predicted scale grouped by the ratio of approximated to actual SDF values. A larger scale can push collided vertices further to compensate for smaller than actual SDF value prediction, and vice versa. It shows that our moving offset based on the predicted scale can compensate for the error from the SDF approximation. }

    \label{fig:scale_analysis}
\end{figure}

We analyze the ratio of the approximated SDF values to the actual ones with the predicted moving offset scale $\alpha$. We show the bar plot for all collision-resolved vertices in \prettyref{fig:scale_analysis}. When the ratio is smaller than one, the approximated SDF value is smaller than the actual one. A larger moving offset scale lets ReFU push the vertices even further and compensate for the error from SDF prediction. Similarly, when the ratio is larger than one, the approximated SDF value is larger than the actual one; a smaller scale can avoid putting the vertex too far away. Among the collision-resolved vertices, the minimum ratio is $0.39$ with a scale of $2.68$; the maximum ratio is $9.13$ with a scale $0.25$. Notice that the mean predicted scale equals $1$ for the group $(2.25, 2.75]$, which shows that our layer learns to push the vertices even further than the distance to the surface to resolve the EE cases.

\section{Local Geometric Comparison with Optimization Post-Processing}

\begin{table}
\caption{\label{table:local_lap} Local Laplacian error for collision resolving region on `shirt male' dataset. Our method can preseve local geometry details than previous methods.}
\centering 
\setlength{\tabcolsep}{4pt}
\begin{tabular}{ccccc}
\toprule[1pt]
 \multicolumn{2}{c}{w/ Opt. Post-Process}   & \multicolumn{3}{c}{w/ ReFU}\\
Approx. SDF & Acc. SDF & Approx. SDF & Acc. SDF & Hybrid\\
\midrule
 $7.02$ & $5.85$ & $5.29$ & $4.24$& $4.36$  \\
\bottomrule[1pt]
\end{tabular}
\end{table}

We use local Laplacian error on the collision resolving regions to show that our method can better preserve small-scale geometric details, as compared to other post-processing methods. For each initial collided vertex, we compute its 1-ring neighborhood Laplacian error. We summarize the mean error on the `shirt male' dataset in \prettyref{table:local_lap}. Our method with different settings results in  lower errors compared to the post-processing counterparts.

\section{Comparison with Collision Loss}

\begin{figure}[h]

    \centering
    \includegraphics[width=.65\textwidth]{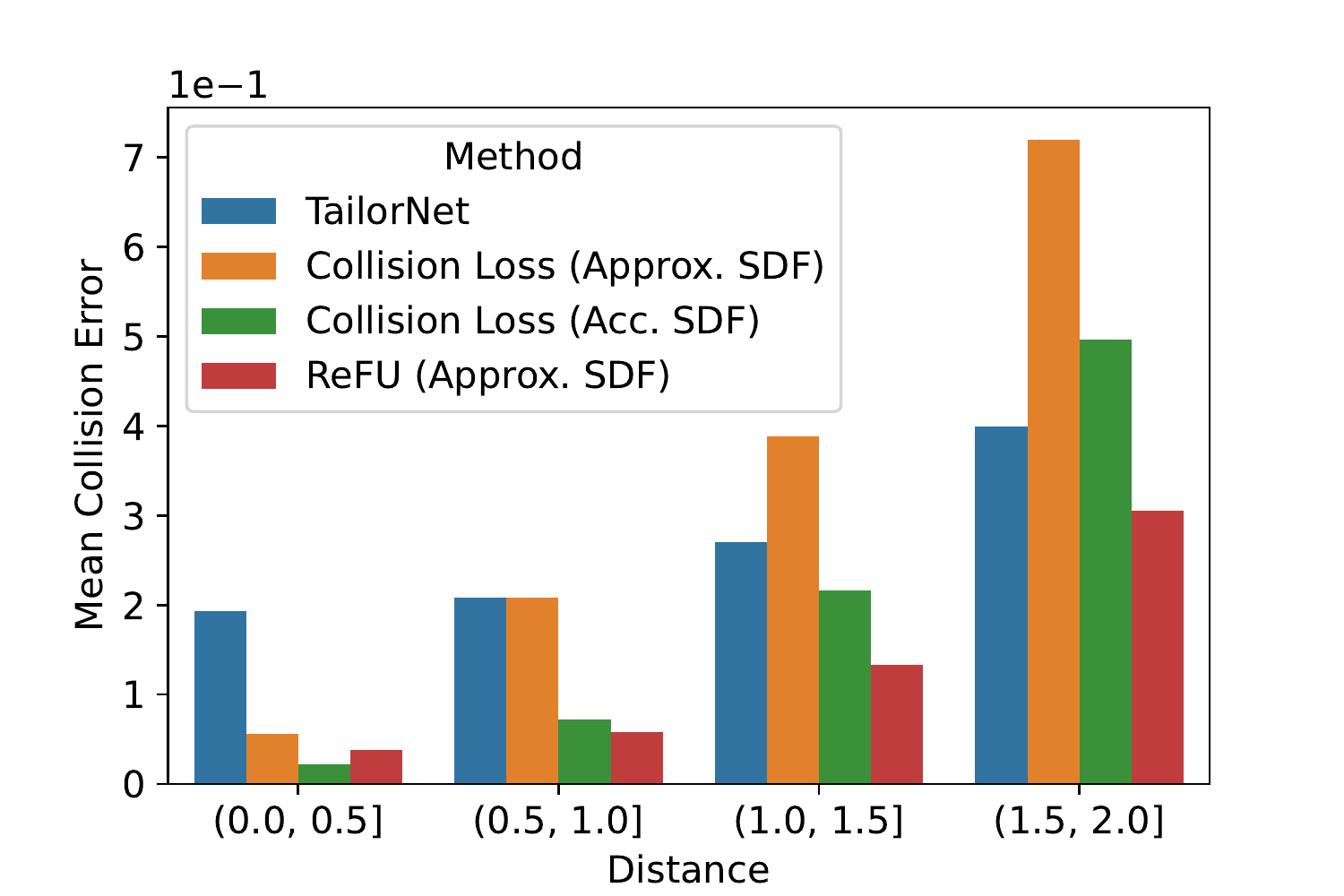}

    \caption{We highlight the benefits of our approach on samples distinct from the training set over methods based on collision loss. Collision loss can only help reduce collisions for samples close to the training set.}
    \label{fig:unseen}

\end{figure}

As we mentioned in the main paper, adding collision loss introduces even more collisions for testing samples that are farther away from the training set. For each sample in the test set, we compute the minimal Euclidean distance to the training set samples in the parameter space (pose, shape, and style). In~\prettyref{fig:unseen}, we show the mean collision error for ``Shirt Male'' grouped by the distance. The soft constraint can only reduce collisions for samples near the training set and even introduces more errors for samples far away. In contrast, ReFU can still resolve some collisions for samples with great differences from the seen training ones.

\section{Visualized Comparison Results}

\begin{figure*}
\centering
\begin{subfigure}[b]{0.2\textwidth}
\centering
\includegraphics[width=\textwidth]{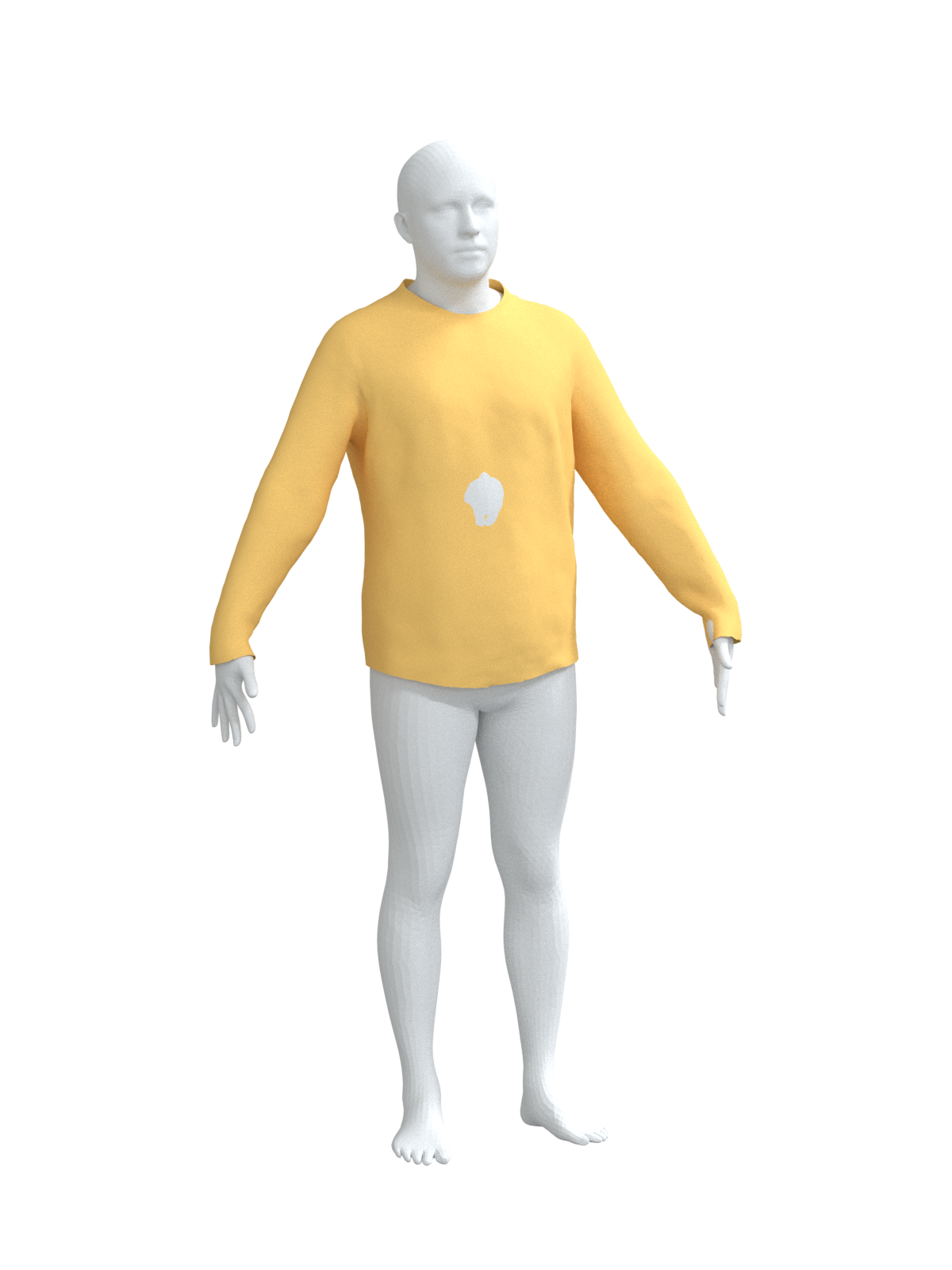}
\end{subfigure}
\hfill
\begin{subfigure}[b]{0.2\textwidth}
\centering
\includegraphics[width=\textwidth]{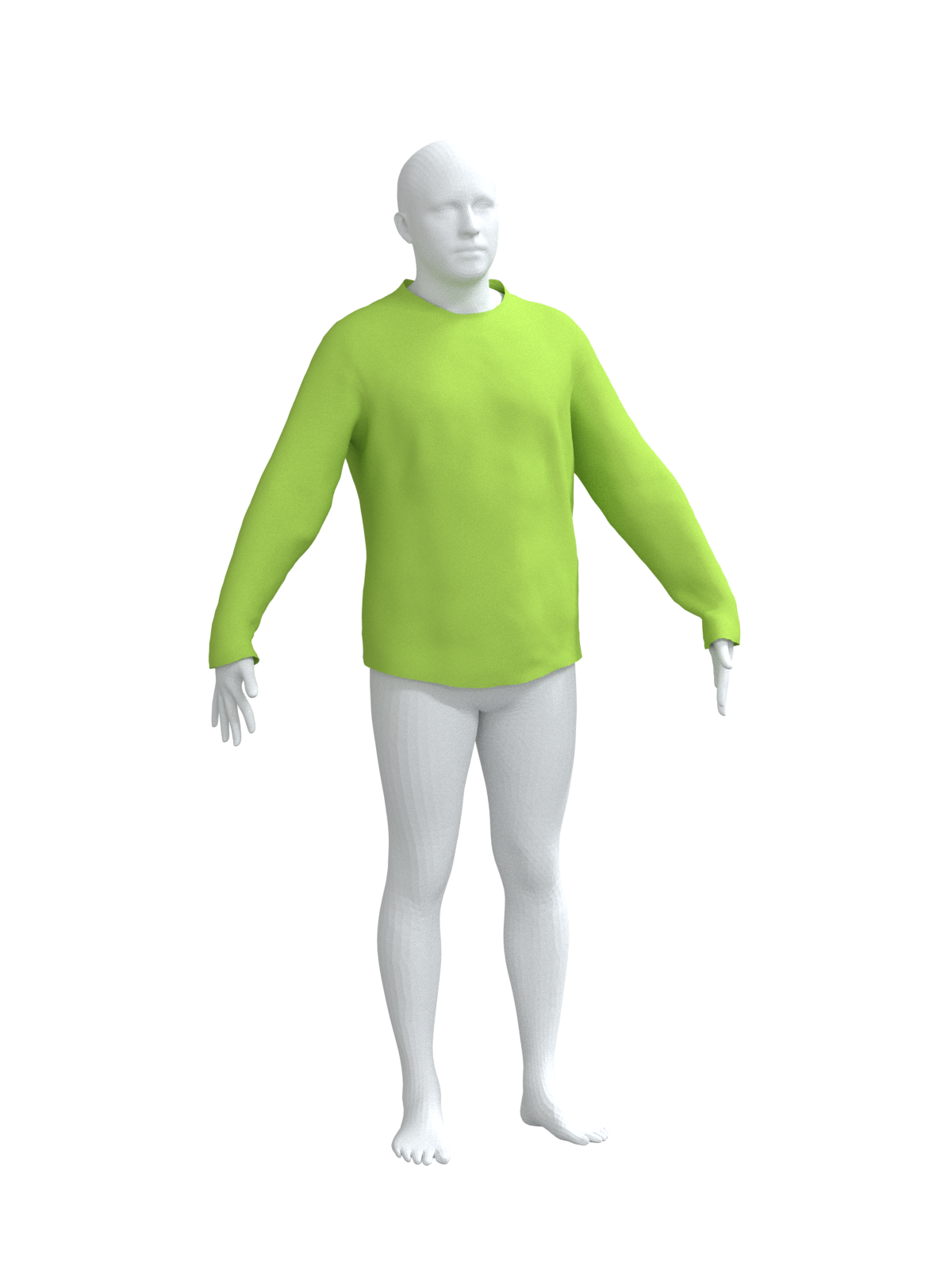}
\end{subfigure}
\hfill
\begin{subfigure}[b]{0.2\textwidth}
\centering
\includegraphics[width=\textwidth]{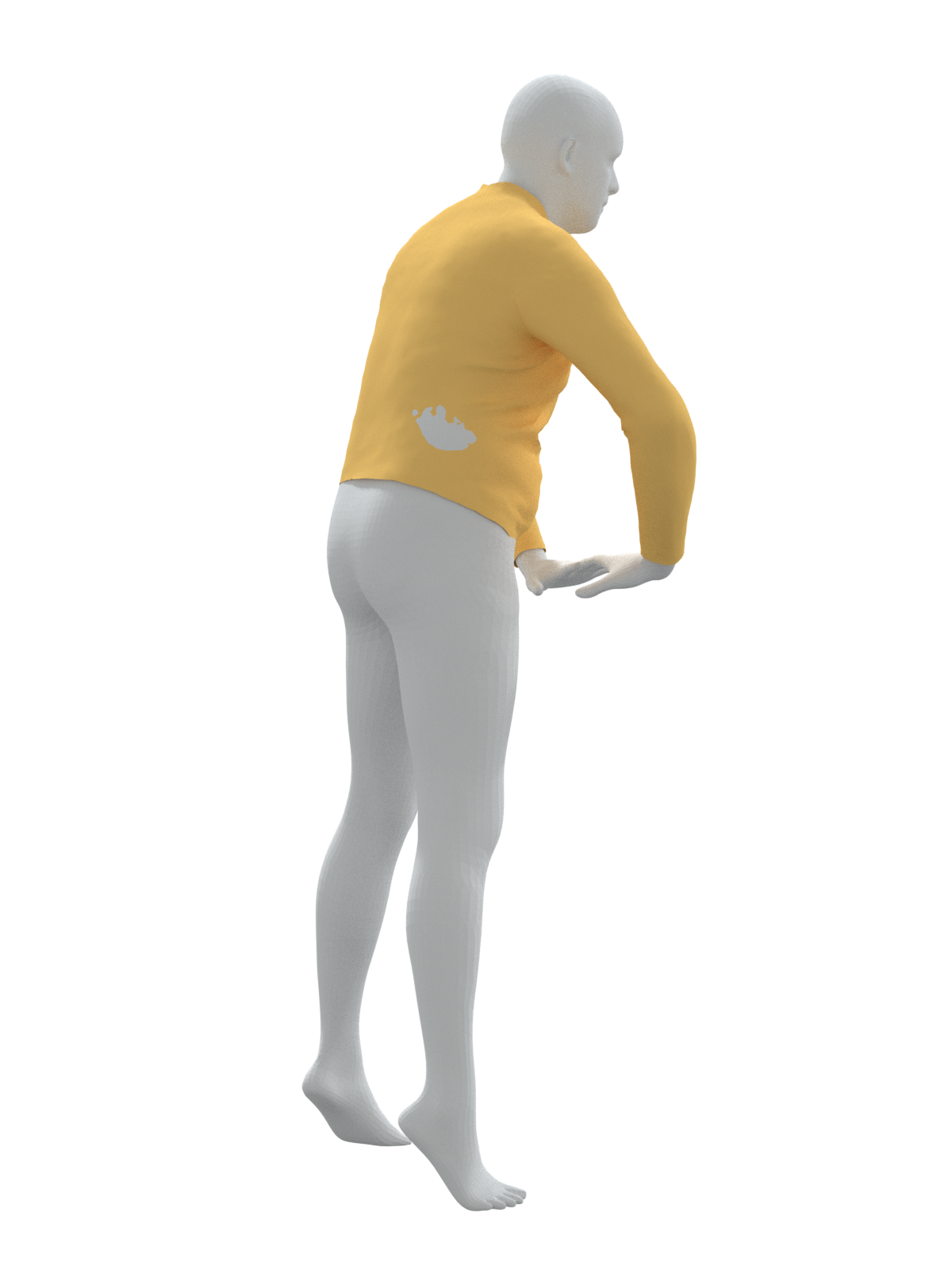}
\end{subfigure}
\hfill
\begin{subfigure}[b]{0.2\textwidth}
\centering
\includegraphics[width=\textwidth]{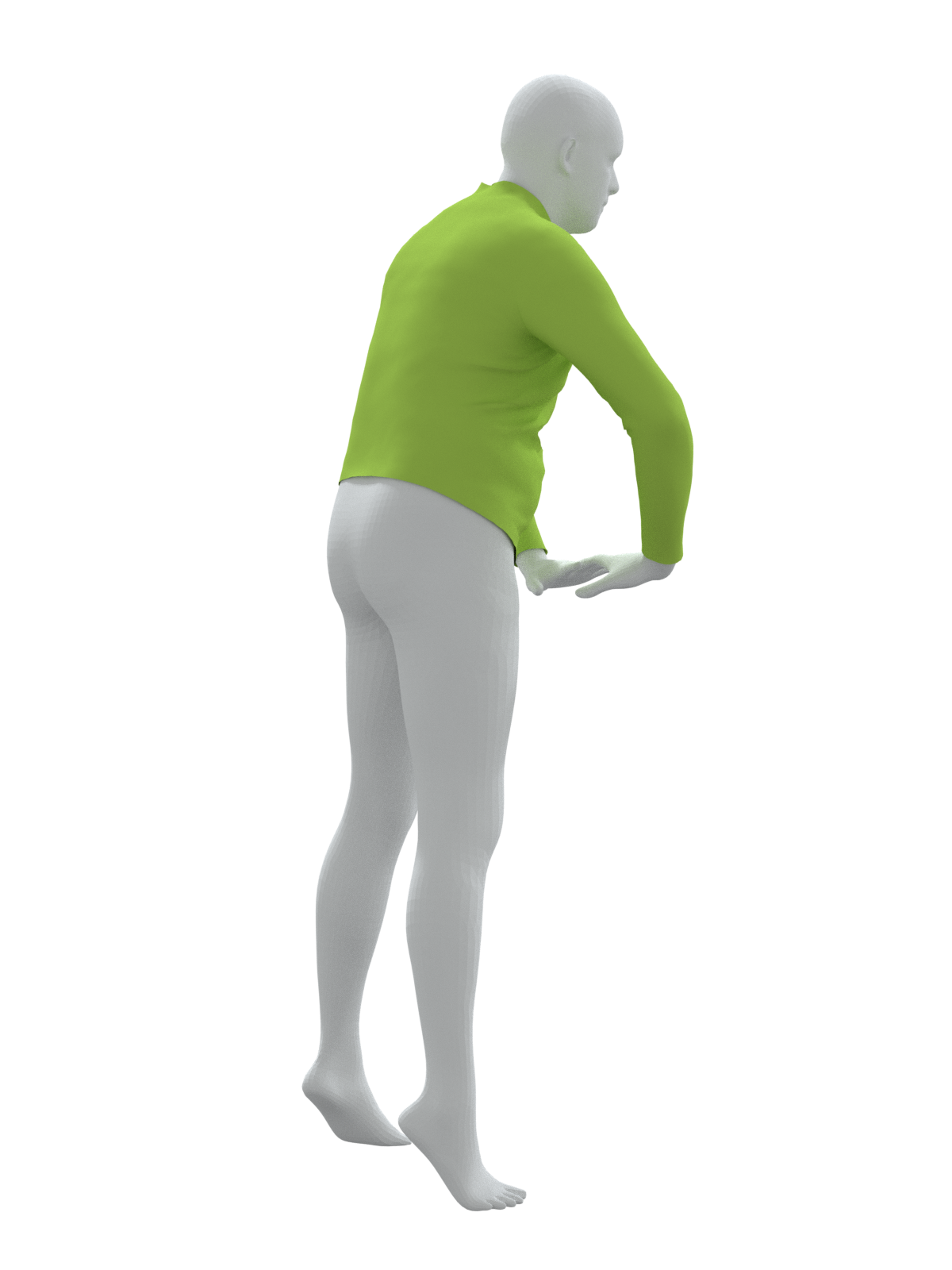}
\end{subfigure}
\begin{subfigure}[b]{0.2\textwidth}
\centering
\includegraphics[width=\textwidth]{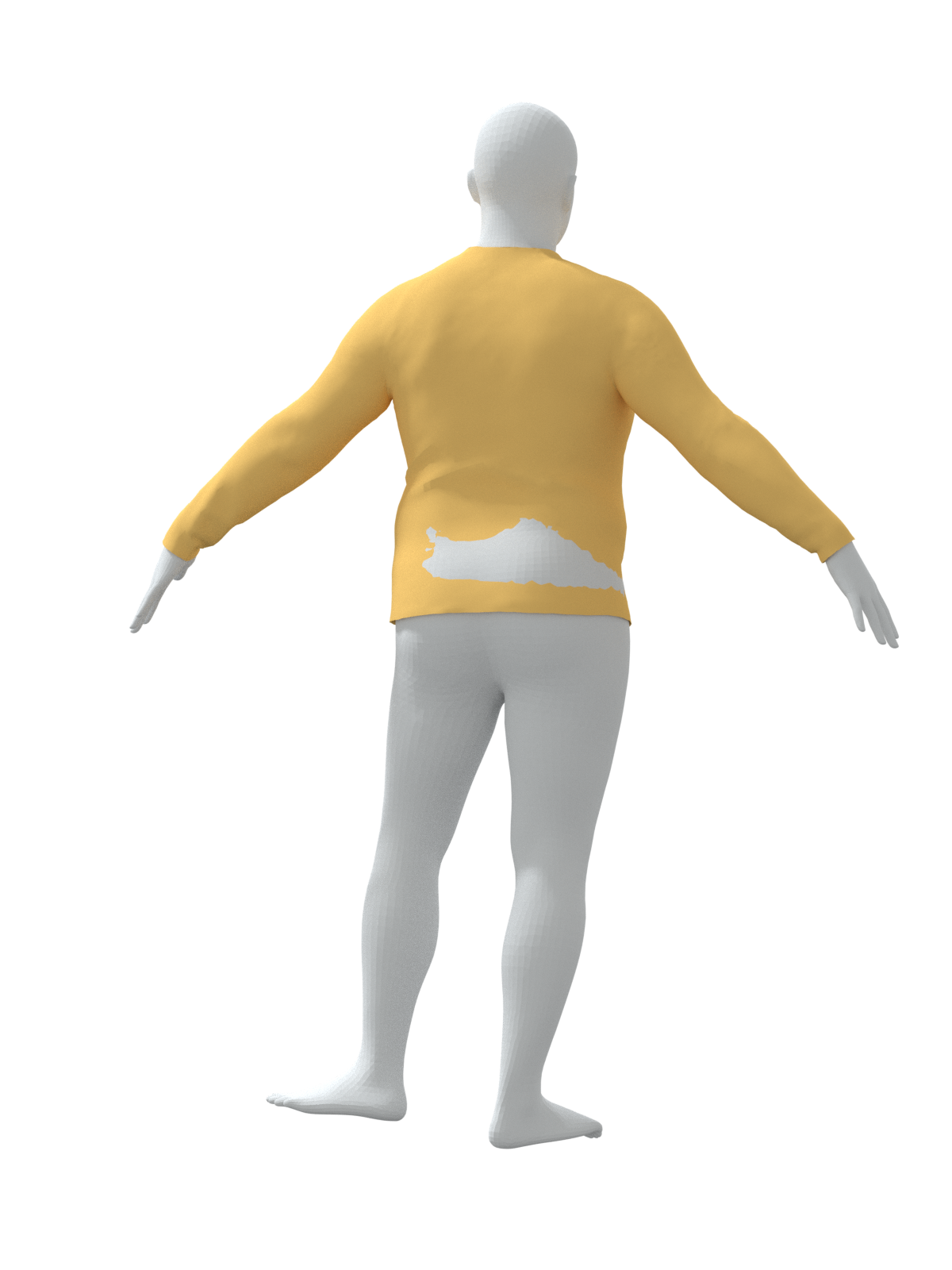}
\end{subfigure}
\hfill
\begin{subfigure}[b]{0.2\textwidth}
\centering
\includegraphics[width=\textwidth]{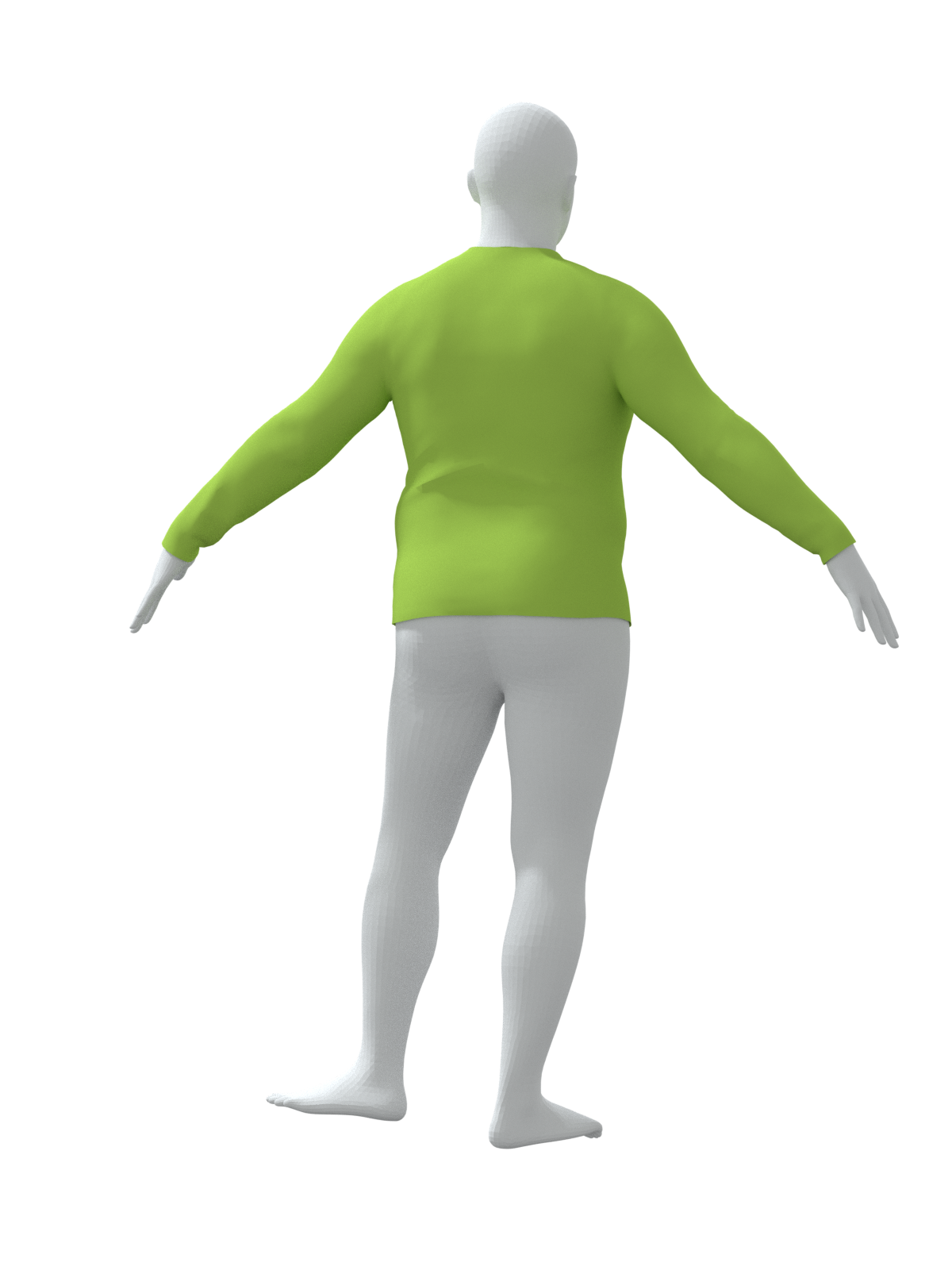}
\end{subfigure}
\hfill
\begin{subfigure}[b]{0.2\textwidth}
\centering
\includegraphics[width=\textwidth]{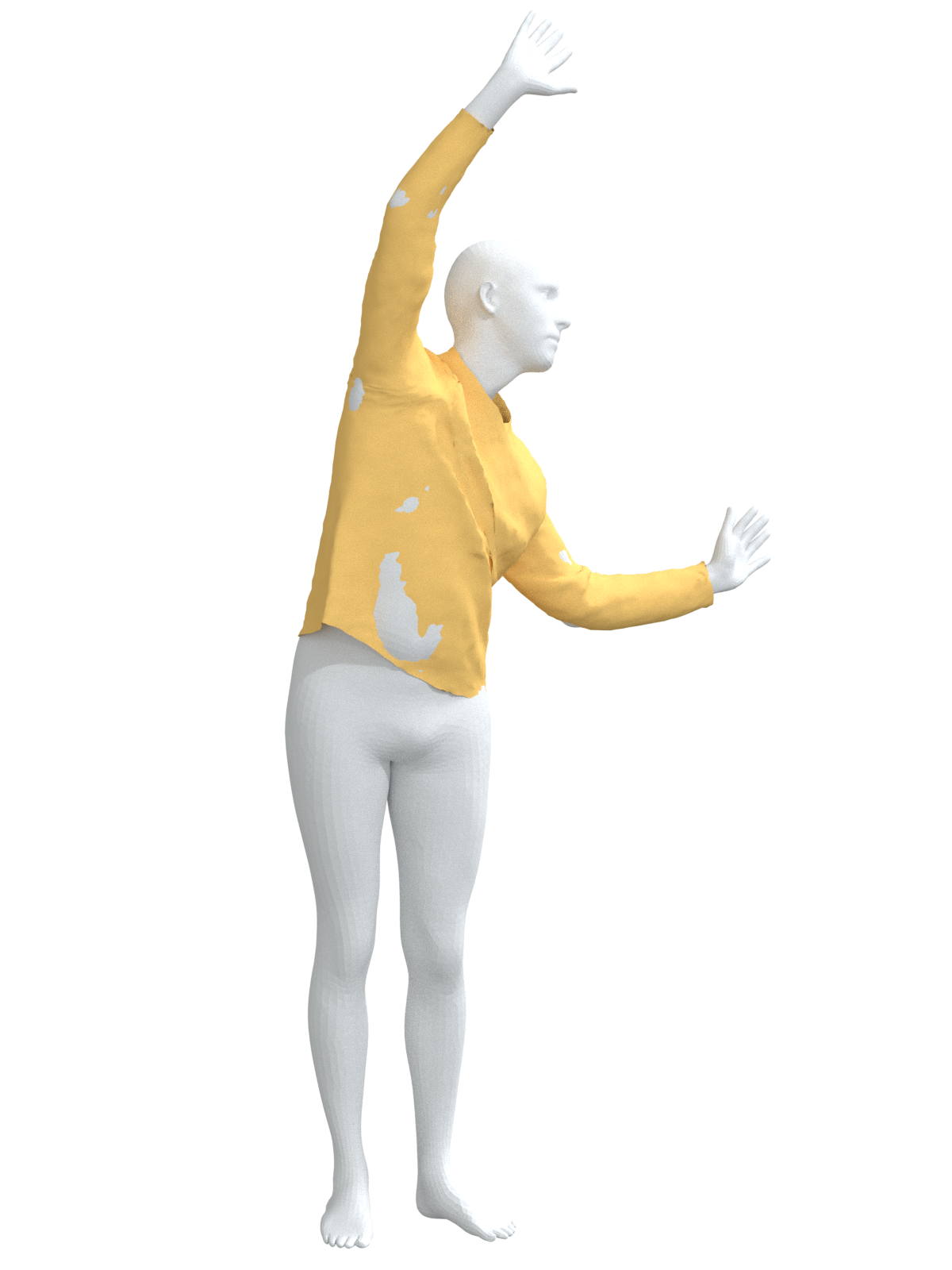}
\end{subfigure}
\hfill
\begin{subfigure}[b]{0.2\textwidth}
\centering
\includegraphics[width=\textwidth]{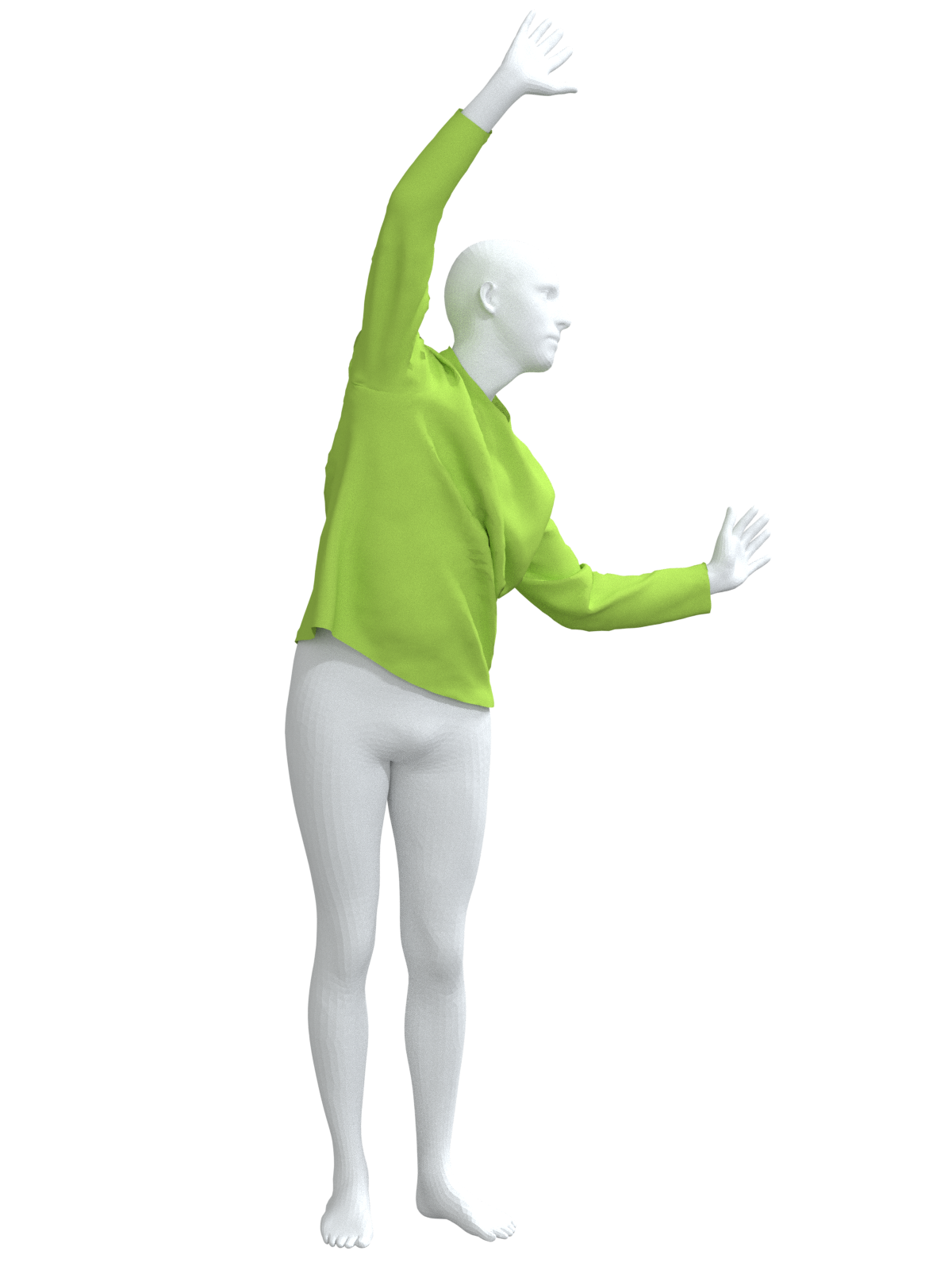}
\end{subfigure}
\begin{subfigure}[b]{0.2\textwidth}
\centering
\includegraphics[width=\textwidth]{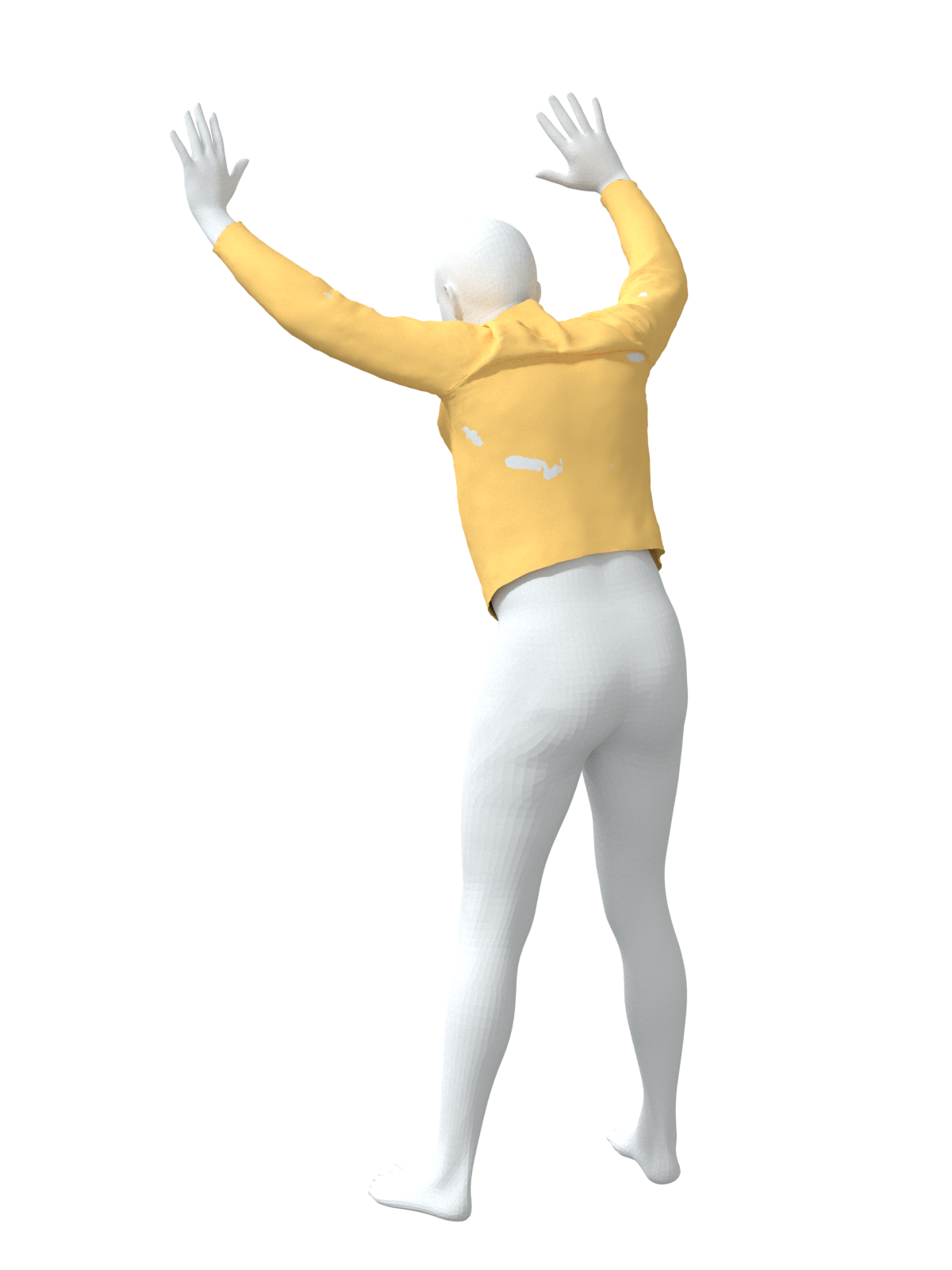}
\end{subfigure}
\hfill
\begin{subfigure}[b]{0.2\textwidth}
\centering
\includegraphics[width=\textwidth]{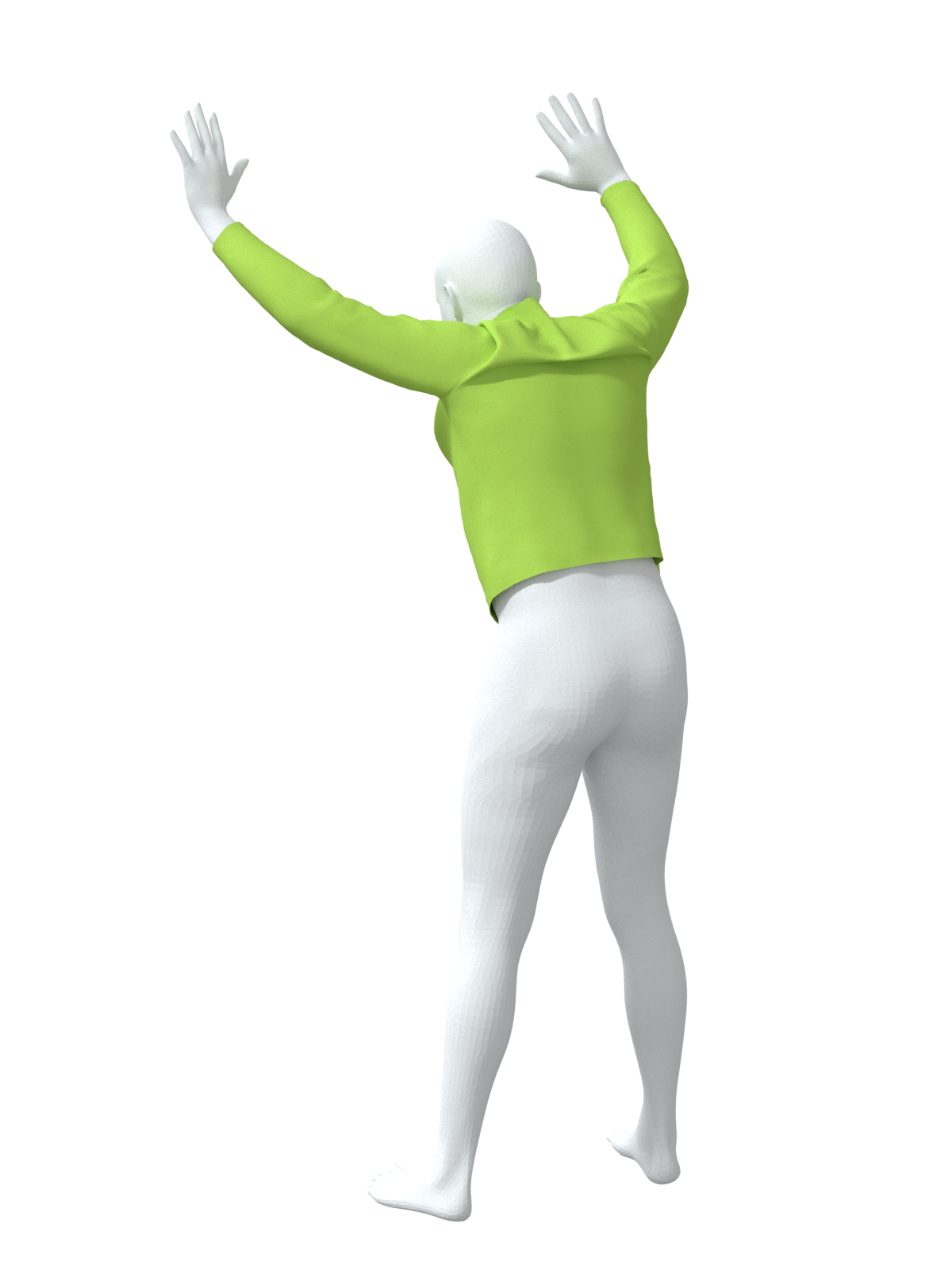}
\end{subfigure}
\hfill
\begin{subfigure}[b]{0.2\textwidth}
\centering
\includegraphics[width=\textwidth]{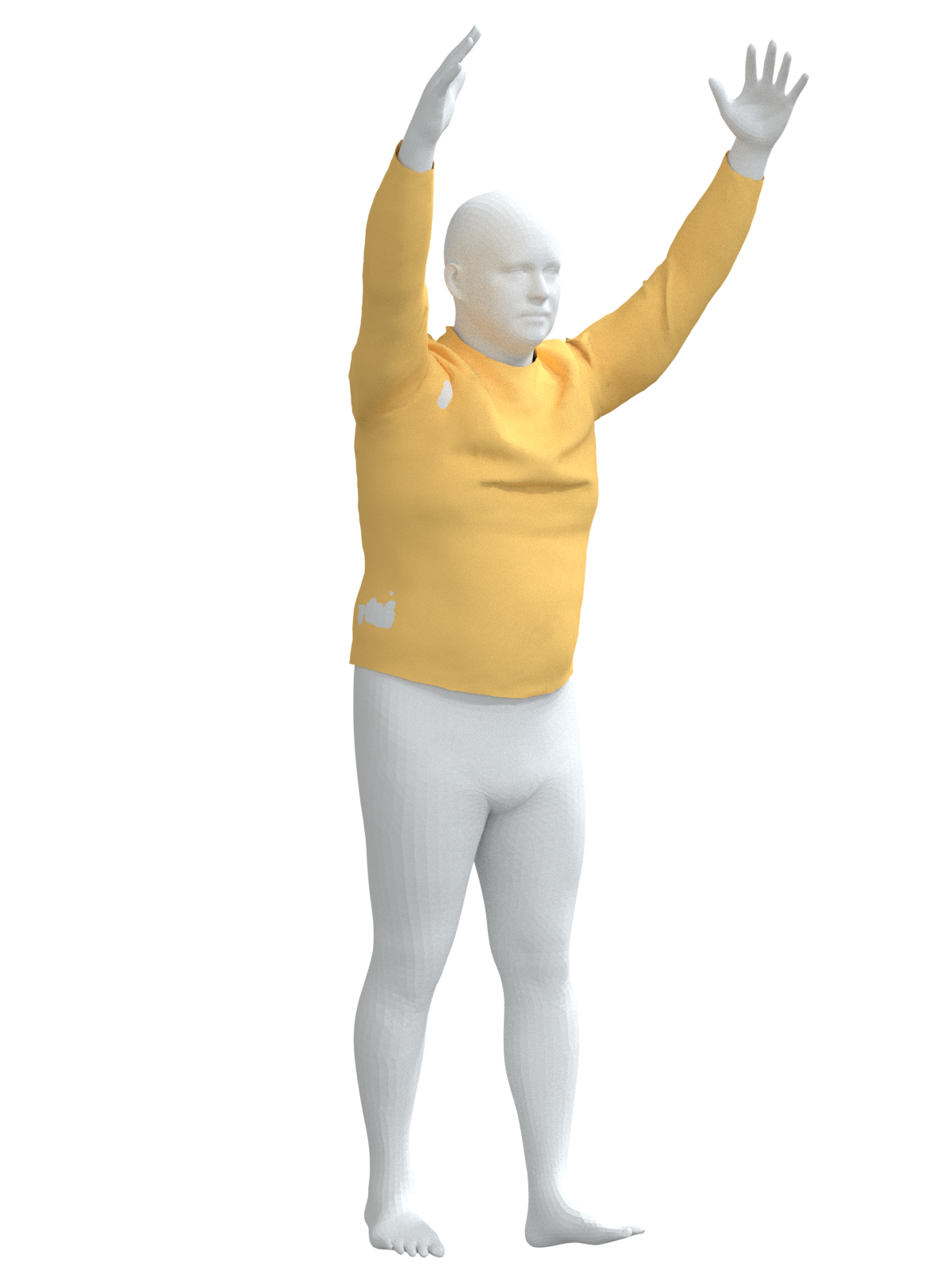}
\end{subfigure}
\hfill
\begin{subfigure}[b]{0.2\textwidth}
\centering
\includegraphics[width=\textwidth]{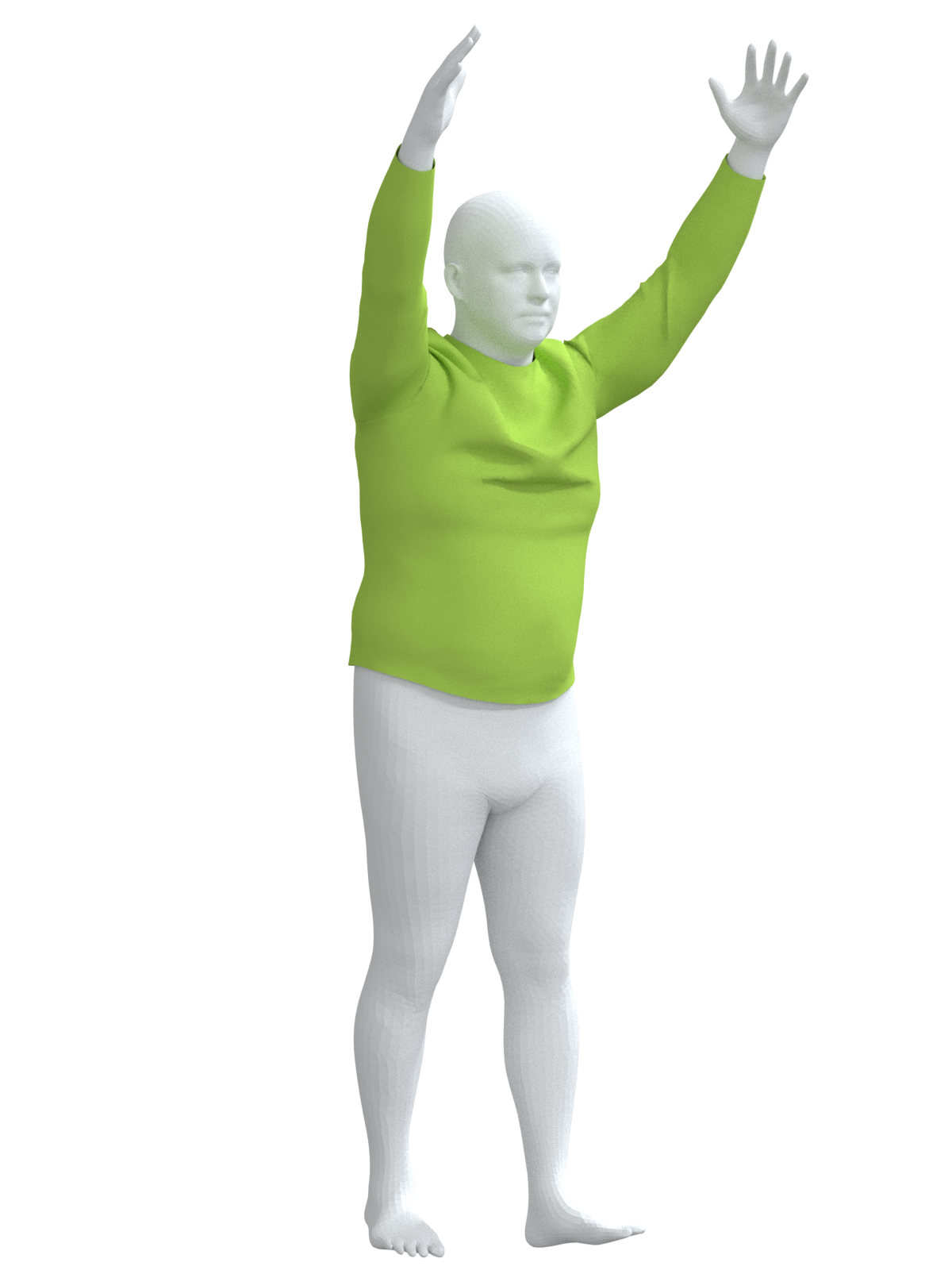}
\end{subfigure}
\caption{\label{fig:add_shirt} Additional examples from Shirt Male dataset, showing collisions resolved by applying our ReFU in TailorNet.}
\end{figure*}

\begin{figure*}
\centering
\begin{subfigure}[b]{0.2\textwidth}
\centering
\includegraphics[width=\textwidth]{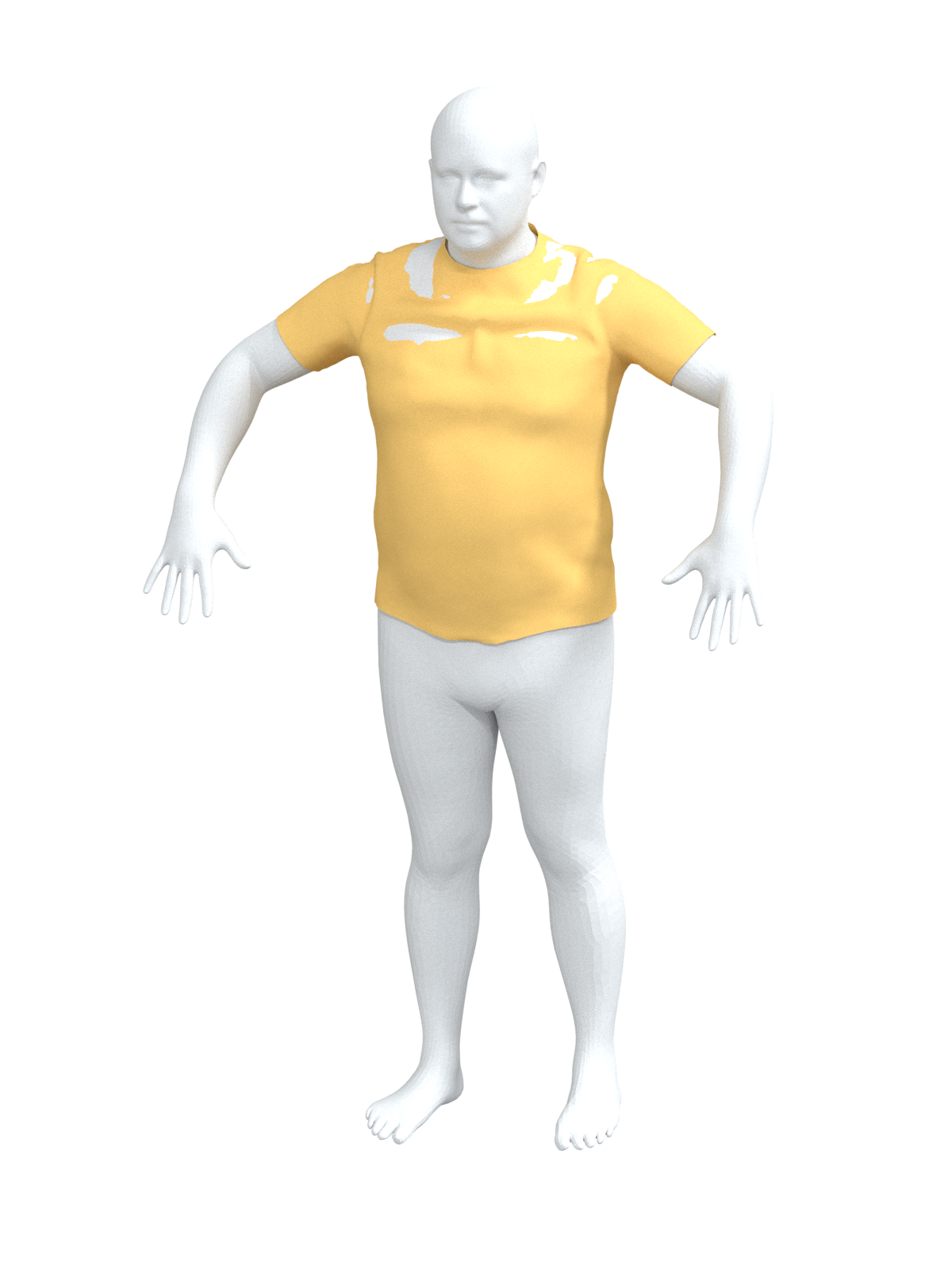}
\end{subfigure}
\hfill
\begin{subfigure}[b]{0.2\textwidth}
\centering
\includegraphics[width=\textwidth]{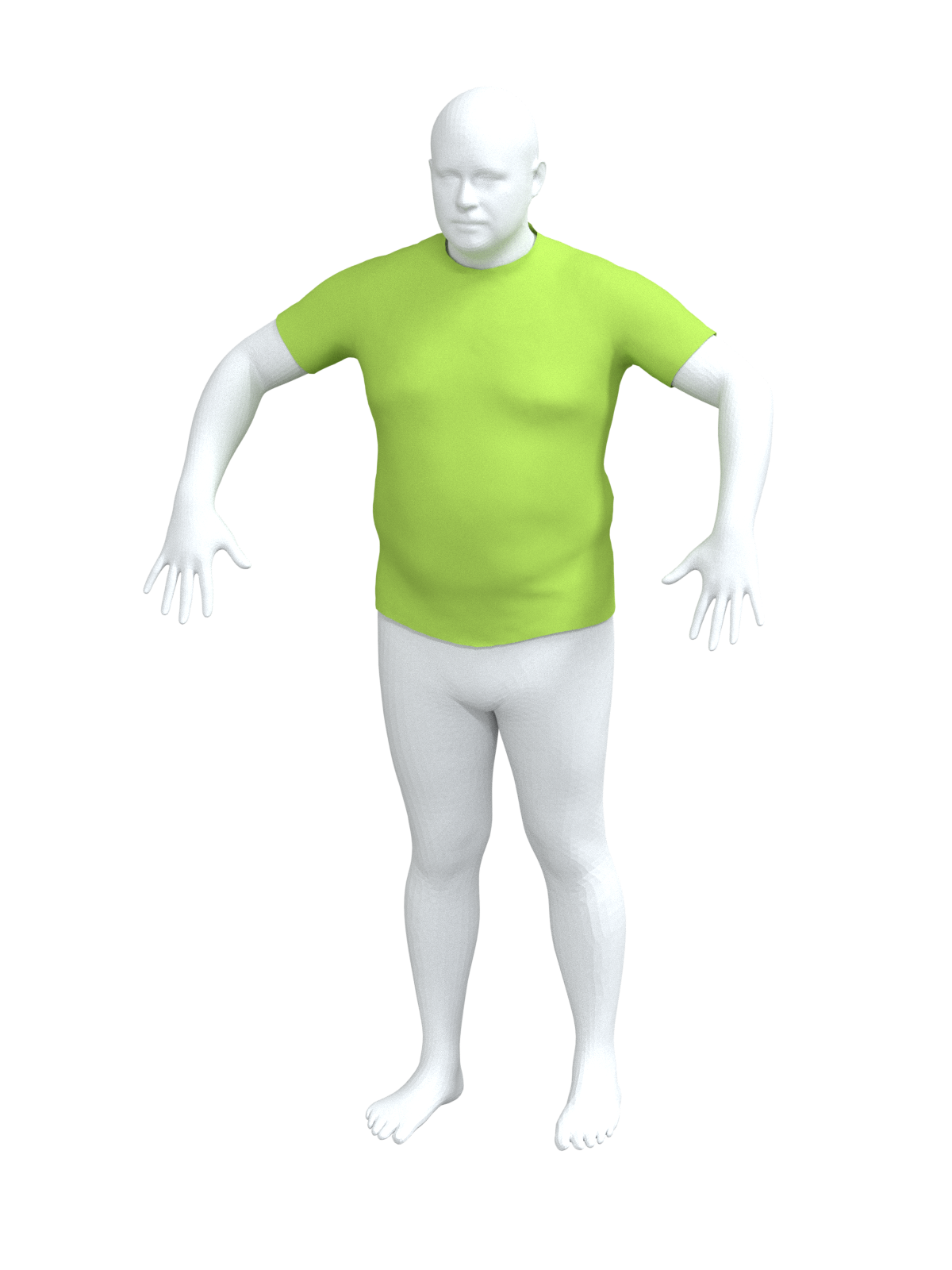}
\end{subfigure}
\hfill
\begin{subfigure}[b]{0.2\textwidth}
\centering
\includegraphics[width=\textwidth]{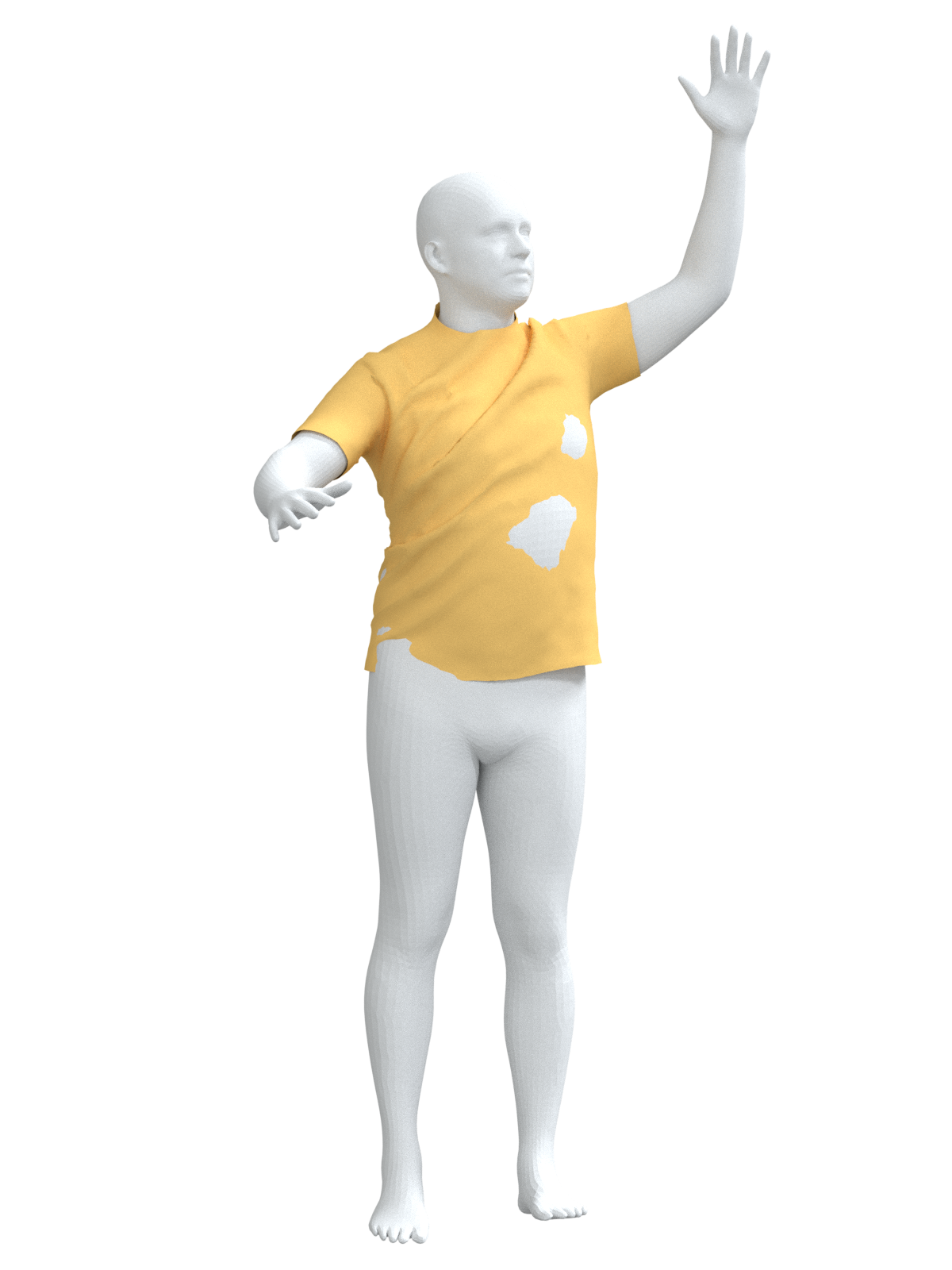}
\end{subfigure}
\hfill
\begin{subfigure}[b]{0.2\textwidth}
\centering
\includegraphics[width=\textwidth]{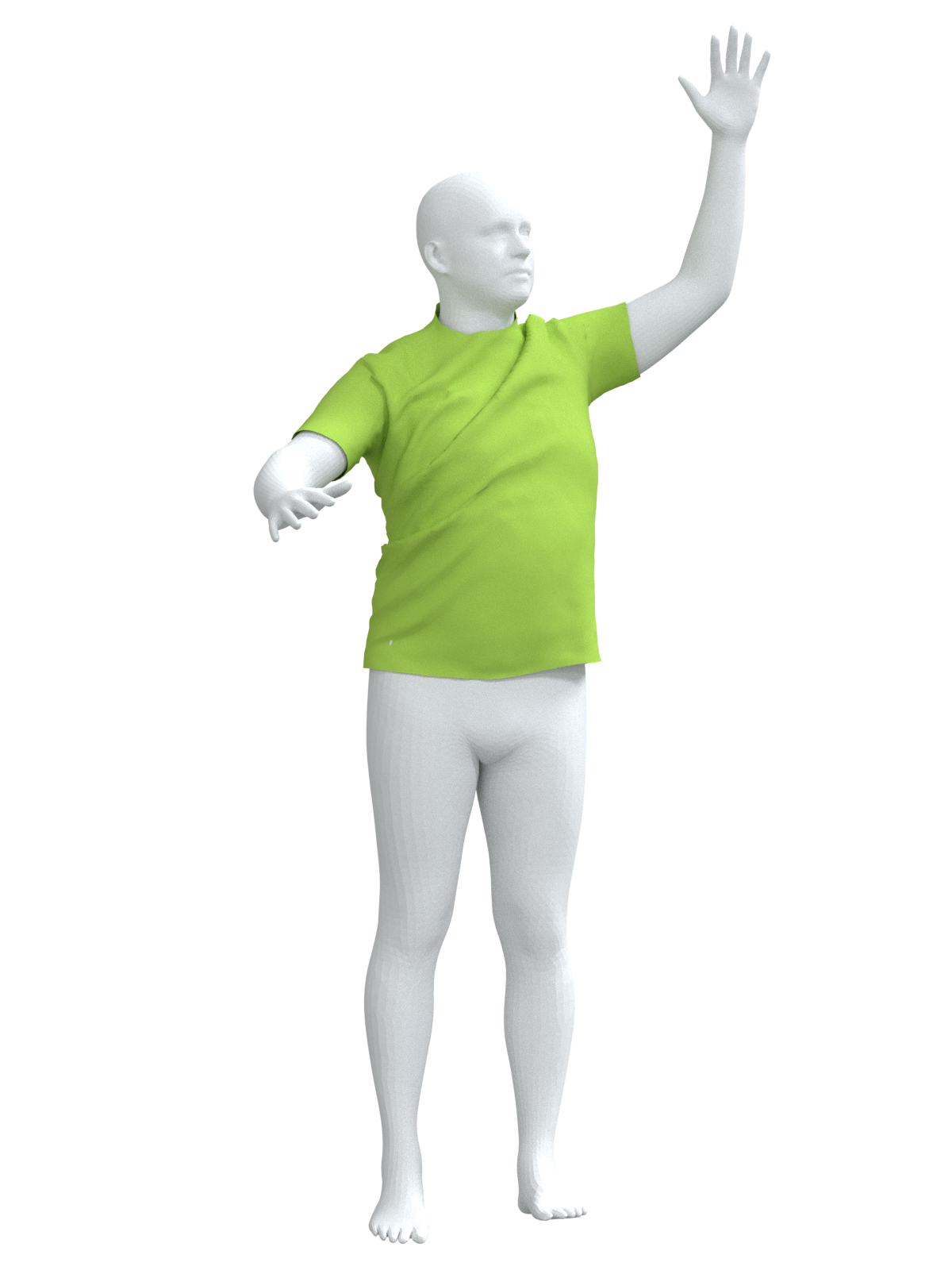}
\end{subfigure}
\begin{subfigure}[b]{0.2\textwidth}
\centering
\includegraphics[width=\textwidth]{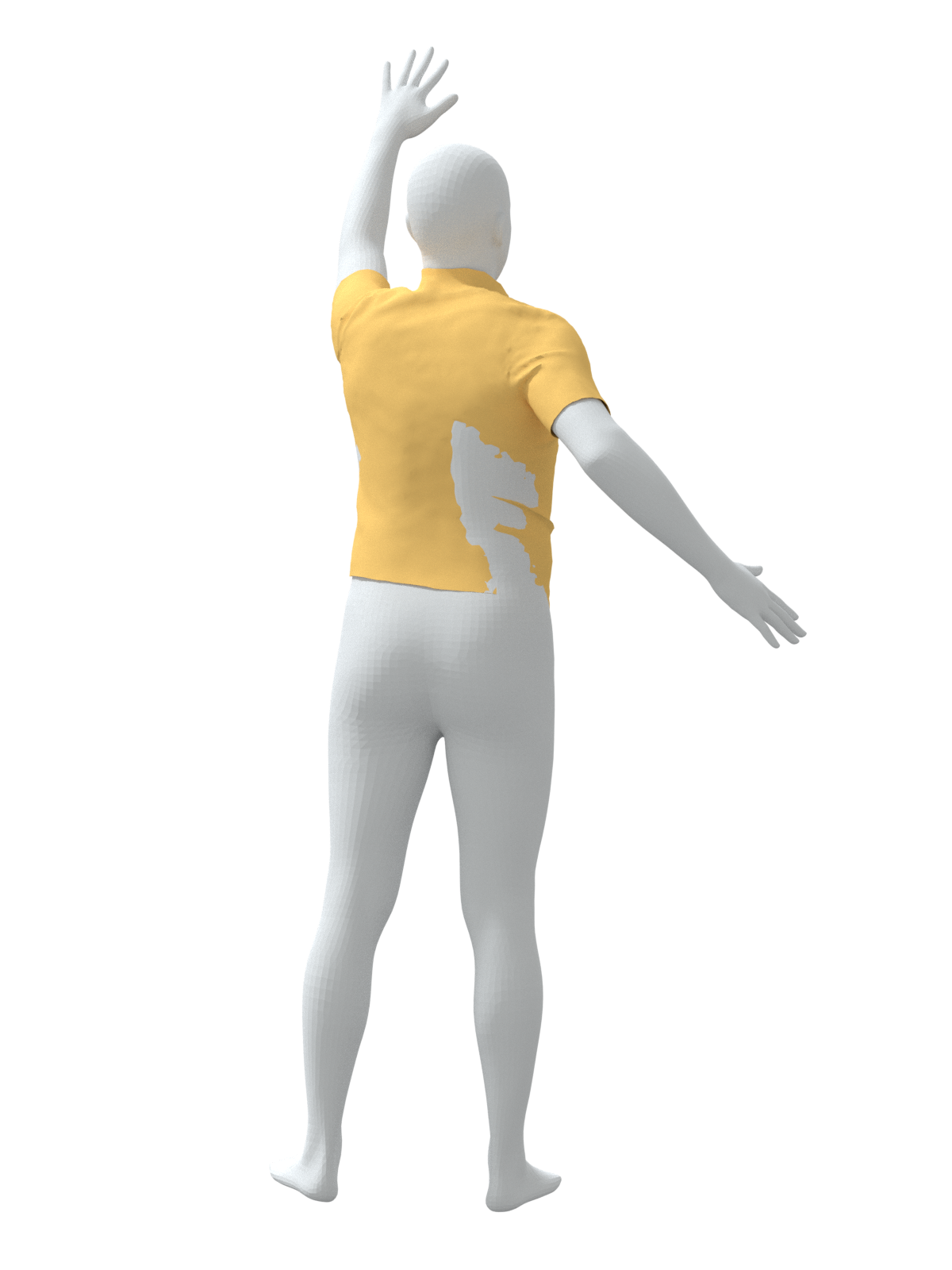}
\end{subfigure}
\hfill
\begin{subfigure}[b]{0.2\textwidth}
\centering
\includegraphics[width=\textwidth]{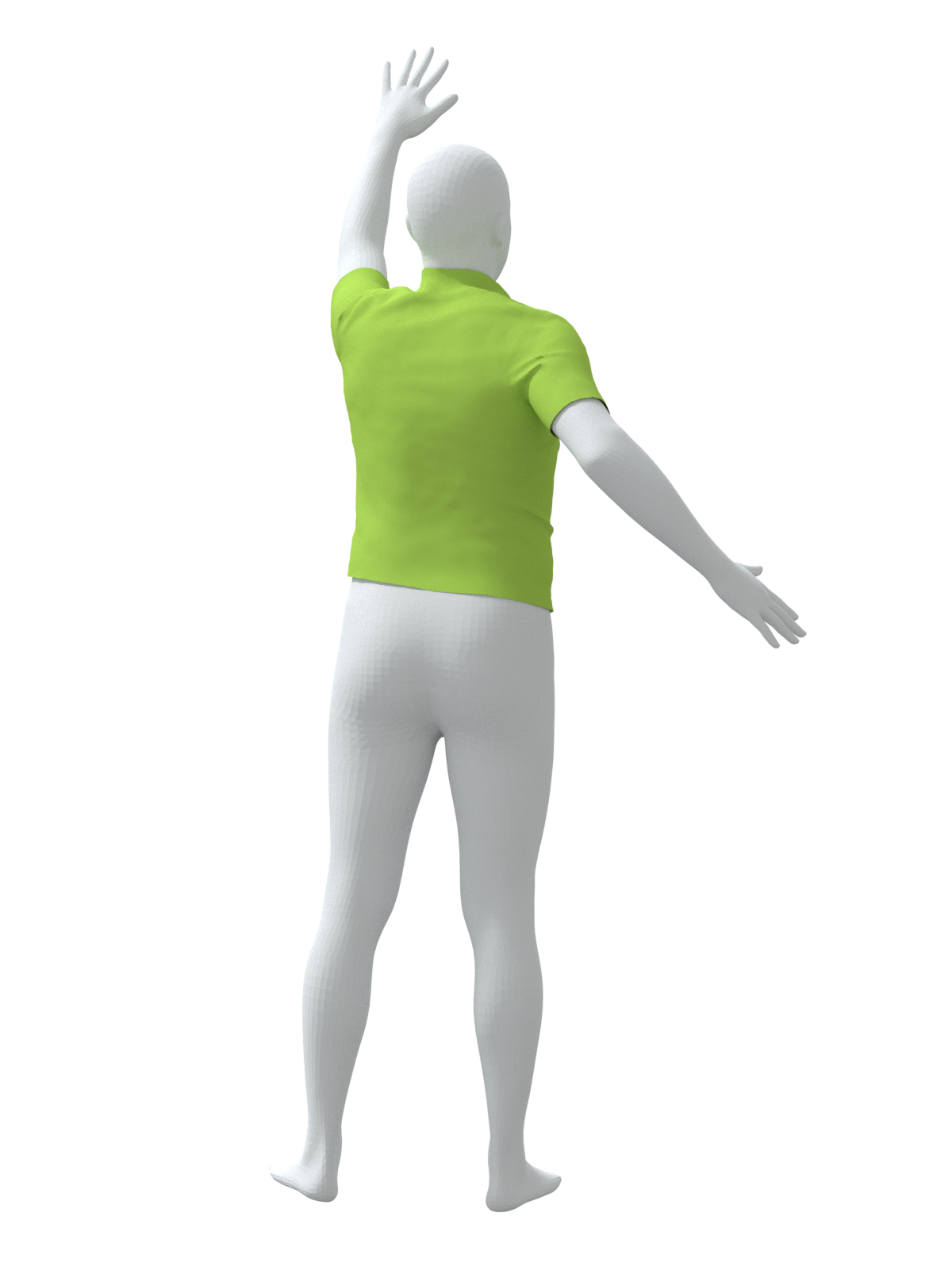}
\end{subfigure}
\hfill
\begin{subfigure}[b]{0.2\textwidth}
\centering
\includegraphics[width=\textwidth]{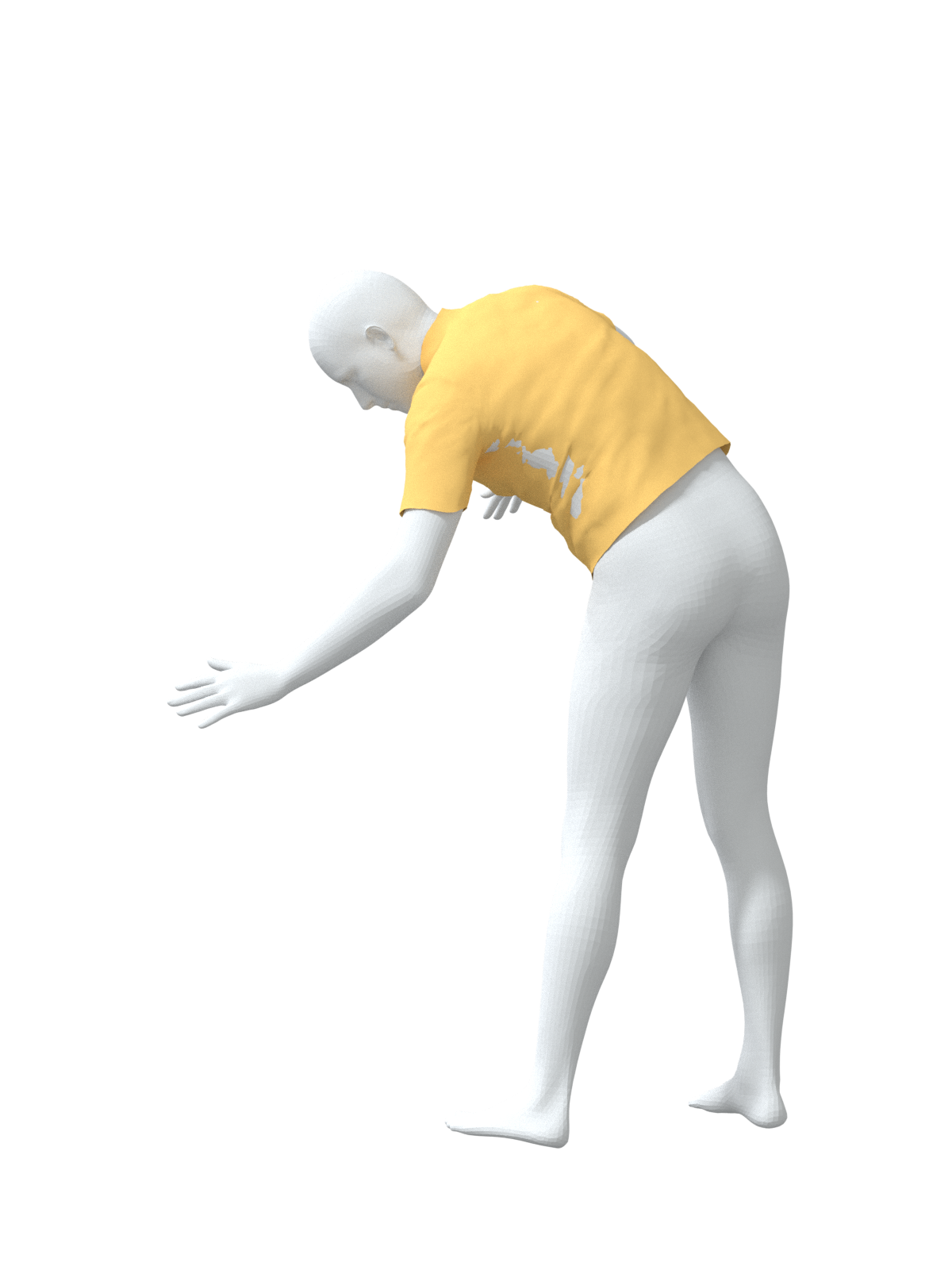}
\end{subfigure}
\hfill
\begin{subfigure}[b]{0.2\textwidth}
\centering
\includegraphics[width=\textwidth]{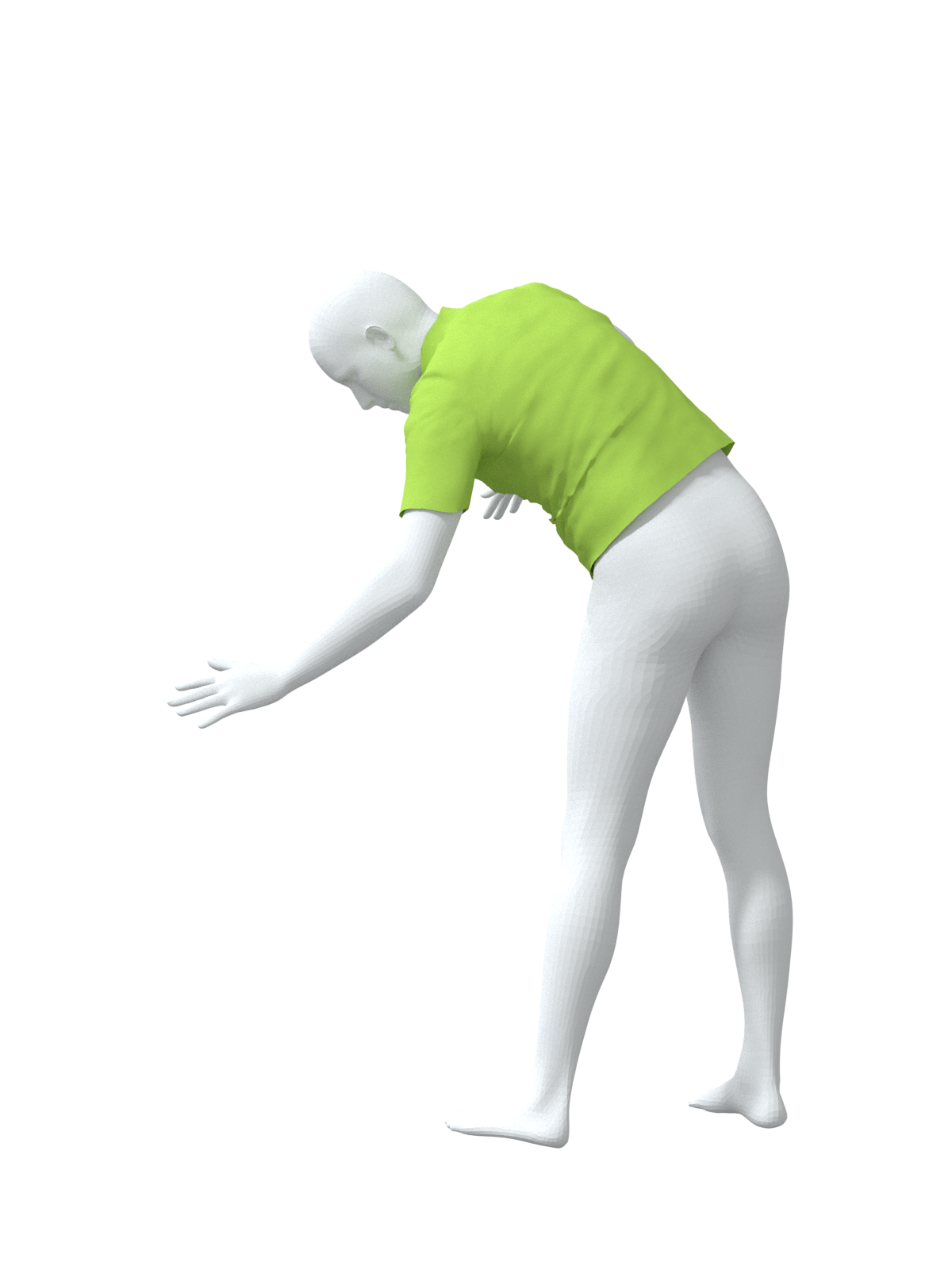}
\end{subfigure}
\begin{subfigure}[b]{0.2\textwidth}
\centering
\includegraphics[width=\textwidth]{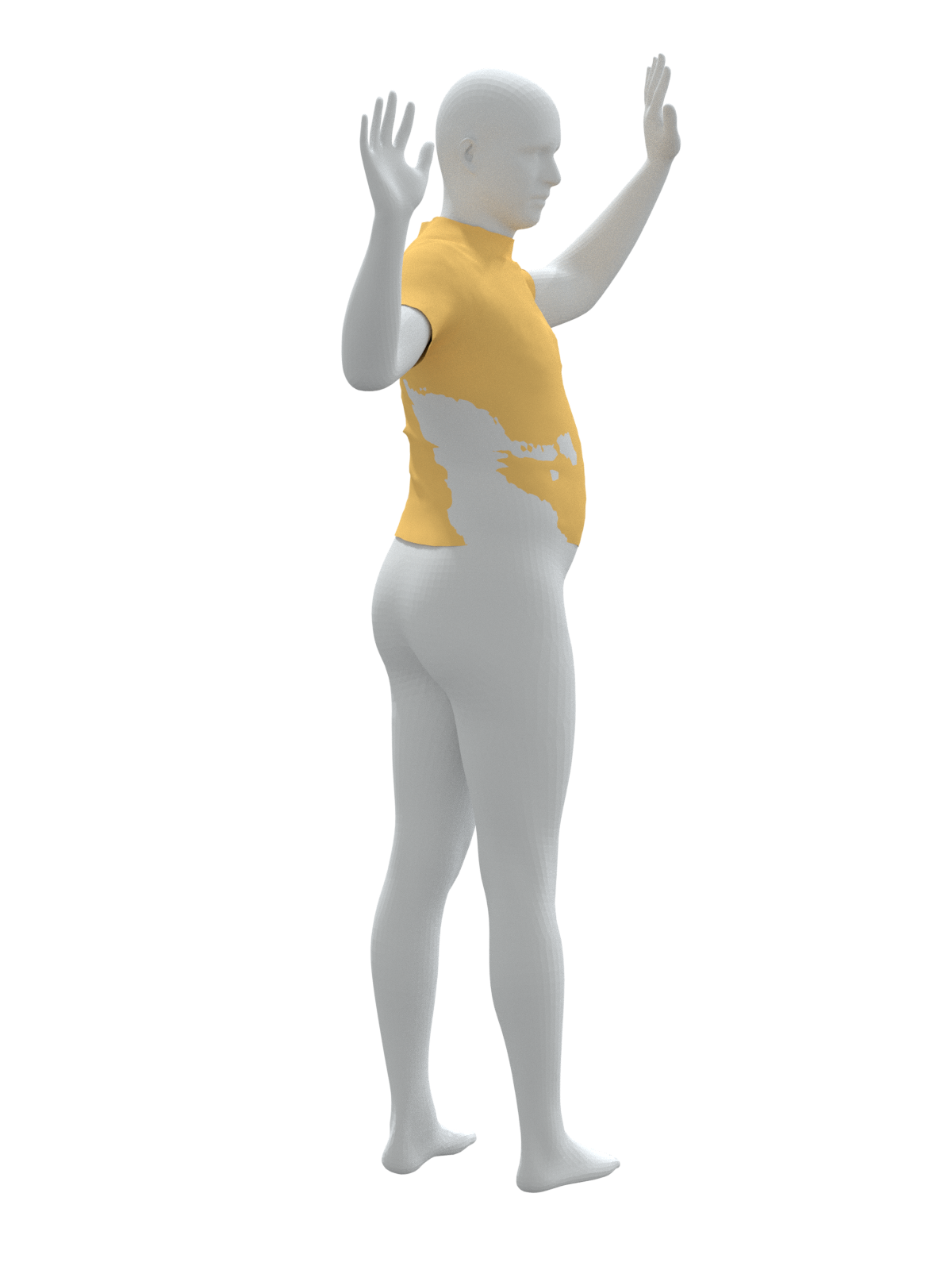}
\end{subfigure}
\hfill
\begin{subfigure}[b]{0.2\textwidth}
\centering
\includegraphics[width=\textwidth]{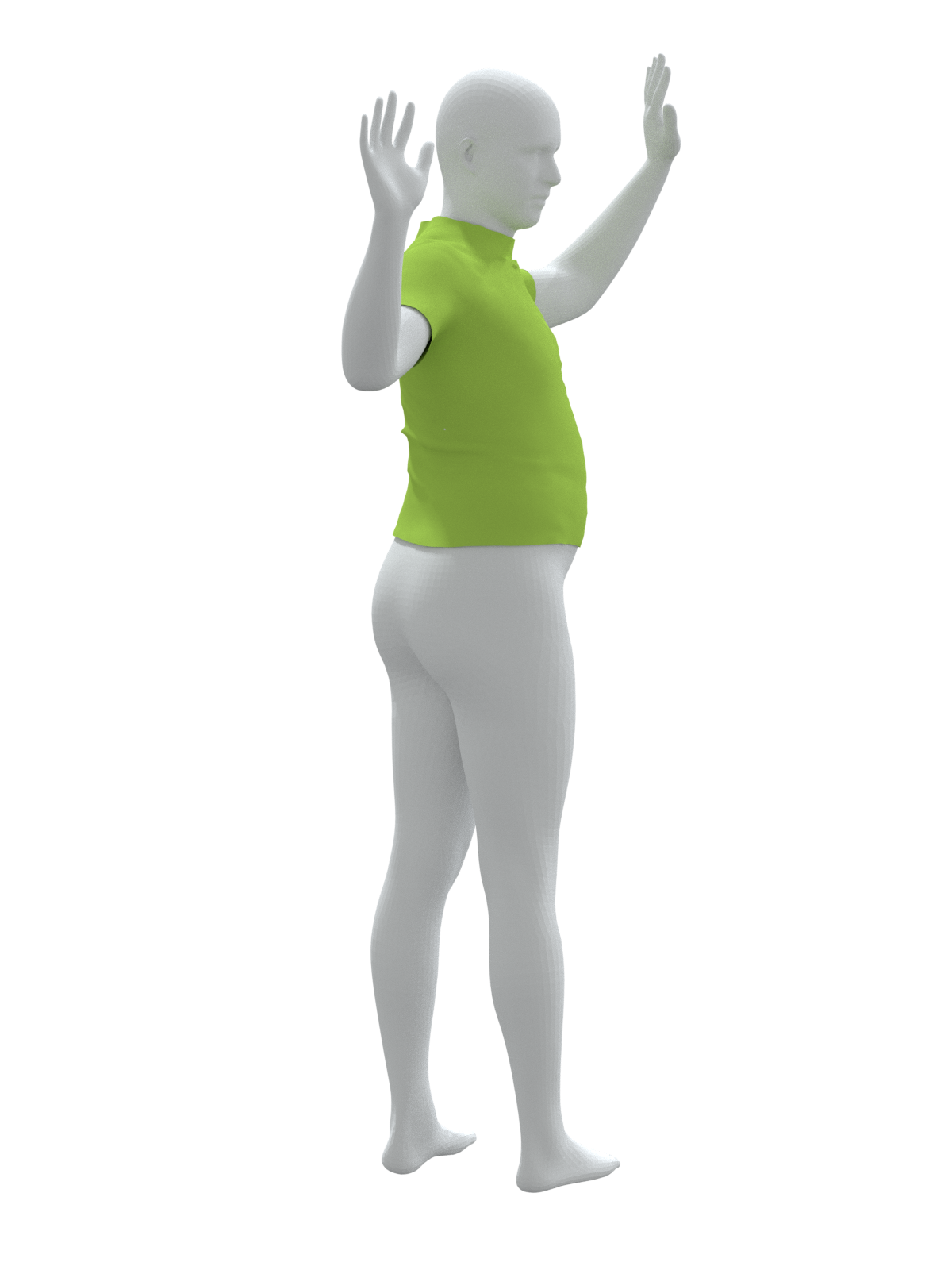}
\end{subfigure}
\hfill
\begin{subfigure}[b]{0.2\textwidth}
\centering
\includegraphics[width=\textwidth]{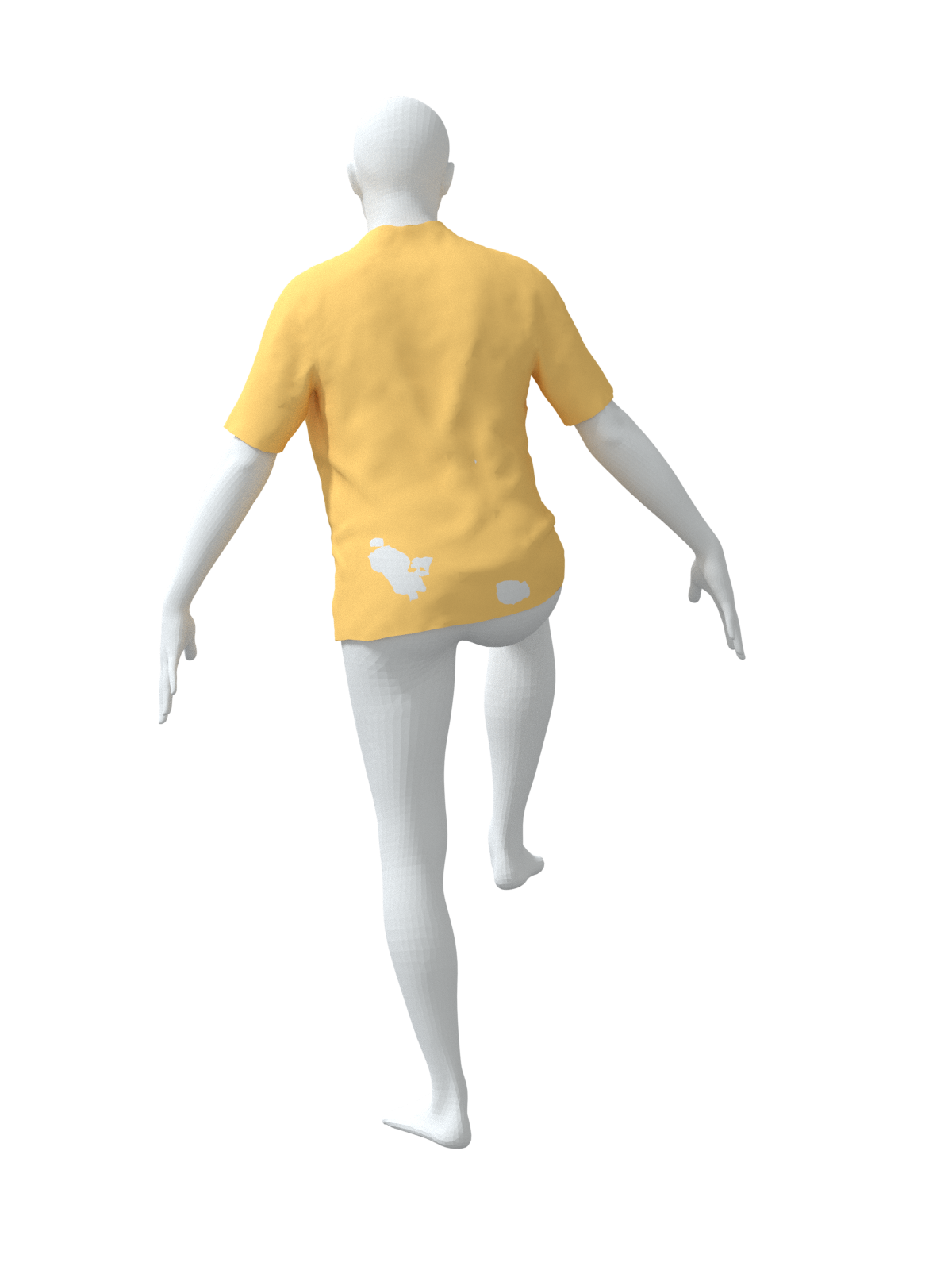}
\end{subfigure}
\hfill
\begin{subfigure}[b]{0.2\textwidth}
\centering
\includegraphics[width=\textwidth]{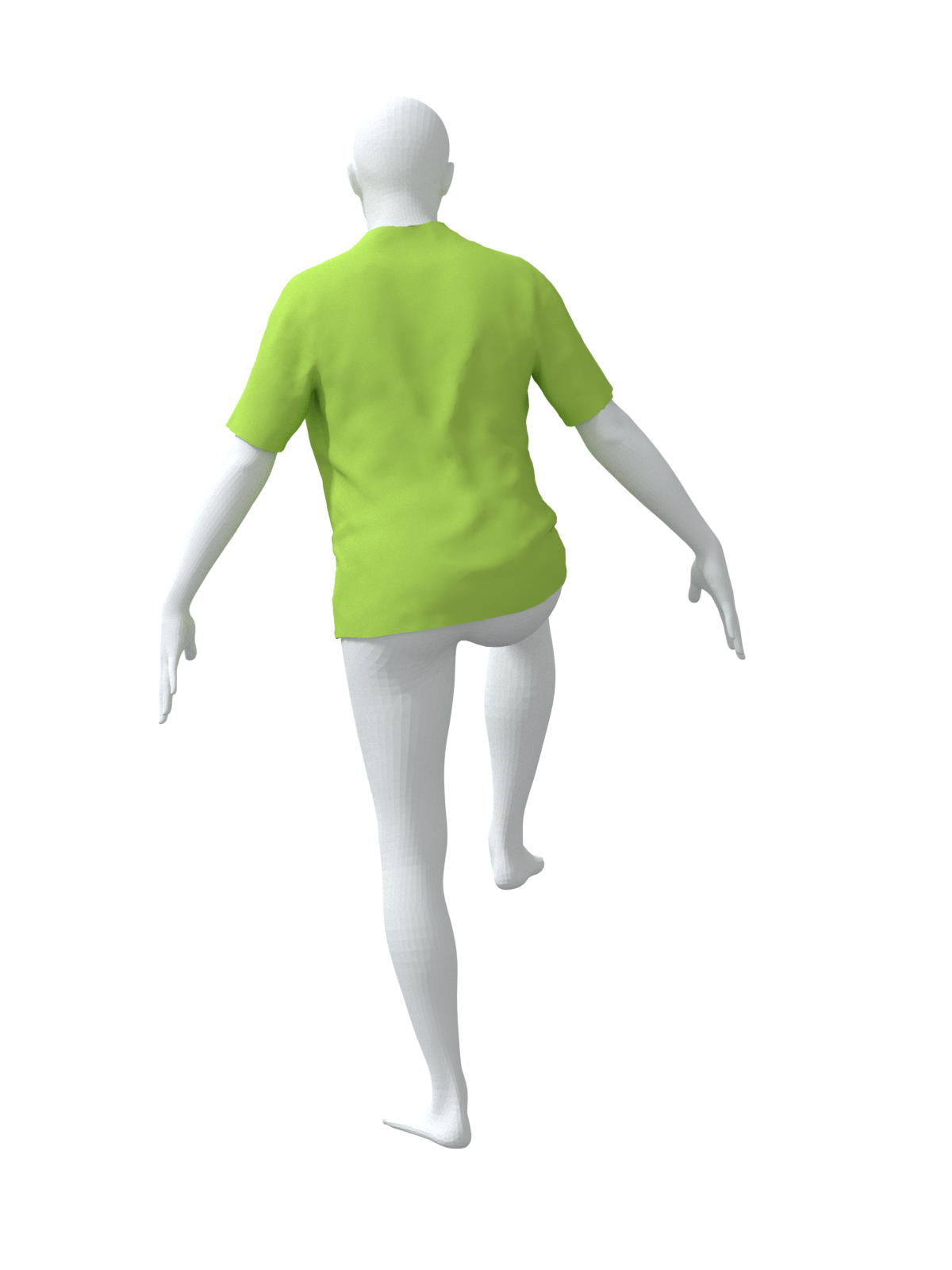}
\end{subfigure}
\caption{\label{fig:add_tshirt} Additional examples from T-Shirt Male dataset, showing collisions resolved by applying our ReFU in TailorNet.}
\end{figure*}

\begin{figure*}
\centering
\begin{subfigure}[b]{0.2\textwidth}
\centering
\includegraphics[width=\textwidth]{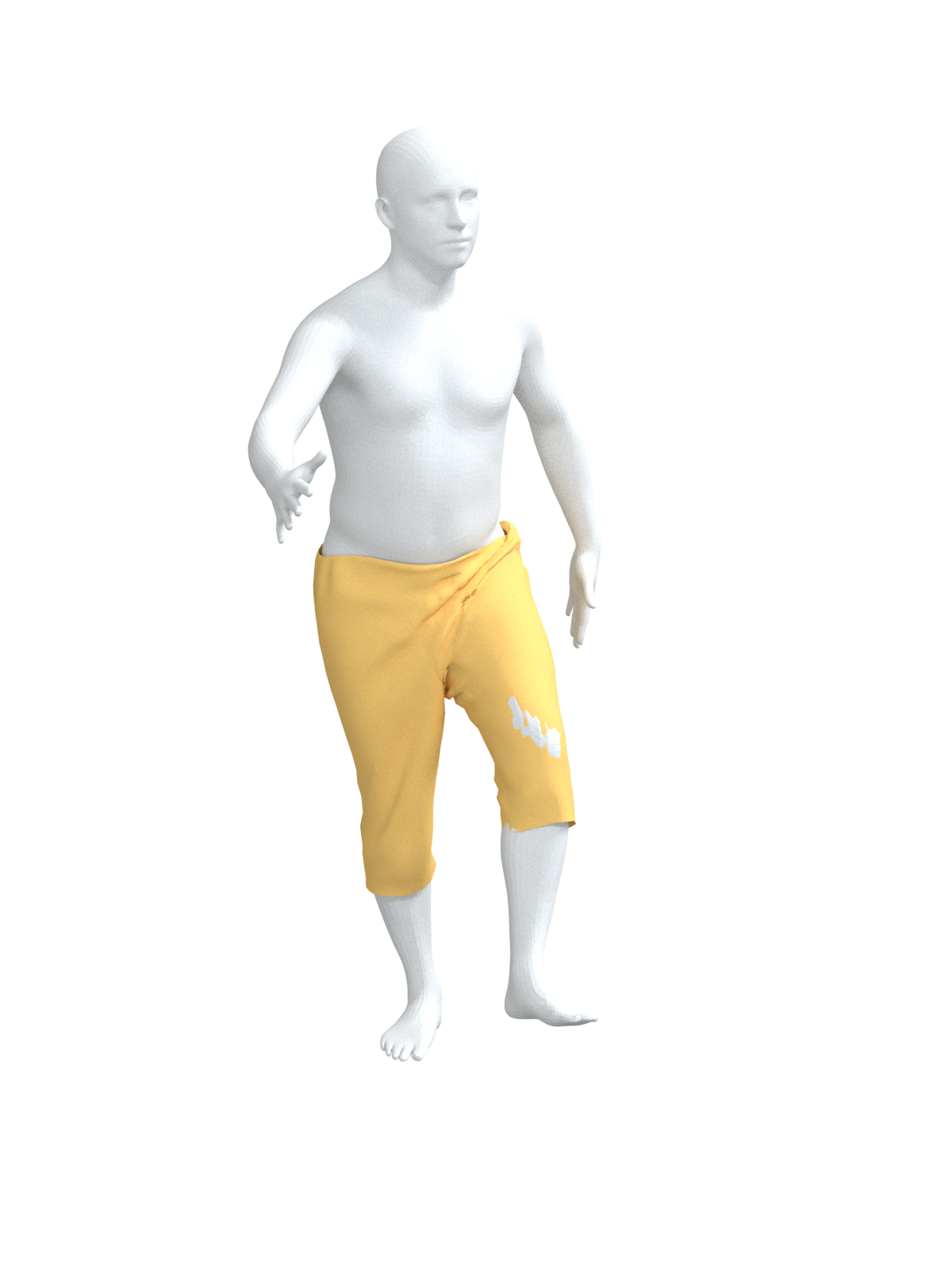}
\end{subfigure}
\hfill
\begin{subfigure}[b]{0.2\textwidth}
\centering
\includegraphics[width=\textwidth]{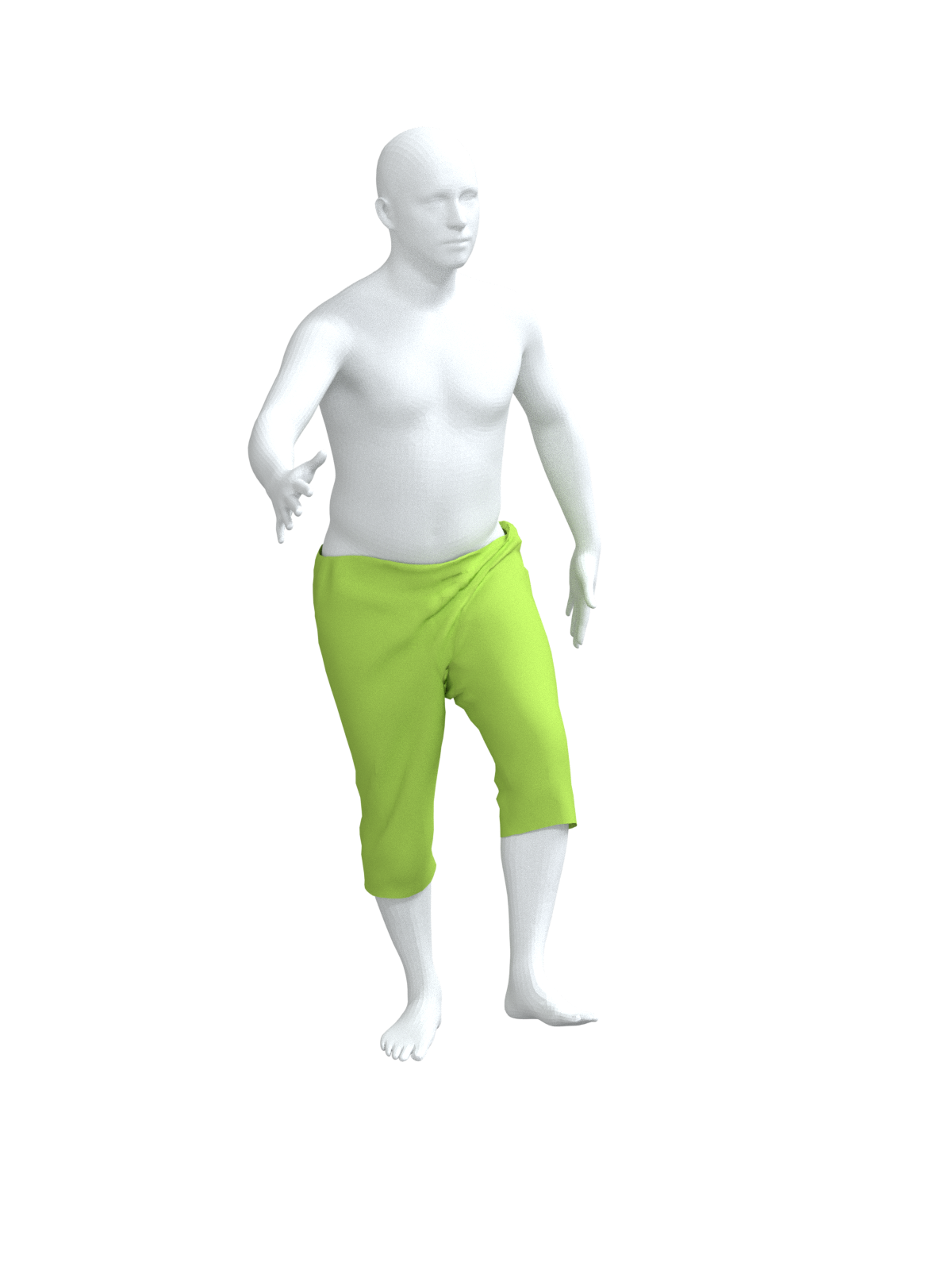}
\end{subfigure}
\hfill
\begin{subfigure}[b]{0.2\textwidth}
\centering
\includegraphics[width=\textwidth]{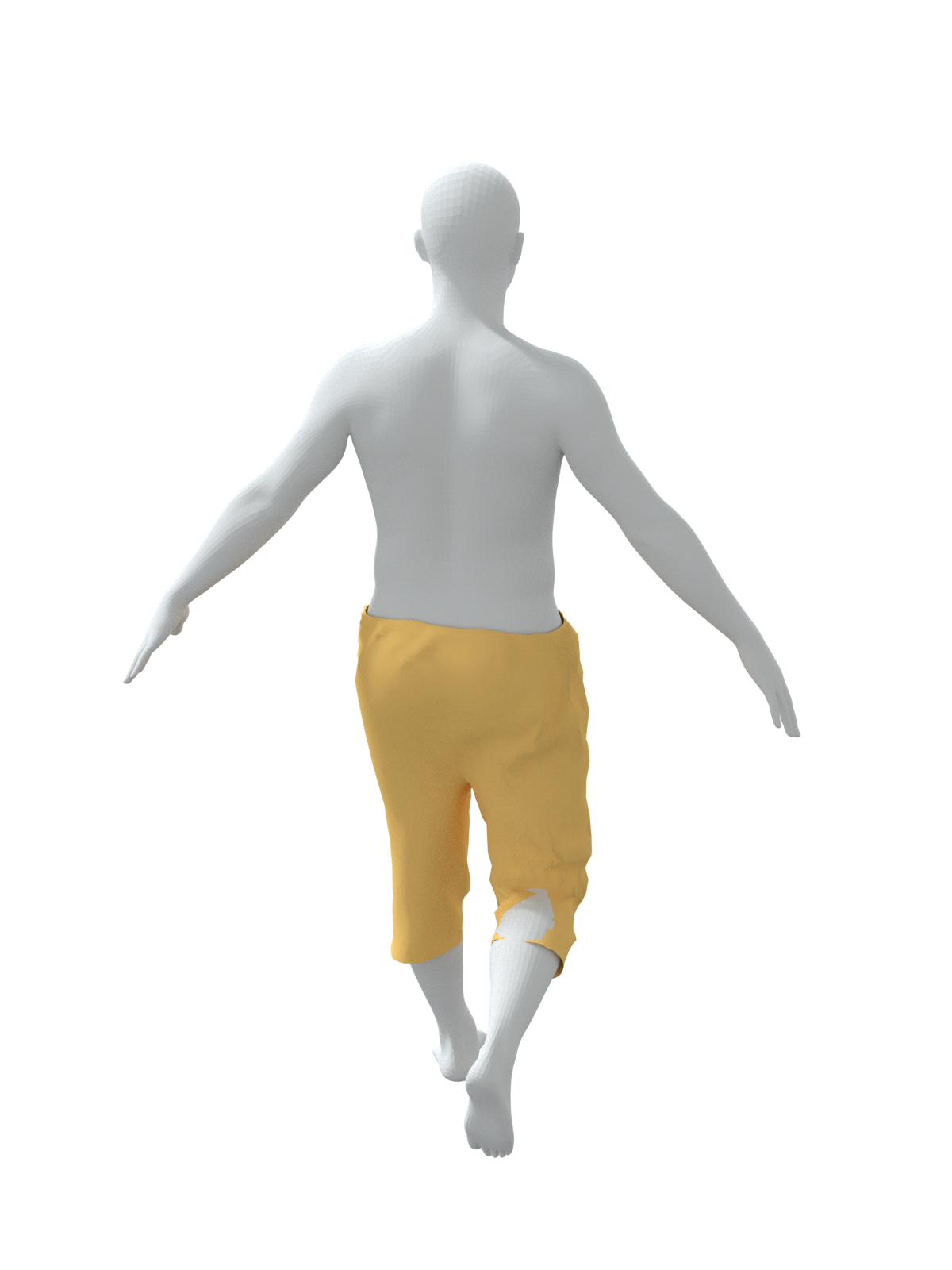}
\end{subfigure}
\hfill
\begin{subfigure}[b]{0.2\textwidth}
\centering
\includegraphics[width=\textwidth]{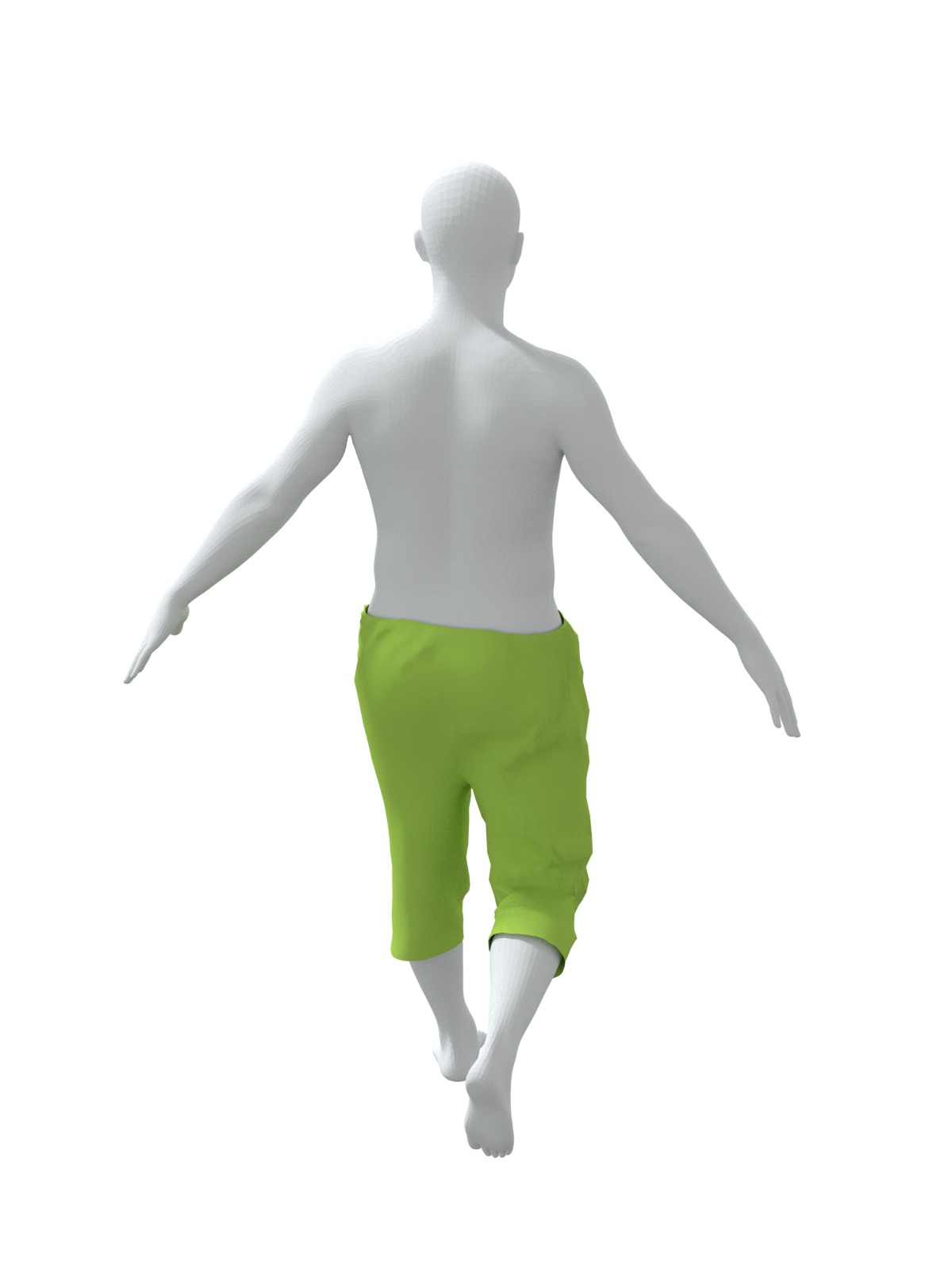}
\end{subfigure}
\begin{subfigure}[b]{0.2\textwidth}
\centering
\includegraphics[width=\textwidth]{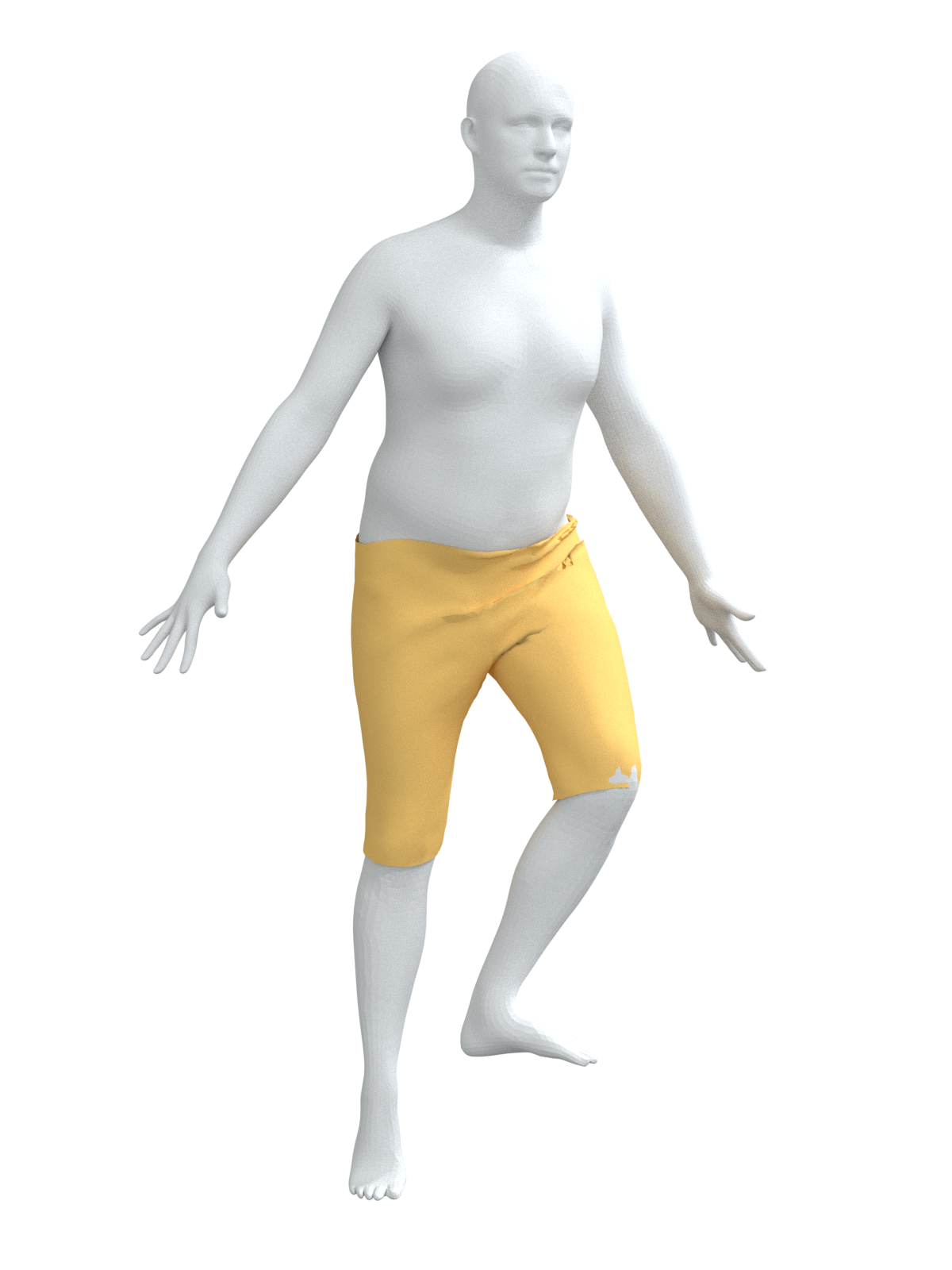}
\end{subfigure}
\hfill
\begin{subfigure}[b]{0.2\textwidth}
\centering
\includegraphics[width=\textwidth]{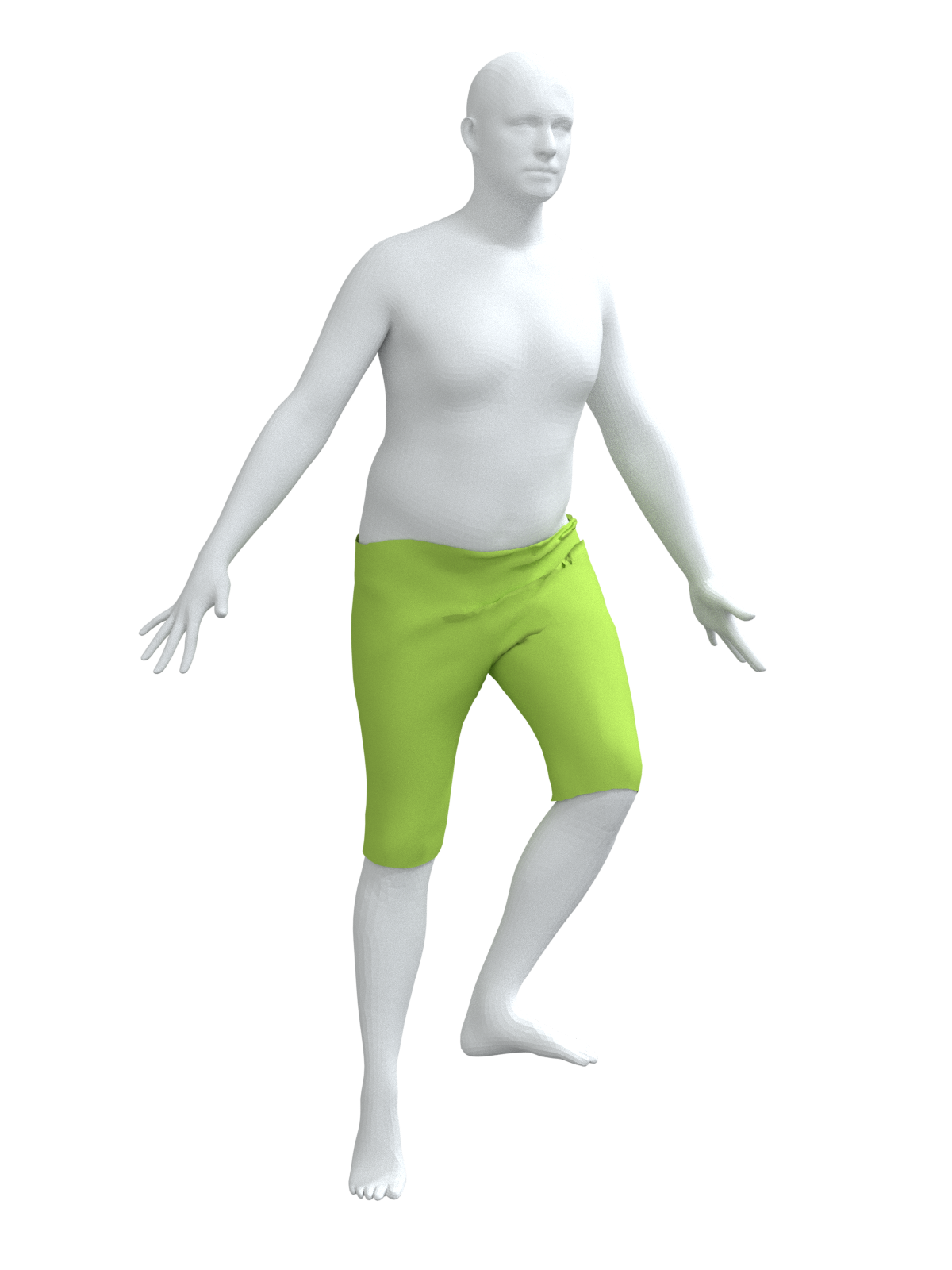}
\end{subfigure}
\hfill
\begin{subfigure}[b]{0.2\textwidth}
\centering
\includegraphics[width=\textwidth]{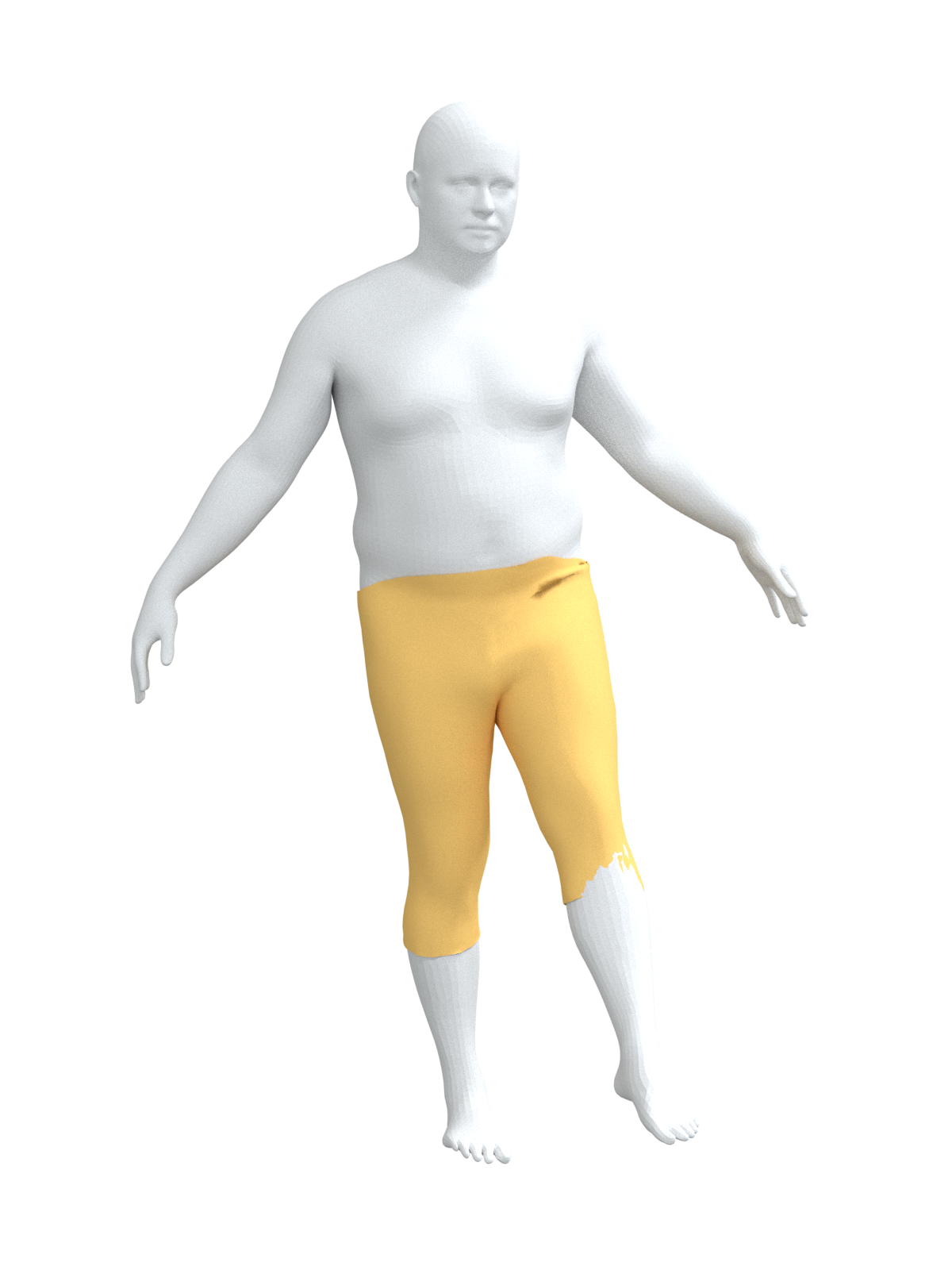}
\end{subfigure}
\hfill
\begin{subfigure}[b]{0.2\textwidth}
\centering
\includegraphics[width=\textwidth]{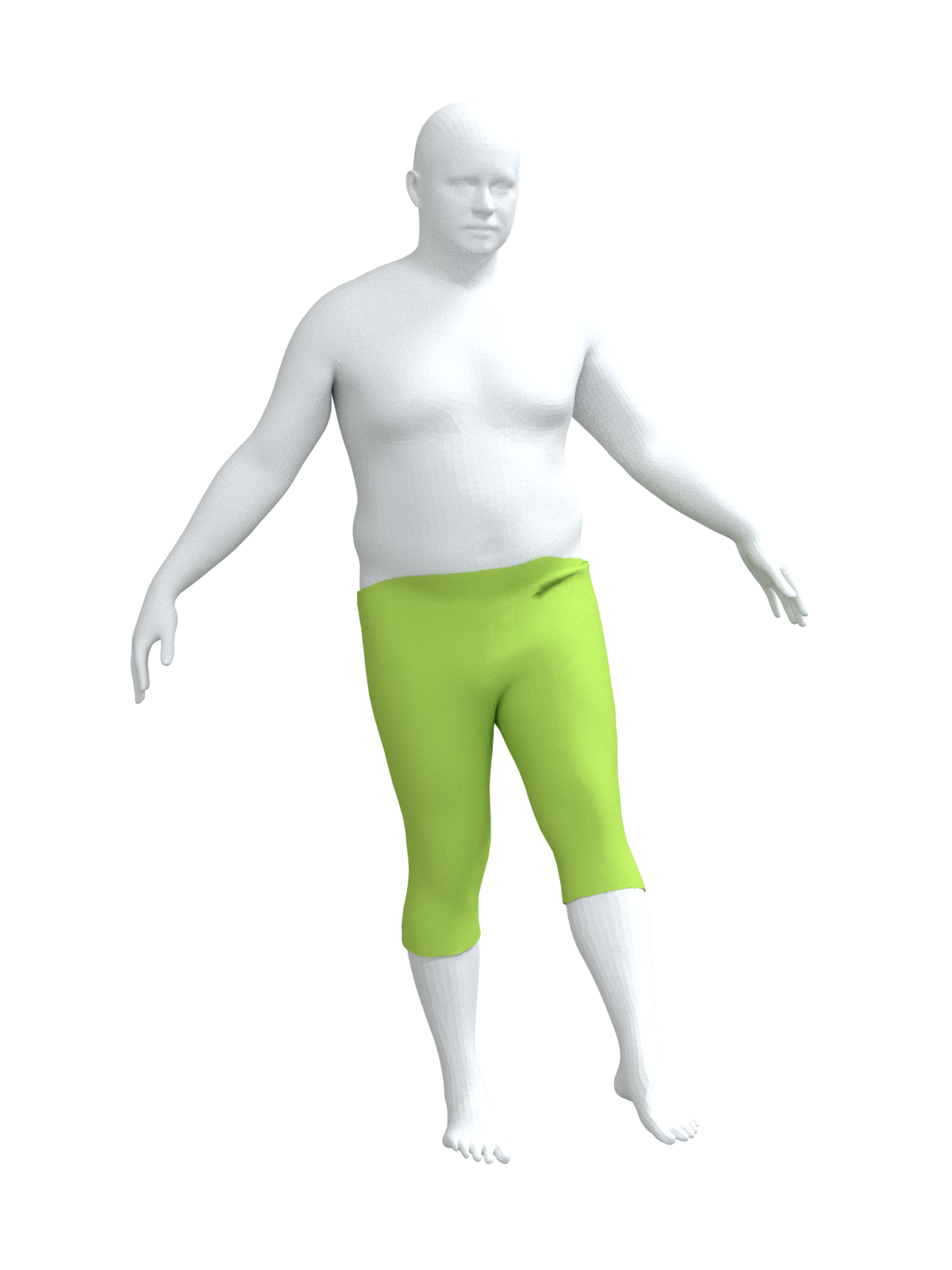}
\end{subfigure}
\begin{subfigure}[b]{0.2\textwidth}
\centering
\includegraphics[width=\textwidth]{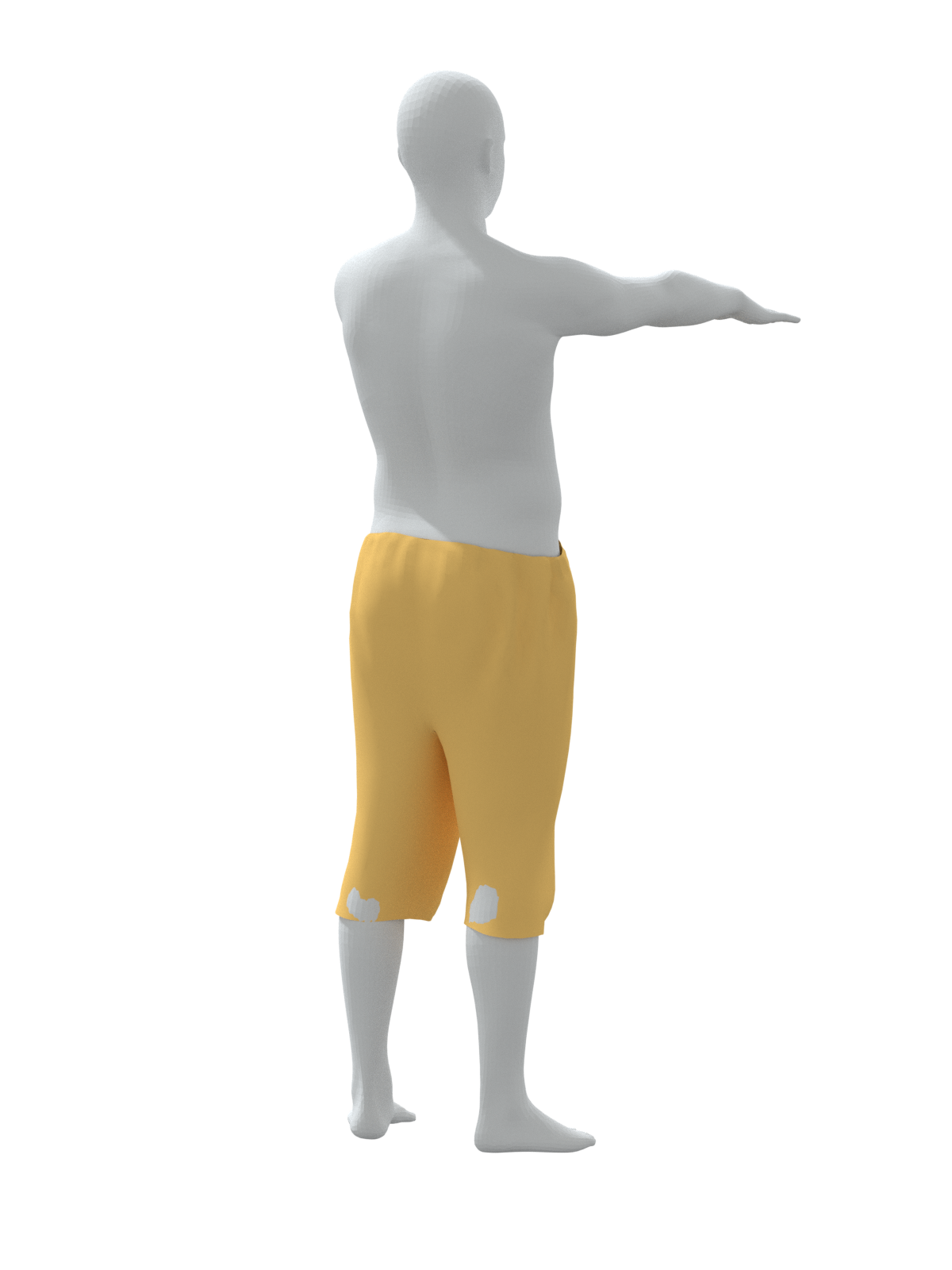}
\end{subfigure}
\hfill
\begin{subfigure}[b]{0.2\textwidth}
\centering
\includegraphics[width=\textwidth]{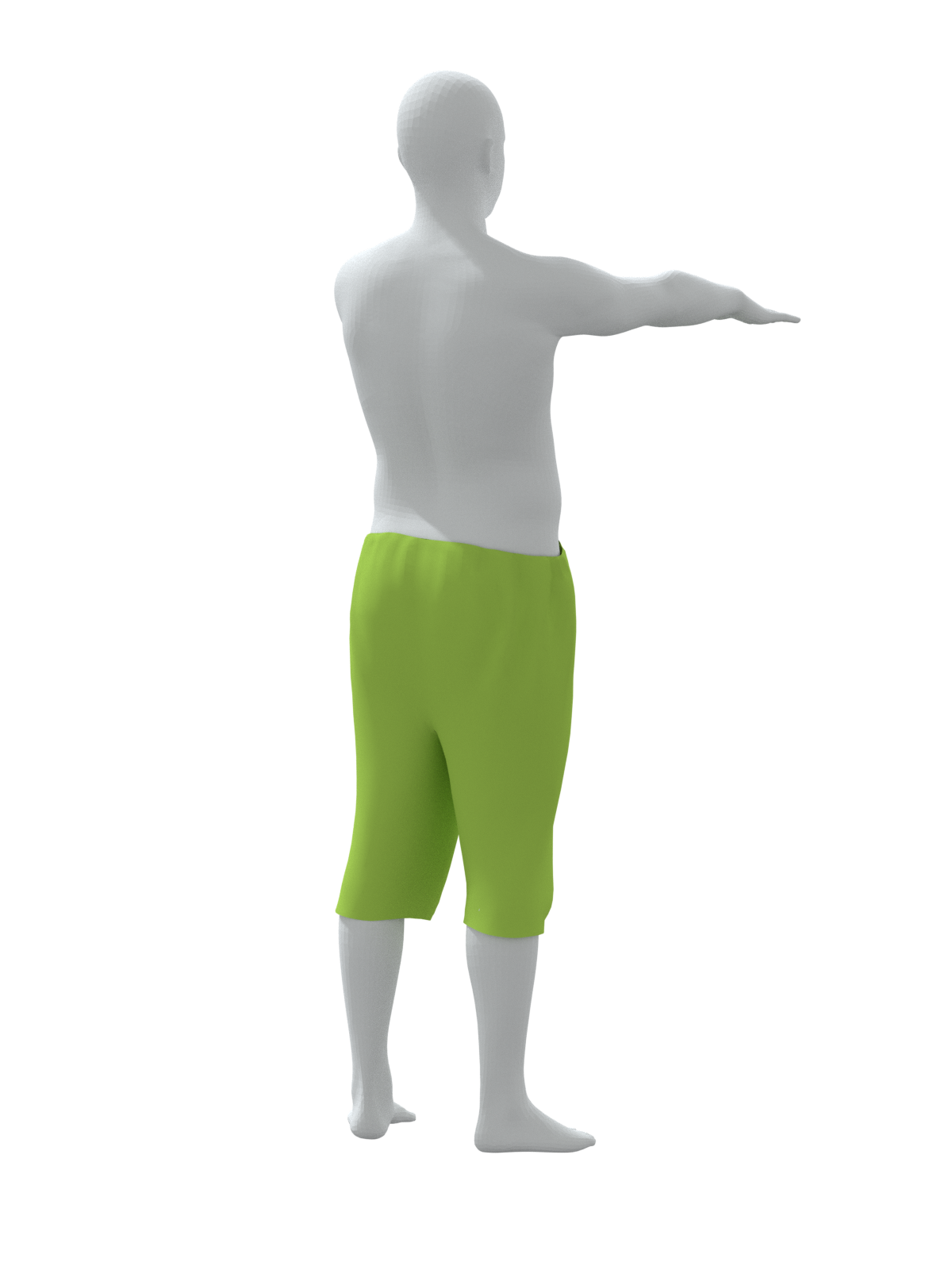}
\end{subfigure}
\hfill
\begin{subfigure}[b]{0.2\textwidth}
\centering
\includegraphics[width=\textwidth]{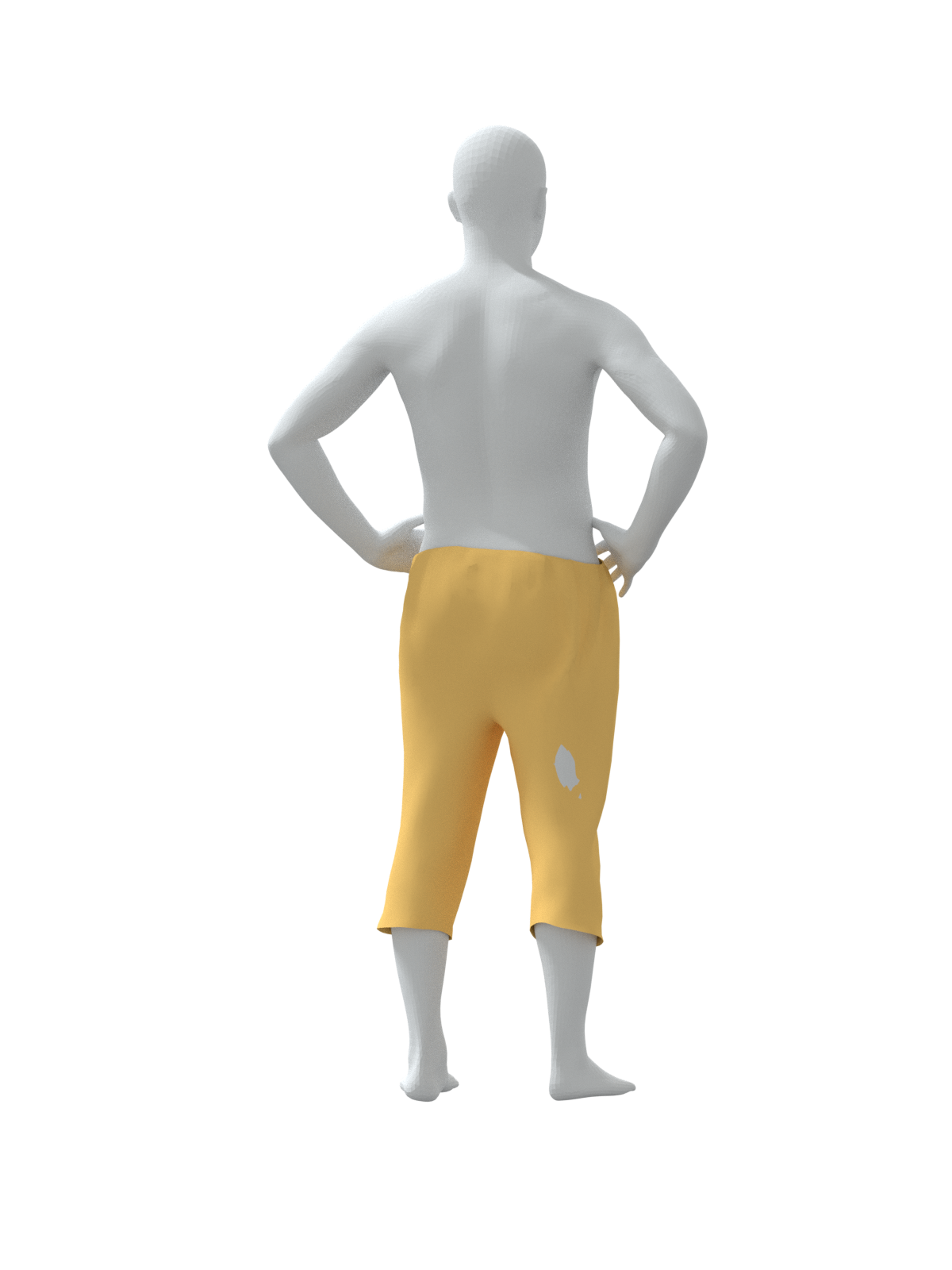}
\end{subfigure}
\hfill
\begin{subfigure}[b]{0.2\textwidth}
\centering
\includegraphics[width=\textwidth]{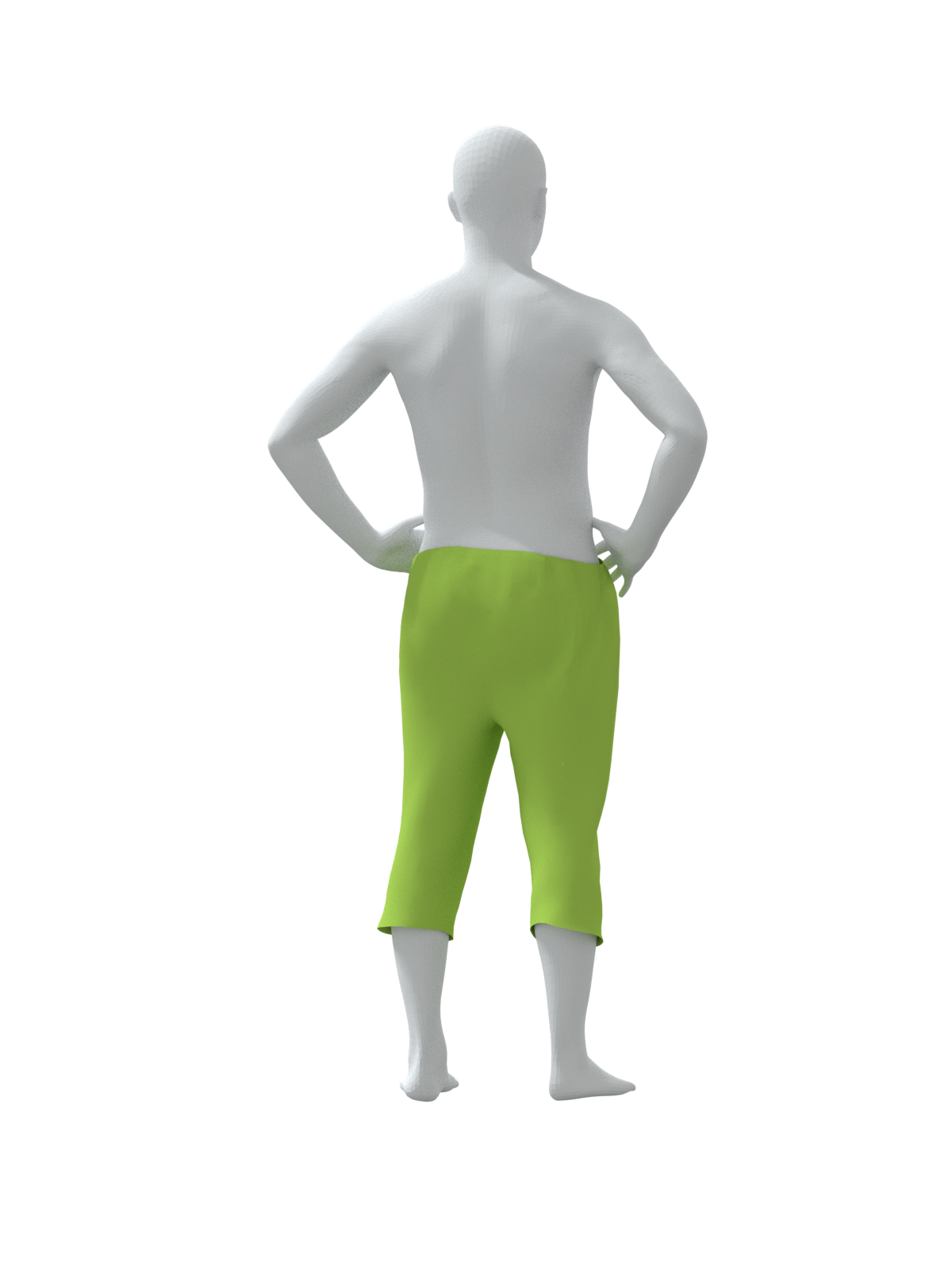}
\end{subfigure}
\caption{\label{fig:add_pant} Additional examples from Short-pant Male dataset, showing collisions resolved by applying our ReFU in TailorNet.}
\end{figure*}

\begin{figure*}
\centering
\begin{subfigure}[b]{0.2\textwidth}
\centering
\includegraphics[width=\textwidth]{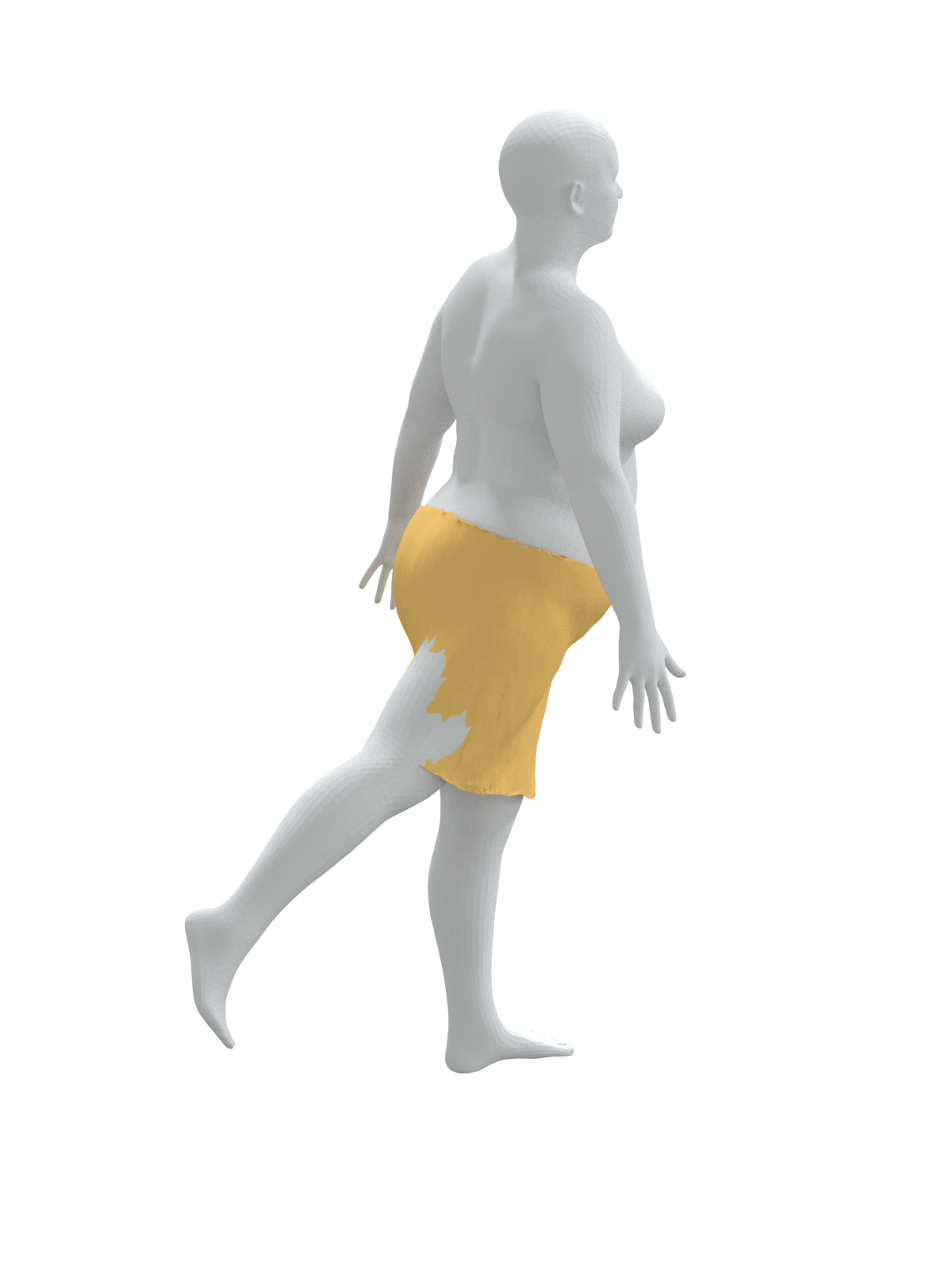}
\end{subfigure}
\hfill
\begin{subfigure}[b]{0.2\textwidth}
\centering
\includegraphics[width=\textwidth]{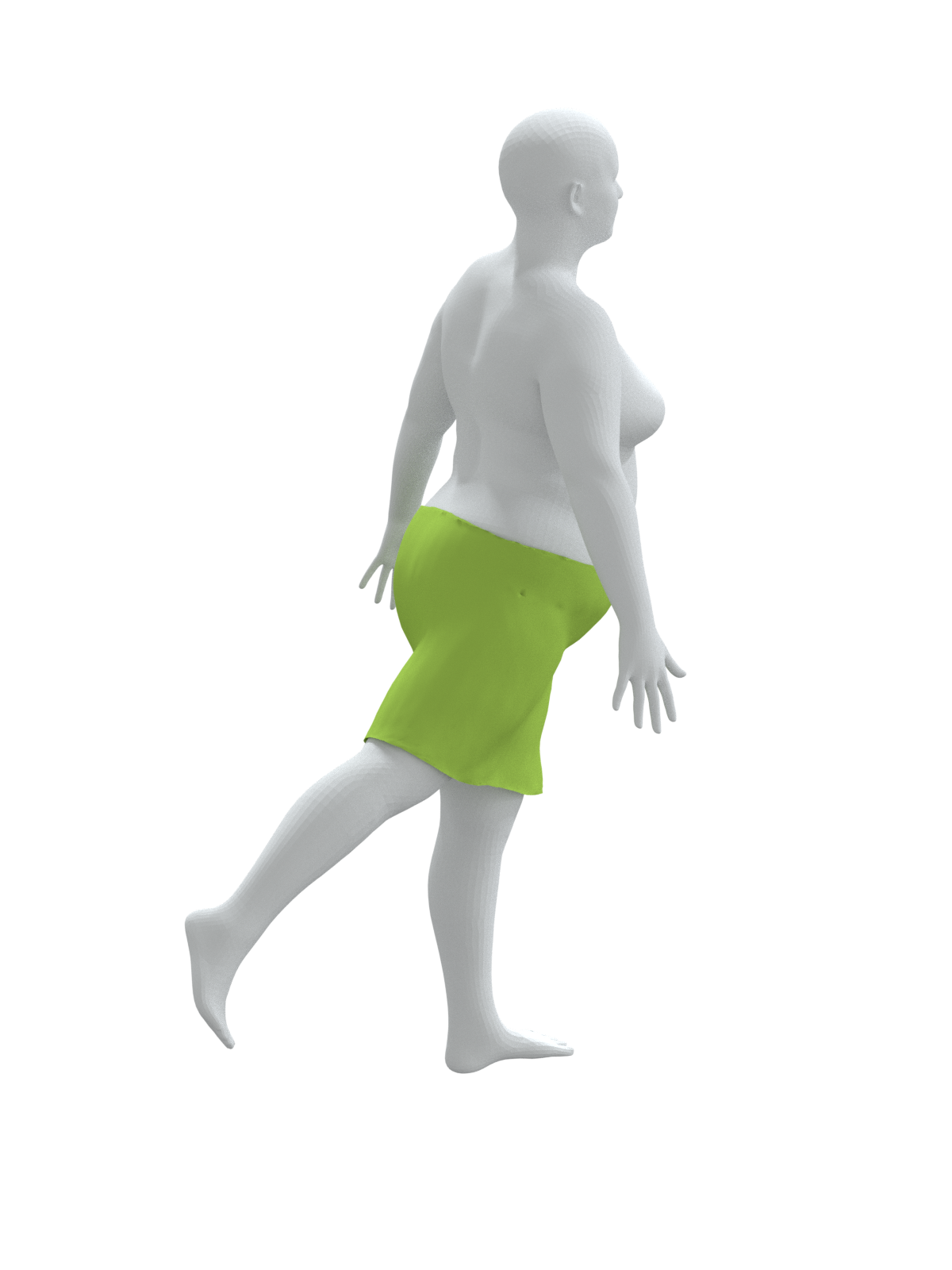}
\end{subfigure}
\hfill
\begin{subfigure}[b]{0.2\textwidth}
\centering
\includegraphics[width=\textwidth]{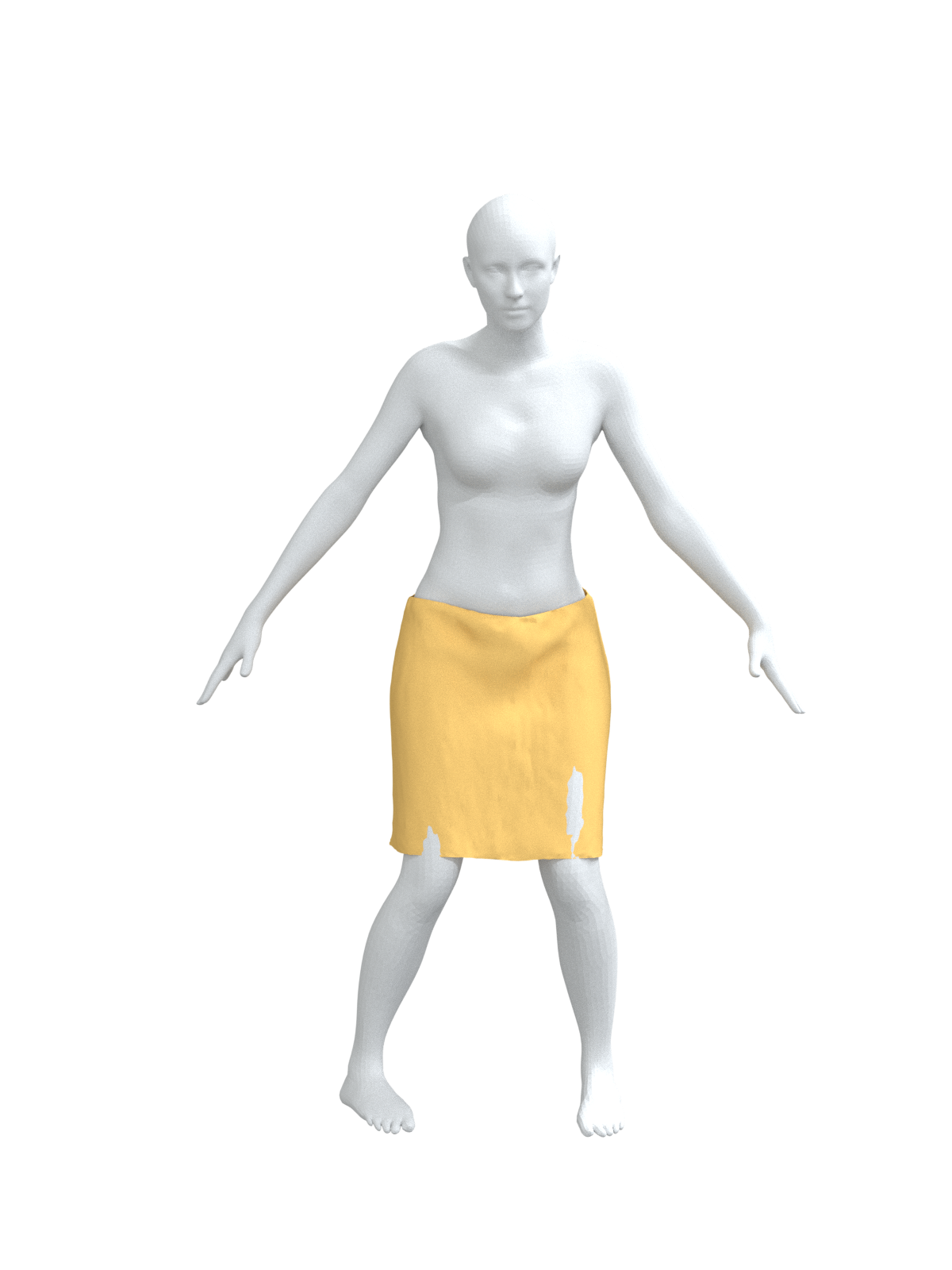}
\end{subfigure}
\hfill
\begin{subfigure}[b]{0.2\textwidth}
\centering
\includegraphics[width=\textwidth]{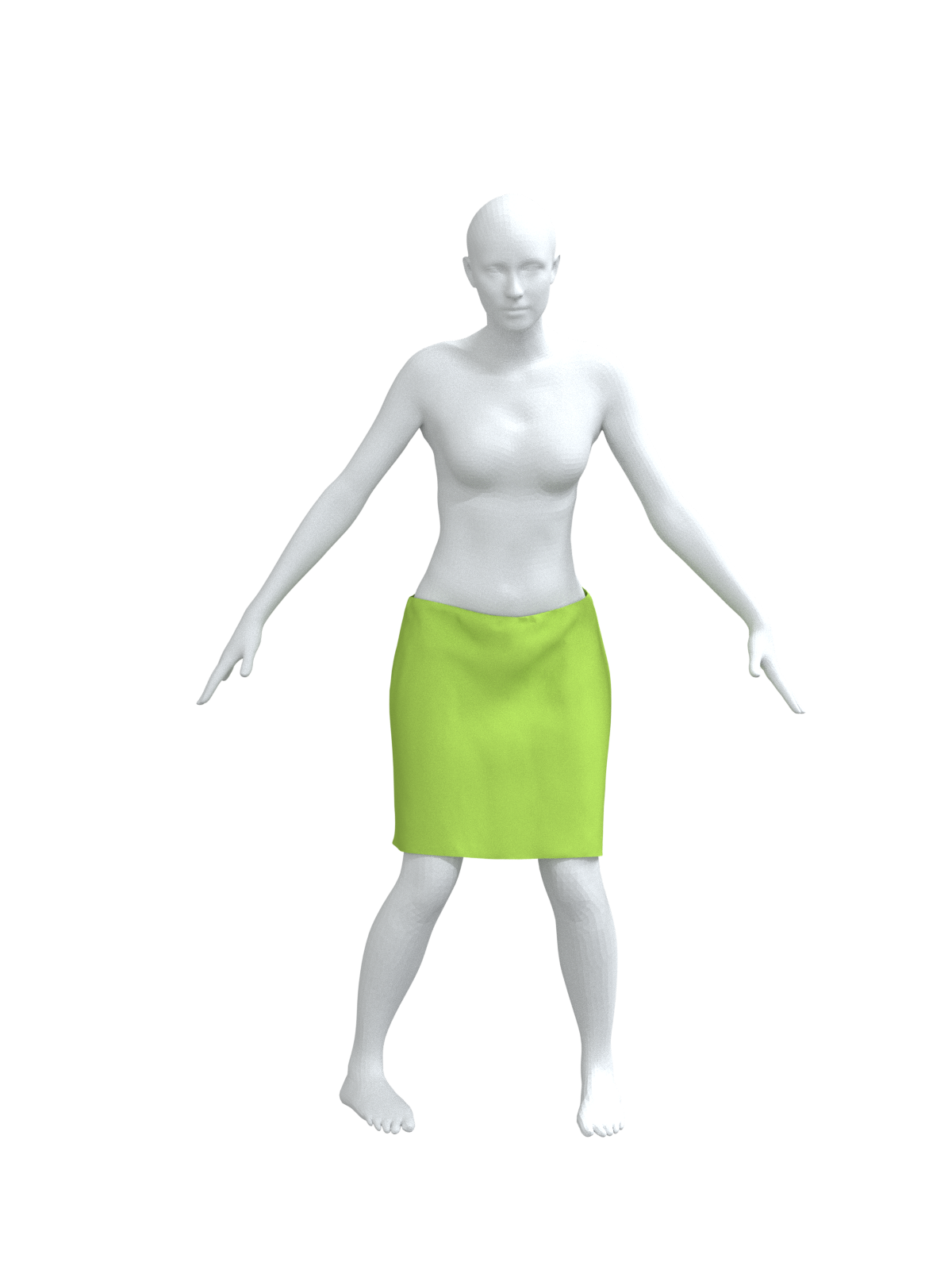}
\end{subfigure}
\begin{subfigure}[b]{0.2\textwidth}
\centering
\includegraphics[width=\textwidth]{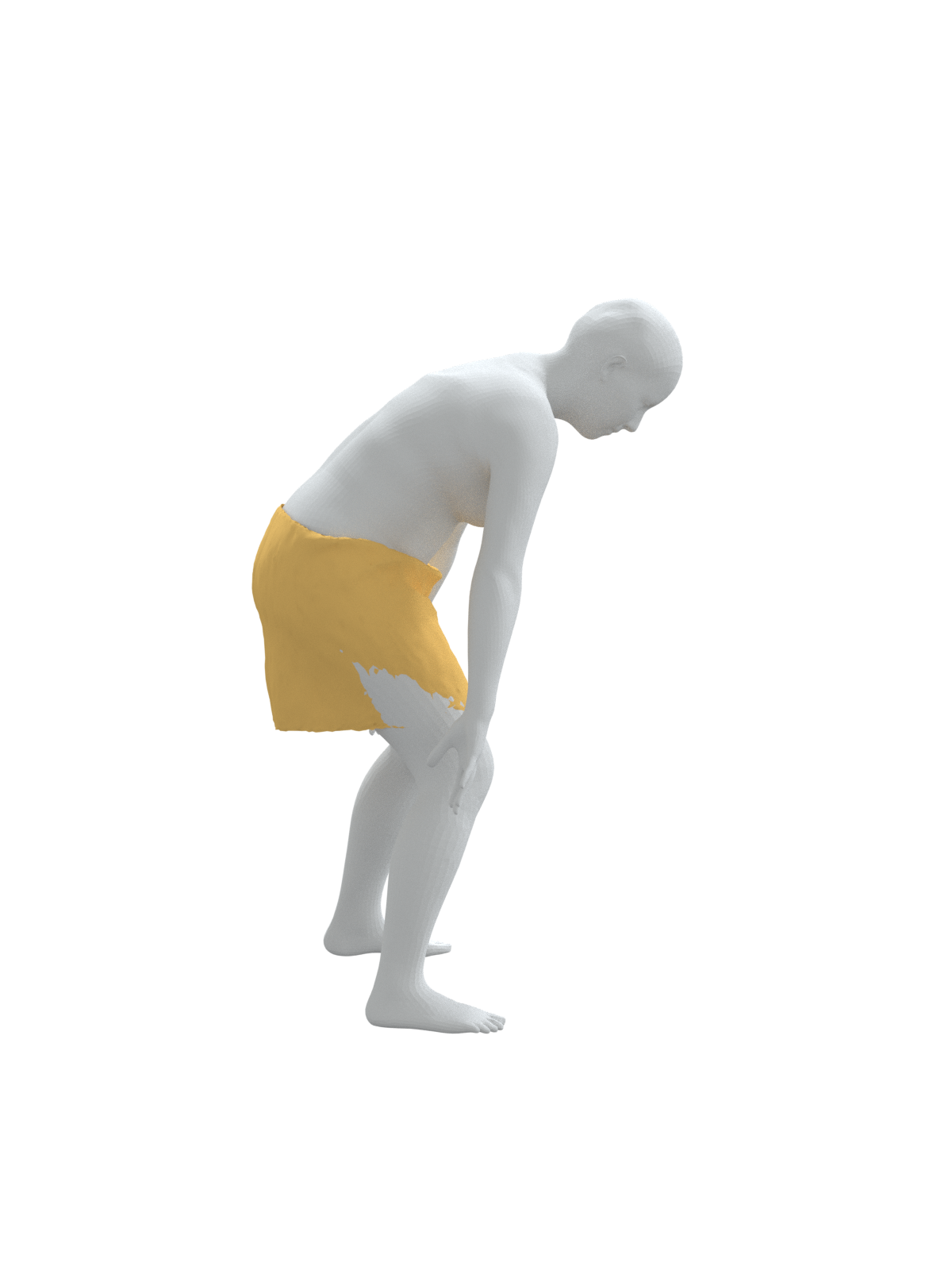}
\end{subfigure}
\hfill
\begin{subfigure}[b]{0.2\textwidth}
\centering
\includegraphics[width=\textwidth]{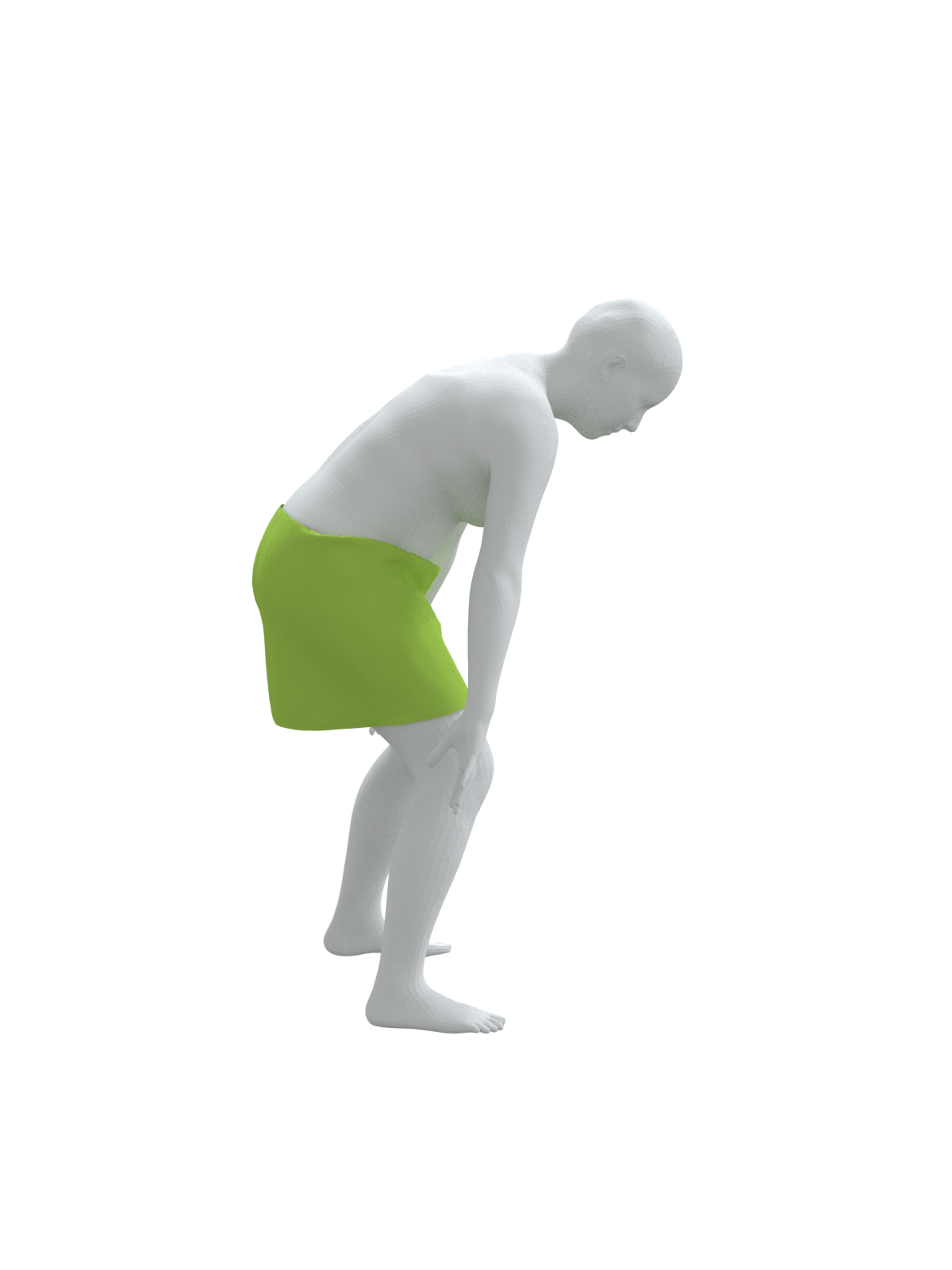}
\end{subfigure}
\hfill
\begin{subfigure}[b]{0.2\textwidth}
\centering
\includegraphics[width=\textwidth]{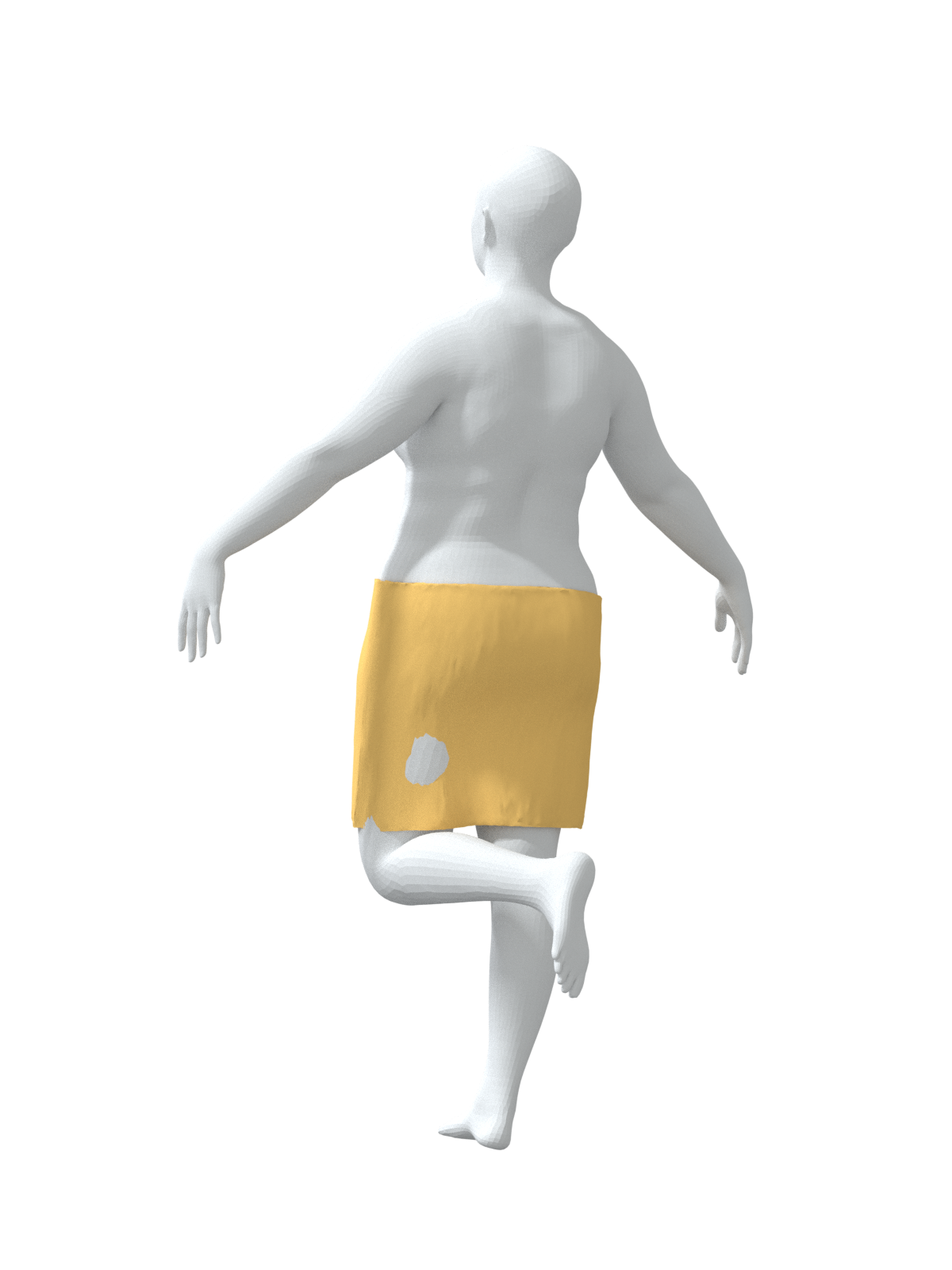}
\end{subfigure}
\hfill
\begin{subfigure}[b]{0.2\textwidth}
\centering
\includegraphics[width=\textwidth]{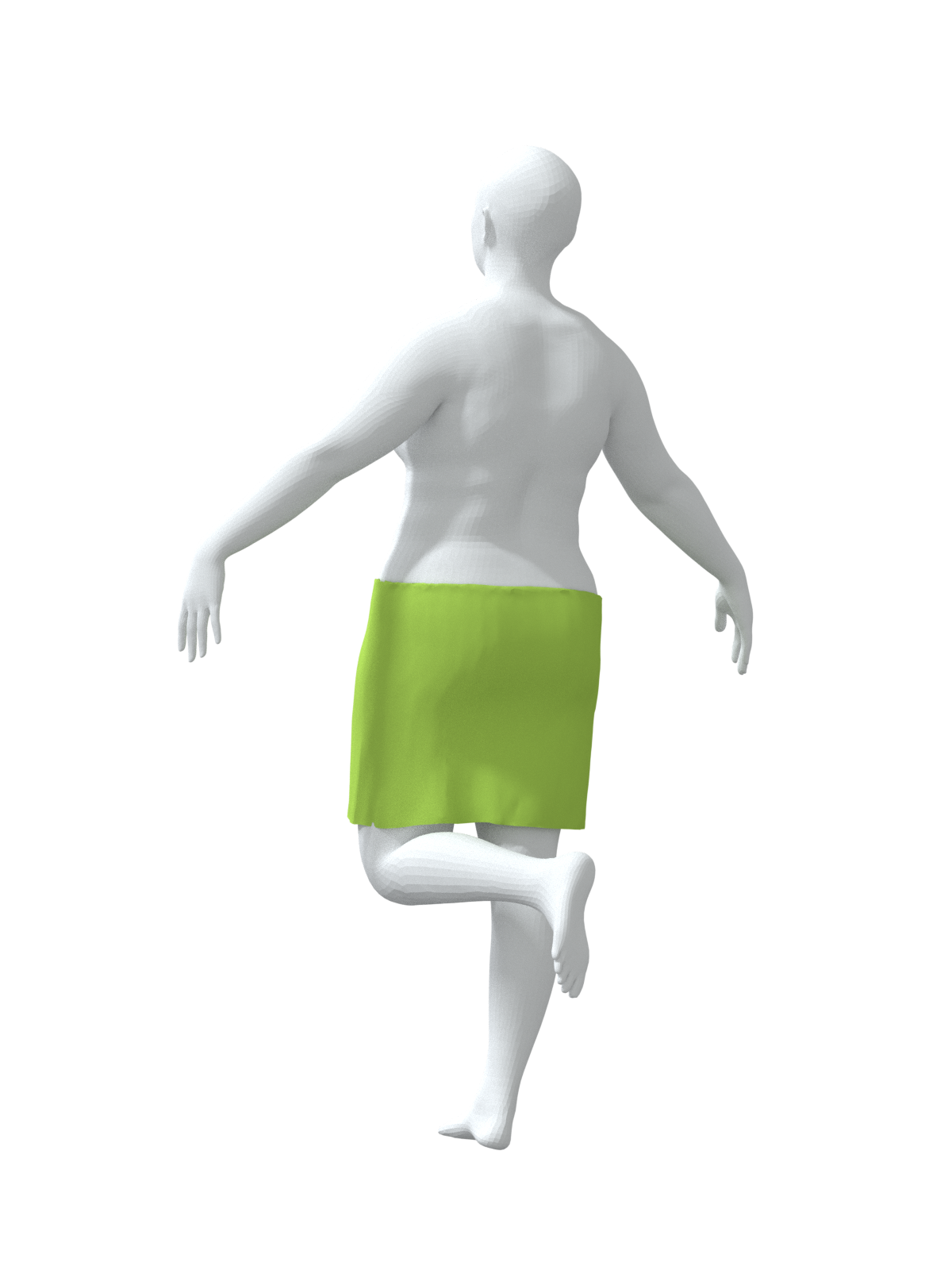}
\end{subfigure}
\begin{subfigure}[b]{0.2\textwidth}
\centering
\includegraphics[width=\textwidth]{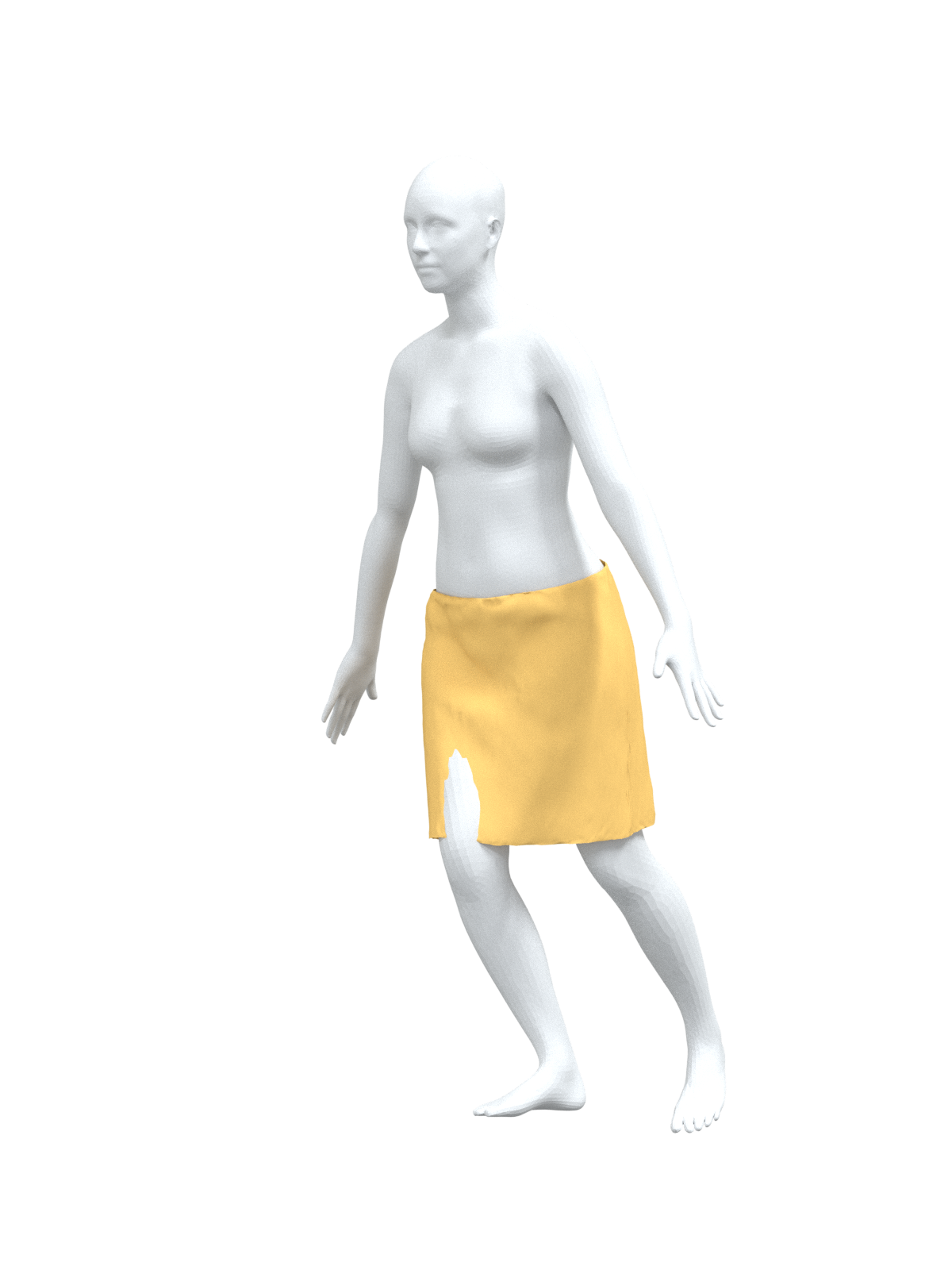}
\end{subfigure}
\hfill
\begin{subfigure}[b]{0.2\textwidth}
\centering
\includegraphics[width=\textwidth]{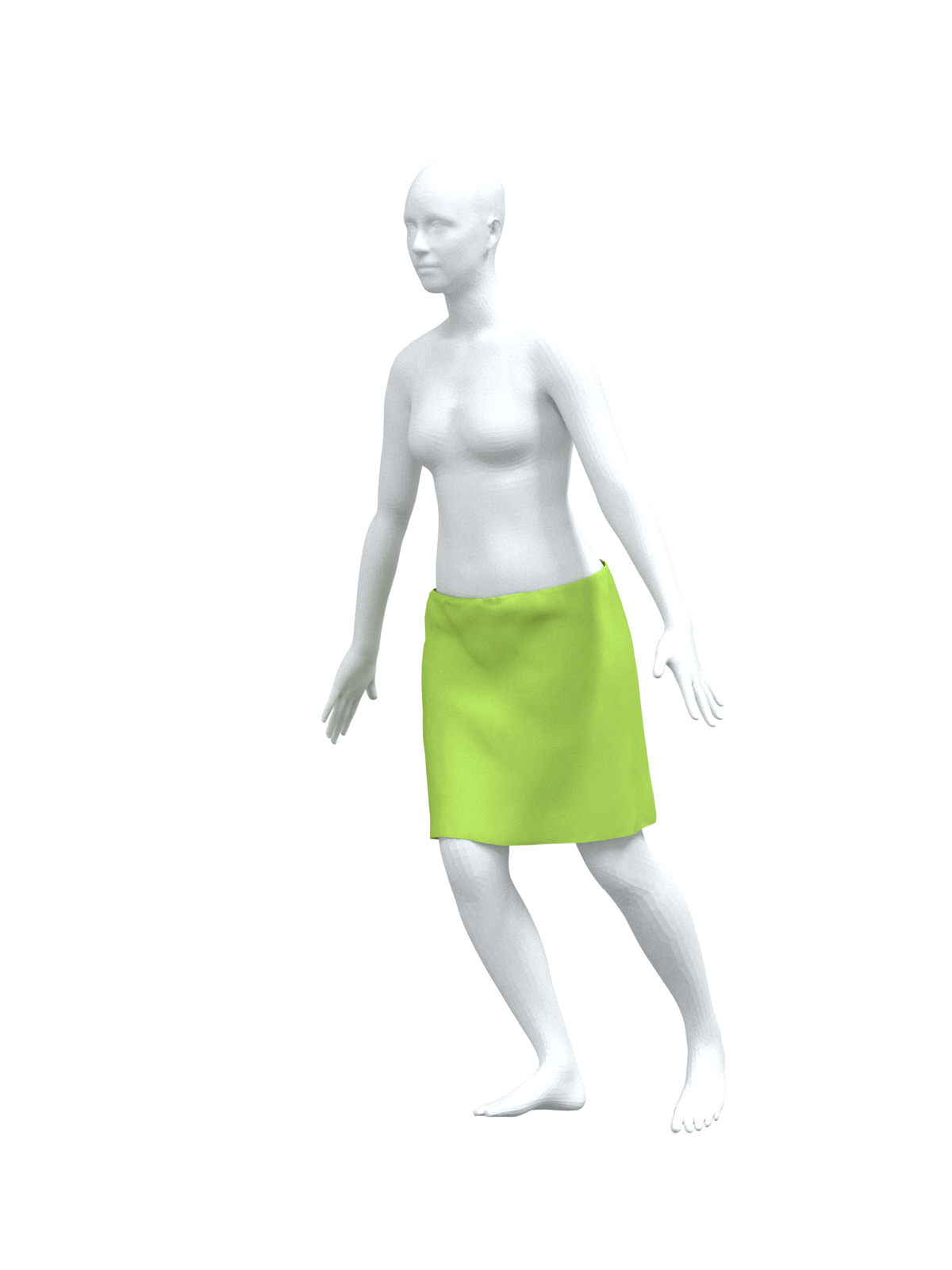}
\end{subfigure}
\hfill
\begin{subfigure}[b]{0.2\textwidth}
\centering
\includegraphics[width=\textwidth]{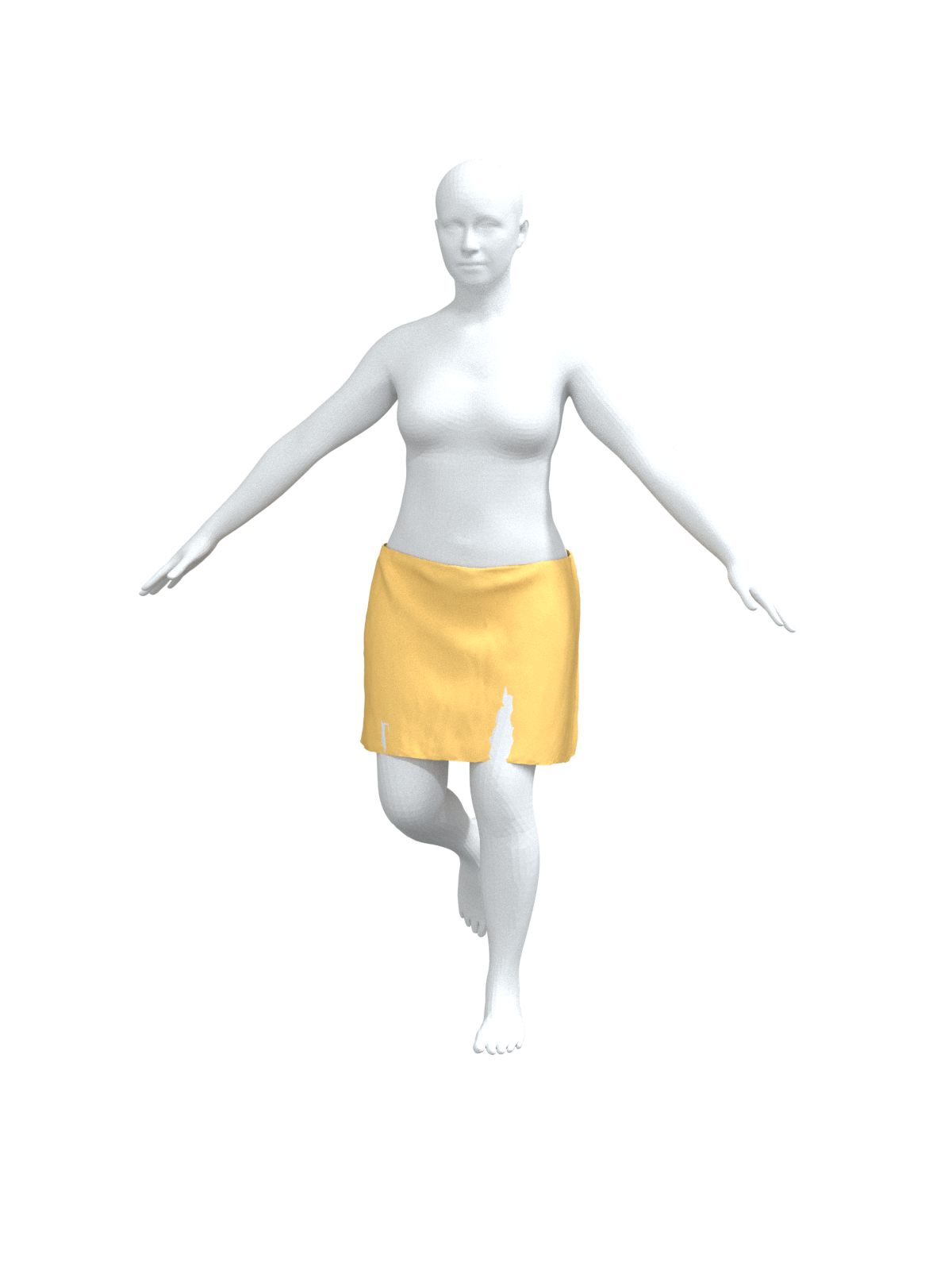}
\end{subfigure}
\hfill
\begin{subfigure}[b]{0.2\textwidth}
\centering
\includegraphics[width=\textwidth]{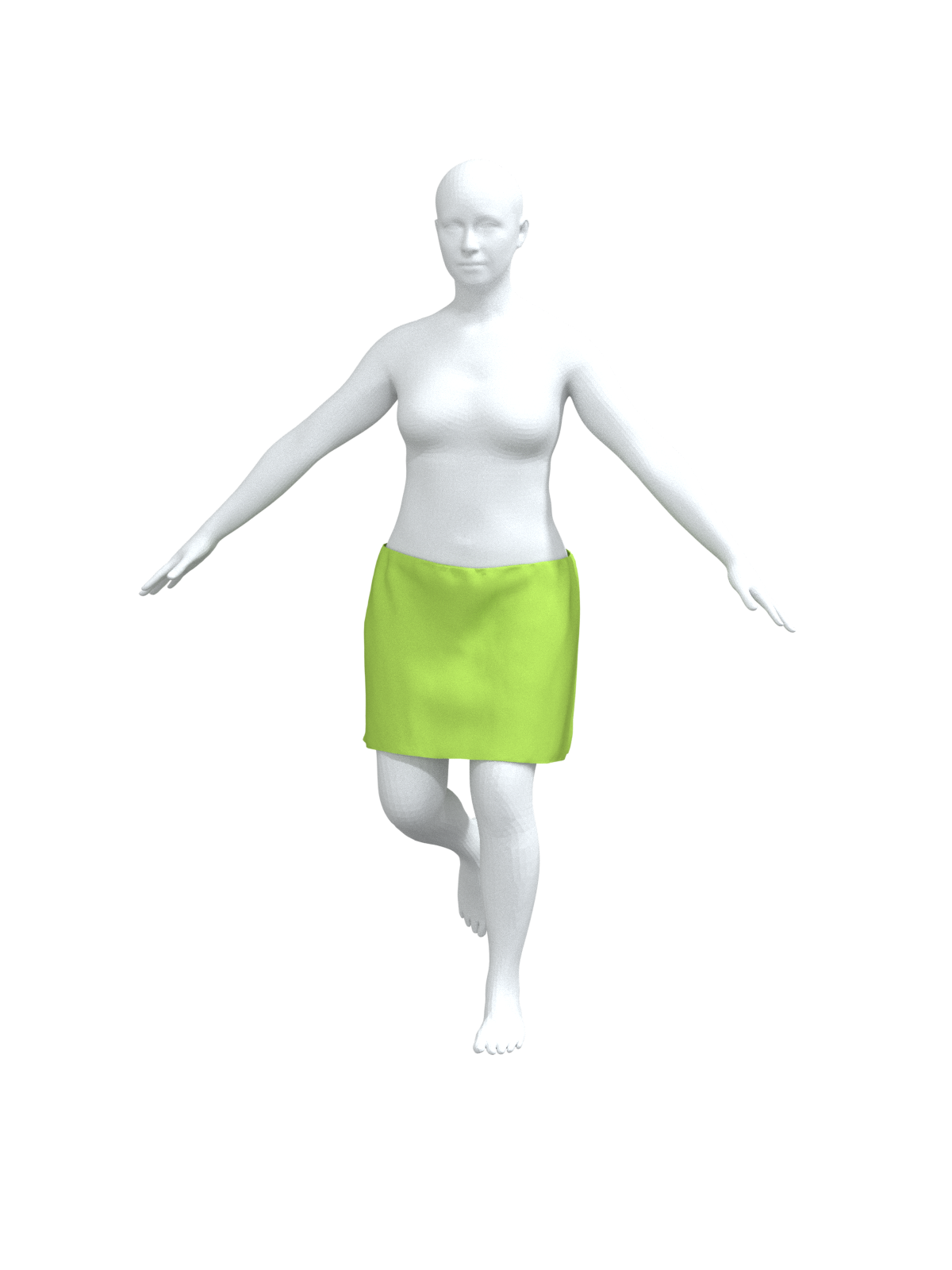}
\end{subfigure}
\caption{\label{fig:add_skirt} Additional examples from Skirt Female dataset, showing collisions resolved by applying our ReFU in TailorNet.}
\end{figure*}

We include more visulizations for the results generated with or without our ReFU layer, in \prettyref{fig:add_shirt}, \prettyref{fig:add_tshirt}, \prettyref{fig:add_pant} and \prettyref{fig:add_skirt}.

\end{document}